\documentclass[a4paper,fleqn,usenatbib]{mnras}
\usepackage{newtxtext,newtxmath}
\usepackage[T1]{fontenc}
\usepackage{ae,aecompl}

\usepackage{amssymb,amsmath,amsfonts}
\usepackage{color}
\usepackage{graphicx,tabularx}
\usepackage[english] {babel}
\usepackage{astron}
\usepackage{natbib}
\usepackage{pdflscape}

\usepackage[toc,page]{appendix}
\usepackage{placeins}
\usepackage{float}
\usepackage{footmisc}
\usepackage{url}
\usepackage{hyperref}

\title[Isolated DM-deprived galaxies in hydro-simulations]{Isolated dark matter deprived galaxies in hydrodynamical simulations: real objects or artefacts?}

\author[C. Saulder et al.]{
Christoph Saulder$^{1}$\thanks{E-mail: csaulder@kias.re.kr},
Owain Snaith$^{2}$,
Changbom Park$^{1}$,
and Clotilde Laigle$^{3}$\\
$^{1}$Korea Institute for Advanced Study, 85 Hoegiro, Dongdaemun-gu, 02455 Seoul, Republic of Korea\\
$^{2}$GEPI, Observatoire de Paris, Universit\'e PSL, CNRS, 5 Place Jules Janssen, 92190 Meudon, France \\
$^{3}$Department of Physics, University of Oxford, Parks Rd, Oxford OX1 3PJ, United Kingdom}
\date{Accepted XXX. Received YYY; in original form ZZZ}
\pubyear{2019}

\begin{document}
\label{firstpage}
\pagerange{\pageref{firstpage}--\pageref{lastpage}}
\maketitle

\begin{abstract}
We searched for isolated dark matter deprived galaxies within several state-of-the-art hydrodynamical simulations: Illustris, IllustrisTNG, EAGLE, and Horizon-AGN and found a handful of promising objects in all except Horizon-AGN. While our initial goal was to study their properties and evolution, we quickly noticed that all of them were located at the edge of their respective simulation boxes. After carefully investigating these objects using the full particle data, we concluded that they are not merely caused by a problem with the algorithm identifying bound structures. We provide strong evidence that these oddballs were created from regular galaxies that get torn apart due to unphysical processes when crossing the edge of the simulation box. We show that these objects are smoking guns indicating an issue with the implementation of the periodic boundary conditions of the particle data in Illustris, IllustrisTNG, and EAGLE, which was eventually traced down to be a minor bug occurring for a very rare set of conditions. 
\end{abstract}

\section{Introduction}
Large-scale hydrodynamical simulations that consider the gravitational interplay between cold dark matter and the baryonic physics of the gaseous and stellar components have been around for about a decade \citep{MareNostrum}. Over the last couple of years they have become a common tool for studying the evolution and properties of galaxies in a cosmological context. There currently is a multitude of hydrodynamical simulations and each of them addresses a wide range of scientific questions. Prominent large-scale simulations are MassiveBlack-II \citep{MassiveBlack2}, Magneticum Pathfinder \citep{Teklu:2015,Remus:2015}, Illustris \citep{Illustris}, EAGLE \citep{EAGLE_sim}, Horizon-AGN \citep{HorizonAGN}, and IllustrisTNG \citep{IllustrisTNG_1}. Additionally, there are several zoom-in simulations that take a closer look at individual galaxies an environmental context e.g.: NIHAO \citep{NIHAO}, Hydrangea \citep{Hydrangea}, and FIRE \citep{FIRE}. We preliminarily focus on simulations that provide easily accessible public data, such as Illustris \citep{Illustris} and EAGLE \citep{EAGLE_sim} or simulations to which we gained access via collaborations to IllustrisTNG \citep{IllustrisTNG_1}, which in the meantime became public as well as Horizon-AGN \citep{HorizonAGN}. These simulations aim to reproduce the overall distribution and appearance of the most prominent populations of galaxies. Within these large datasets, we searched for galaxies with unusual stellar-to-halo mass ratios. 

Within the last decade the stellar-to-halo mass relation \citep{Moster:2010} has become a topic of much discussion and research. Numerous papers studied its formation and evolution (e.g. \citet{Leauthaud:2012,Tinker:2012,Tinker:2013,Tinker:2017,Legrand:2018,Cowley:2019}), its morphological dependence \citep{Rodrigues:2015}, and in how far the simulations match the observations \citep{Moster:2013,Munshi:2013,Zu:2015,Shan:2017}. With modern hydrodynamical simulations, one can study the formation of the the stellar-to-halo mass relation through the distribution and redistribution of dark matter in different environments \citep{Niemiec:2017}. 

The topic of dark matter deprived galaxies gained some attention in the last year due to the observational discovery of dwarf galaxy without notable amount of dark matter \citep{vanDokkum:2018a,vanDokkum:2018b}. This caused some controversy about the mass estimates and distance measurement to this peculiar galaxy \citep{Blakeslee:2018,vanDokkum:2018c,Emsellem:2018,Fensch:2018,Trujillo:2019}. Recently, a second candidate for this type of galaxy was discovered in the same galaxy group \citep{vanDokkum:2019}. However in this paper, we will not focus on dwarf galaxies, but more massive objects that are properly resolved in current large-scale hydro-simulations. A couple larger, potentially dark matter deprived, galaxies have been found observationally by studying the kinematics of their halos \citep{Salinas:2012,Lane:2015} and strong-lensing studies of clusters \citep{Monna:2017}. Modern hydrodynamical simulations provide a possibility to study potential outliers of the stellar-to-halo mass relation and identify dark matter deprived galaxies. For non-central galaxies in clusters, processes of dark matter stripping have been studied in Illustris \citep{Niemiec:2018} and EAGLE \citep{Jing:2018}, which are able to create dark matter deprived galaxies. We will discuss massive non-central galaxies like this in separate paper (Saulder et al., in preparation), because we want to focus on even more exotic cases. Some of the highly dark matter deficient galaxies in Illustris \citep{Yu:2018} were found to be isolated objects. In this paper, we follow-up on this claim and search for similar objects in other hydrodynamical simulations as well. By studying the history of these galaxies (which we dubbed 'oddballs') and their environment, we initially wanted to understand if these objects could correspond to real objects that one might detect in surveys, \text{but they quickly turned out to be} just artefacts of the simulations. Hence, we focused on collecting all clues and evidence to aid in the identification of the issue causing them.

Since all the simulations have finite volumes (typically boxes) and particles may move beyond their initial limits, a method was developed to solve the problems arising from this, which is called periodic boundary conditions. Particles leaving the defined boundaries of the box on one side will appear entering the box from the opposite side. Also the forces implemented in the simulation will reach beyond the boundaries following the same principle. 

This paper is structured in the following way: in Section \ref{sec_data}, we present a description of the various datasets that we obtained from several different hydrodynamical simulations. Our methods of identifying oddballs are explained in Section \ref{sec_method}. We present the main results of our work in Section \ref{sec_results} and discuss their implications in Section \ref{sec_discussion}. A brief summary and conclusions are provided in Section \ref{sec_summary}. Since the figures required to illustrate the environment, particle distribution, and evolution of the individual oddballs consume a lot of space, we separated them from the main body of the paper and placed them in Appendix \ref{app_plots}.

\section{Data}
\label{sec_data}
\begin{table*}
\begin{center}
\begin{tabular}{c|cccccc|ccccc}
Simulation & $\Omega_{\textrm{m}}$ & $\Omega_{\textrm{b}}$ & $\Omega_{\Lambda}$ & $H_{0}$& $\sigma_{8}$ & $n_{s}$ & $L_{\textrm{box}}$ & N$_{\textrm{dm}}$ & N$_{\textrm{snap}}$ & $M_{\textrm{dm}}$ & $M_{\textrm{gas,init}}$    \\ \hline
Illustris-1 & 0.2726 & 0.0456 & 0.7274 & 70.4 & 0.809 & 0.963 & 106.5 & 1820$^{3}$ & 134 & $6.3 \cdot 10^6$ & $1.3 \cdot 10^6$ \\
IllustrisTNG100-1 & 0.3089 & 0.0486 & 0.6911 & 67.74 & 0.8159 & 0.9667 & 110.7 & 1820$^{3}$ & 100 & $7.5 \cdot 10^6$ & $1.4 \cdot 10^6$ \\
EAGLE-RefL0100N1504 & 0.307 & 0.04825 & 0.693 & 67.77 & 0.8288 & 0.9611 & 100 & 1504$^{3}$ & 29 & $1.81 \cdot 10^6$ & $9.70 \cdot 10^6$ \\
Horizon-AGN & 0.272 & 0.045 & 0.728 & 70.2 & 0.81 & 0.967 & 142 & 1024$^{3}$ & 56 & $8 \cdot 10^7$ & $1 \cdot 10^7$
\end{tabular}
\end{center}
\caption{Basic parameters of the simulations used for this project. Column 1: name of the specific simulation run; column 2: relative matter (dark + baryonic) density; column 3: relative baryonic matter density; column 4: relative dark energy density; column 5: present-day Hubble parameter in km s$^{-1}$ Mpc$^{-1}$; column 6:  amplitude of the (linear) power spectrum on the scale of 8 $h^{-1}$ Mpc; column 7: primordial spectral index of scalar fluctuations; column 8: side length of the simulation box in co-moving Mpc at z=0 snapshot; column 9: number of dark matter particles; column 10: number of snapshots; column 11: mass of the dark matter particles in $\textrm{M}_{\odot}$; column 12: initial mass of the gas cells in $\textrm{M}_{\odot}$.}
\label{sim_basics}
\end{table*} 
Taking advantage of the growing number of large-scale hydrodynamical simulations, we obtained data from Illustris, IllustrisTNG, EAGLE, and Horizon-AGN. These simulations cover volumes (cubes) with side-lengths of the order of 100 Mpc and for our application, we gathered the corresponding cubes with the highest resolution available from them and searched for dark matter deprived central galaxies within these cubes. 

\subsection{Illustris}
The Illustris project \citep{Illustris_nature,Illustris,Torrey:2015} provides a suite of hydrodynamical simulations using the moving-mesh code {\sc AREPO} \citep{Springel:2010}. The simulations consider gas cooling, a subresolution interstellar medium  model, stochastic star-formation, stellar evolution, gas recycling, chemical enrichment, kinetic stellar feedback driven by supernovae, procedures for supermassive black hole (SMBH) seeding, SMBH accretion and SMBH merging, and related AGN feedback on top of gravitationally interacting dark matter. It uses the cosmological parameters of WMAP-9 \citep{WMAP_9}. Hydrodynamical simulations with three different resolutions were calculated as well as three complementary dark matter only simulations with the same initial conditions. In this paper, we focus only on Illustris-1, which was the highest resolution hydro-simulation, for which the general specifications are listed in Table \ref{sim_basics}. At redshift zero, Illustris-1 resolves gravitational dynamics down to about 710 pc and the cells resolving gas hydrodynamics and baryonic processes can be as small as 48 pc. In post-processing, a friends-of-friends (FOF) algorithm \citep{Davis:1985} was used to identify dark matter halos from the particle data. A version of the {\sc SUBFIND} algorithm \citep{Springel:2001,Dolag:2009}, adapted for hydro-simulations, was applied to identify gravitationally bound structures within them. The merger trees were constructed using different methods, but we used the ones built by the {\sc SubLink} code \citep{Rodriguez:2015}. 

For the Illustris-1 box, we obtained the entire group catalogue and the full particle data for the redshift zero snapshot. Additionally, we accessed the merger-trees for the oddballs (see Section \ref{sec_method} for an exact definition) found in this dataset. Furthermore, we also used the group catalogue and full particle data of snapshot 100 (corresponding to a redshift of 0.58) of the same box, which we used to investigate the evolution of these objects more closely. To ensure good particle sampling that would allow us to derive the overall properties of these galaxies with sufficient quality, we restrict our sample to galaxies containing a stellar mass of at least $10^{9.5} \textrm{M}_{\odot}$ (limits of the same order of magnitude were previously employed by other projects \citep{vdSande:2019,Thob:2019}), which limited us to 14 902 galaxies from the Illustris-1 group catalogue. 

\subsection{IllustrisTNG}
Following-up on the the Illustris project \citep{Illustris_nature,Illustris,Genel:2014,Nelson:2015,Sijacki:2015}, the Illustris-TNG simulations \citep{IllustrisTNG_public,IllustrisTNG_1,Naiman:2018,Pillepich:2018,Springel:2018,Genel:2018,IllustrisTNG_DMfractions} were performed using the Planck-2015 cosmological parameters \citep{Planck_cosmo} with enhanced numerical and astrophysical modelling and improving the identified shortcomings of the previous project. Aside from a variety of updated models, which are described in \citet{Weinberger:2017,IllustrisTNG_1}, idealized magnetohydrodynamics \citep{Pakmor:2016} were added. The Illustris-TNG project produced three complementay sets (within each set there are runs with different resolutions and complementary dark-matter only runs) of simulations TNG50 \citep{IllustrisTNG_50_star,IllustrisTNG_50_bh}, TNG100, and TNG300, which are cubes with side-length $\sim50$, $\sim100$, and $\sim300$ Mpc, respectively. We focused on the highest resolution box of the TNG100 with its parameters provided in Table \ref{sim_basics}. The group catalogue was constructed using {\sc SUBFIND} and the merger trees were connected using the {\sc SubLink} code. 

We obtained the redshift zero group catalogue and particle data for the Illustris TNG100-1 box, as well as merger-trees for the oddballs within the simulation. Additionally, we acquired the group catalogue and particle data for the snapshot 66 (corresponding to a redshift of 0.52) of the same box, which we used to investigate the evolution of these objects more closely. Using the same stellar mass limits as Illustris, we selected 12 501 galaxies for our sample. 

\subsection{EAGLE}
EAGLE \citep{EAGLE_sim,EAGLE_public,EAGLE_particles} is a suite of hydrodynamical simulations that were carried out using a modified version of the N-BodyTree-PM smoothed particle hydrodynamics code {\sc gadget3} \citep{GADGET}. The subgrid physics that were implemented in the code are radiative cooling, star formation, stellar mass loss, energy feed-back from star formation, gas accretion onto SMBH and mergers of SMBH, and AGN feedback and were tested in various precursor projects \citep{Crain:2009,Schaye:2010,LeBrun:2014}. EAGLE assumed the cosmological parameters of Planck-2013 \citep{Planck_cosmo_old} and its largest volume simulations, which is the one we focused on, is a cube with a side length of 100 Mpc with parameters listed in Table \ref{sim_basics}. As with Illustris, the {\sc SUBFIND} algorithm was used to detect bound structures, however merger trees were constructed using the {\sc D-Trees} algorithm \citep{Jiang:2014}. 

For this project, we made use of the group catalogue of snapshots five to 28 (corresponding to redshifts 7.05 to zero) of the RefL0100N1504 run of the EAGLE simulation. Additionally, we limited our sample to galaxies more massive than $10^{9.5} \textrm{M}_{\odot}$ in stellar mass at the present-day snapshot, which yielded 7 313 galaxies. Furthermore, we used the full particle data of the present-day snapshot (number 28) as well as snapshot 18 (corresponding to a redshift of 1.26, at which was before the oddballs (except for one) in EAGLE were formed) for our analysis. 

\subsection{Horizon-AGN}
The Horizon-AGN simulation \citep{HorizonAGN,Dubois:2016,Kaviraj:2017} was performed using the adaptive mesh refinement Eulerian hydrodynamics code {\sc RAMSES} \citep{RAMSES}. Horizon-AGN considers gas heating from an uniform UV background, gas cooling, star formation, feedback from stellar winds, supernovae Type Ia, and TypeII, and formation and growth of black holes as well as heating and jets caused by them. The simulation assumed the cosmological parameters of WMAP-7 \citep{WMAP_7} and its basic parameters are provided in Table \ref{sim_basics}. In contrast to the previous simulations, Horizon-AGN did not use the {\sc SUBFIND} algorithm to detect bound structures, but the {\sc ADAPTAHOP} halo finder \citep{Aubert:2004,Tweed:2009} instead. 

For our project, we made use of the last snapshot of the Horizon-AGN simulation and applied the same mass cut as for the previous simulations to obtain a sample of 48 631 galaxies. This number is notably larger than for the other simulations, which is one hand due to more than a factor of two larger volume and on the other hand due to a larger number of moderately massive galaxies. Since we did not find any oddballs within this sample, no additional datasets were required for the subsequent analysis. 

\section{Method}
We define the oddballs as clear outliers of the (inverted) stellar-to-halo mass relation with stellar masses beyond $10^{9.5} \textrm{M}_{\odot}$ that are located in the centre of their group. 

\label{sec_method}
\subsection{Identification of oddball galaxies}
\begin{figure*}
\begin{center}
\includegraphics[width=0.45\textwidth]{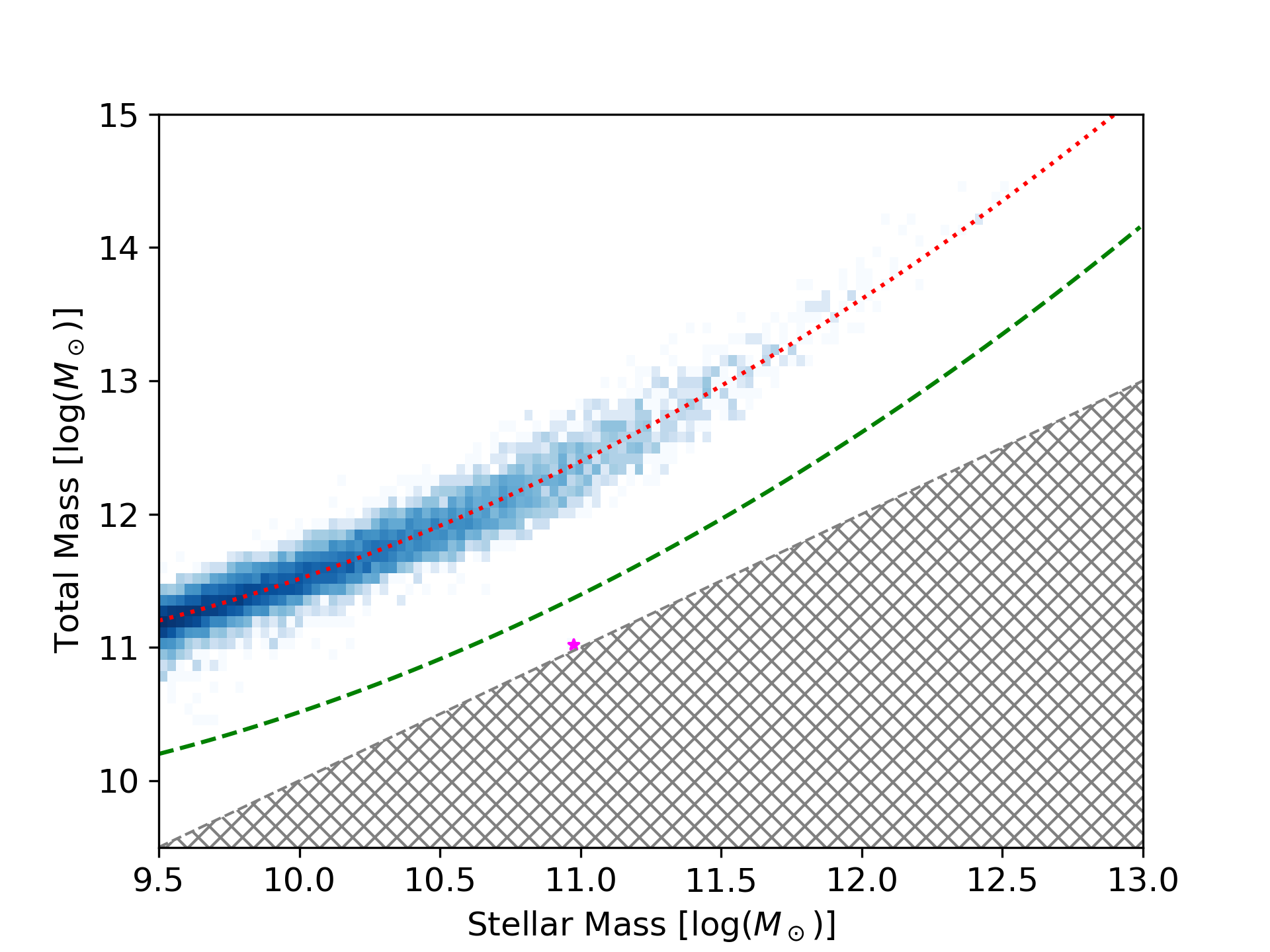}\includegraphics[width=0.45\textwidth]{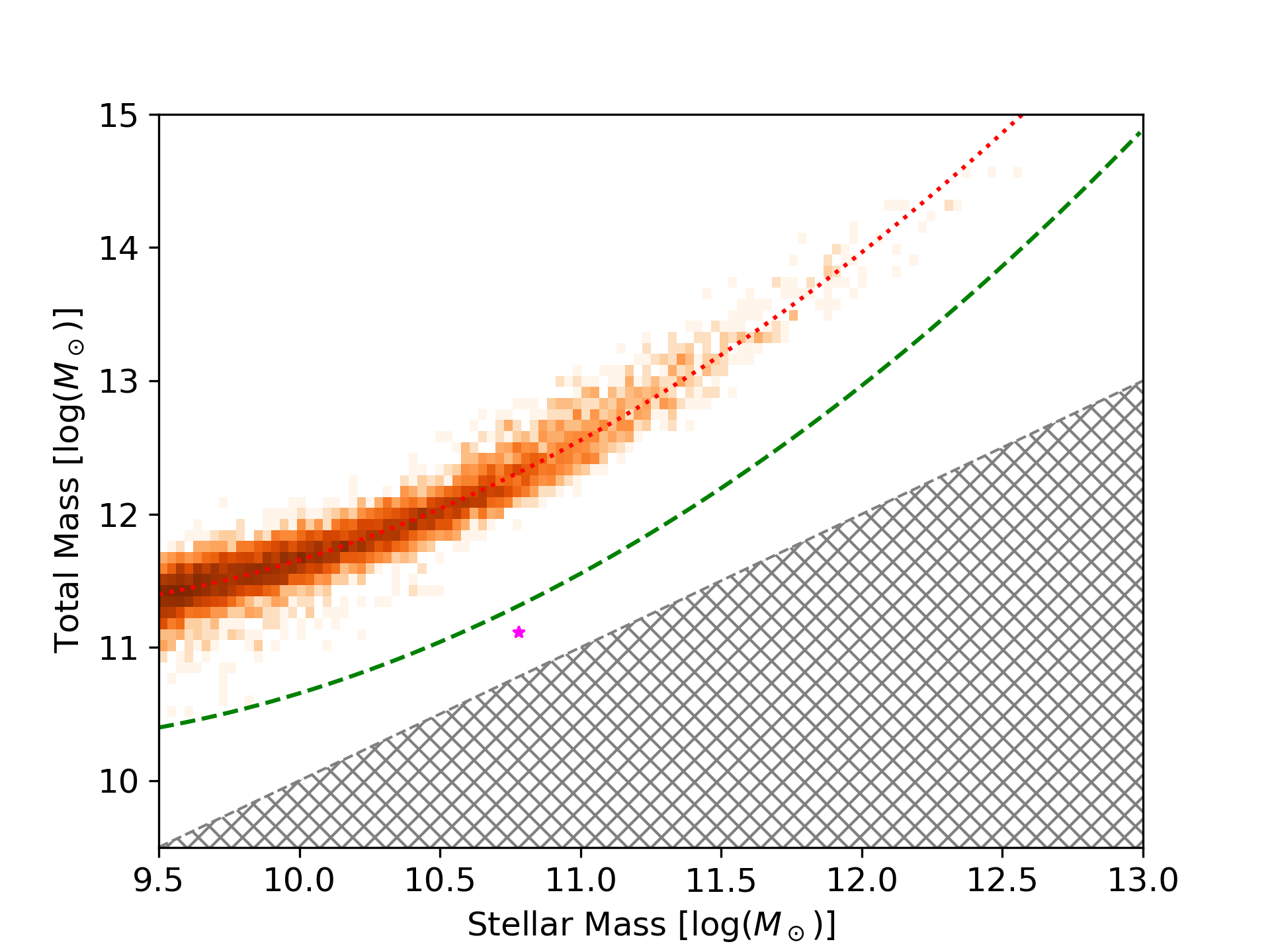}\\
\includegraphics[width=0.45\textwidth]{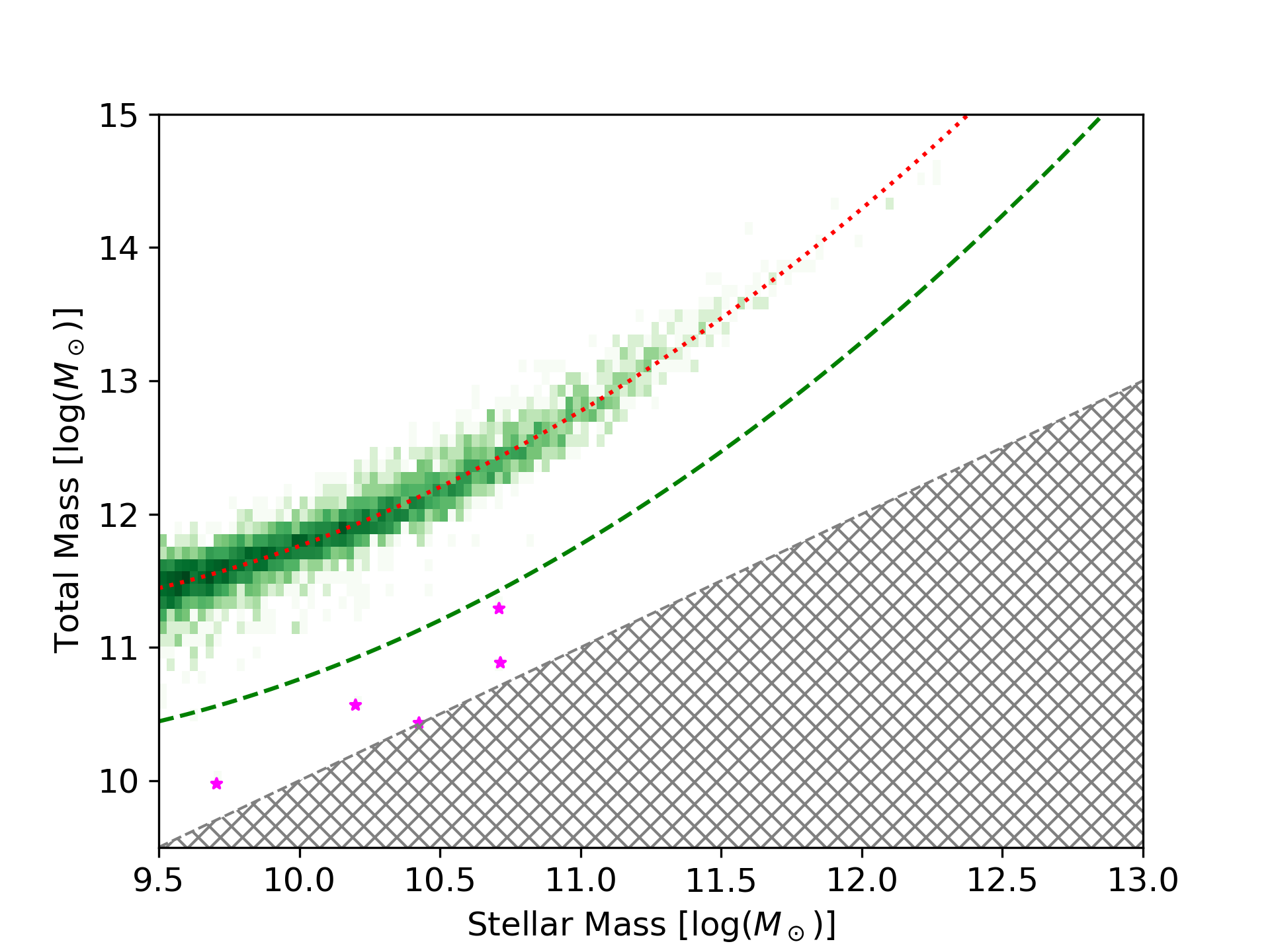}\includegraphics[width=0.45\textwidth]{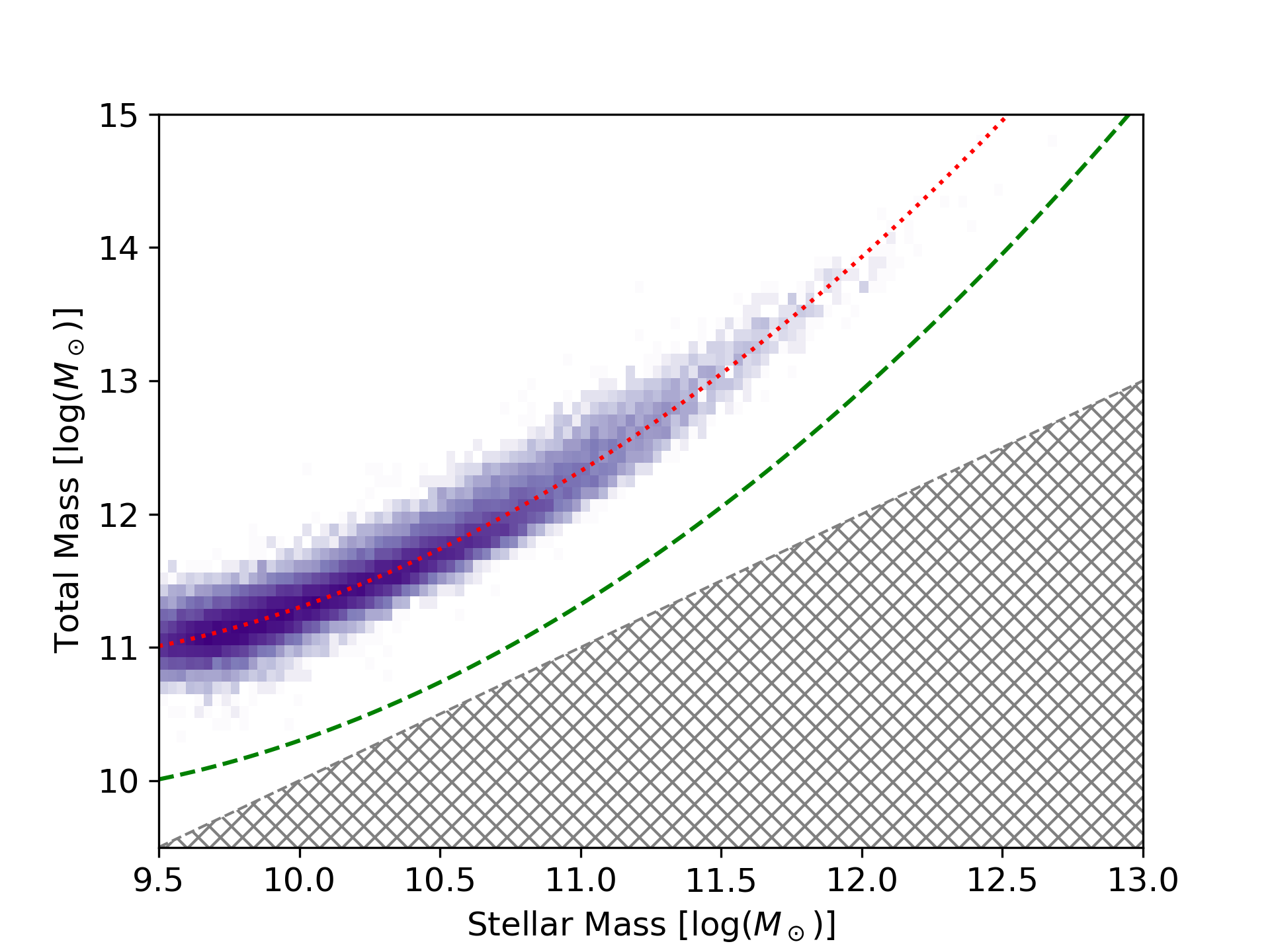}\\
\caption{Selection of oddballs as outliers of the stellar-to-halo mass relation. Upper left panel: Illustris; upper right panel: IllustrisTNG-100; bottom left panel: EAGLE; bottom right panel: Horizon-AGN. The density maps represent the distribution of central galaxies in their respective simulation. Red dotted line: fit for the inverted stellar-to-halo mass relation; Green dashed line: our selection criterium; Magenta stars: oddball galaxies.}
\label{selection_oddballs}
\end{center}
\end{figure*}

Our galaxy samples obtained from all simulations that we considered already had a suitable cut of $10^{9.5} \textrm{M}_{\odot}$ in stellar mass. We did not consider galaxies below this limit, since we wanted our sample to have sufficient particles to derive kinematic and photometric features that might help us to identify objects like them in surveys. 

In the next step, we selected only galaxies that are the dominant object in their group (rank zero). Thereby, we found 9 854 objects in Illustris, 7 237 in IllustrisTNG-100, 4 239 in EAGLE, and 29 868 in Horizon-AGN. Afterwards, we used the sample of central galaxies to fit our version of inverted\footnote{We put the stellar mass on the x-axis and the total halo mass on y-axis. We did this, because of our cut-off in stellar mass.} stellar-to-halo mass relations. Slightly breaking with convention, we used the total halo mass, because we are looking for galaxies that have unusual dark matter halos. 

To identify clear outliers at the more massive end of the stellar-to-halo mass relations, we did not use its full functional form but approximated it with a simple second order polynomial. We defined outliers to be at least one order of magnitude below this fitted relation, which is clearly beyond the scatter of the relation (see Figure \ref{selection_oddballs} for an illustration of the the selection process). Using our criteria, we only selected between one and five objects (or in the case of Horizon-AGN none), but these oddballs are extraordinary in more ways than one might naively expect. 

\subsection{Merger trees}
In order to trace the history of these oddballs and how they came to be or lost most of their dark matter, we used the merger trees provided by the different hydrodynamical simulations. In these case of Illustris and IllustrisTNG, we used the {\sc API} functions of {\sc illustris-python} to get all relevant informations from past snapshots of our objects of interest. In the case of EAGLE, we had to search the earlier snapshots using ID numbers linking the merger tree together. In both cases, we only followed the main branches of the trees, because none of the oddballs showed any indications of significant mergers past the loss of their dark matter halos. 

\subsection{Full particle data}
As a final step in the analysis of the oddballs, we looked into the full particle data surrounding the oddballs. We measured the masses of the stellar particles, gas particles, black hole particles, and the dark matter particles within spherical regions of different radii around the centres of the oddballs obtained from the subhalo catalogues. Using an analogy to observations, we refer to these regions as apertures. When doing so, we also considered the periodic boundary conditions of the simulation boxes. Additionally, we visualized the density profiles and inspected it by eye to look for any outstanding irregularities. 

We also used particle data from past snapshots to better understand how the oddballs were formed and how the particles got redistributed over time. 

\section{Results}
\label{sec_results}
\subsection{Overview}
\begin{table*}
\begin{center}
\begin{tabular}{cc|cccc|cccc|cc}
simulation & GalaxyID & $M_{\textrm{halo,SF}}$ & $M_{\textrm{halo,P30}}$ & $M_{\textrm{halo,P100}}$ & $M_{\textrm{halo,P200}}$ & $M_{\textrm{*,SF}}$ & $M_{\textrm{*,P30}}$  & $M_{\textrm{*,P100}}$  & $M_{\textrm{*,P200}}$ & $t_{\textrm{loss}}$ & $D_{\textrm{edge}}$ \\ \hline 
Illustris & 476171 & 10.33 & 10.35 & 10.99 & 13.20 & 9.42 & 9.32 & 9.49 & 9.60 & 5.38 & 2.48 \\
TNG & 585369 & 13.0 & 12.79 & 13.11 & 13.23 & 6.03 & 6.00 & 6.03 & 6.03 & 4.92 & 12.59 \\
EAGLE & 60521664 & 0.95 & 0.95 & 1.03 & 1.38 & 0.51 & 0.51 & 0.51 & 0.52 & 7.35 & 6.28 \\
EAGLE & 11419697 & 19.62 & 16.52 & 19.74 & 19.88 & 5.12 & 5.05 & 5.12 & 5.12 & 9.49 & 19.65 \\
EAGLE & 4209797 & 3.71 & 3.70 & 3.89 & 5.35 & 1.58 & 1.58 & 1.58 & 1.58 & 2.32 & 5.12 \\
EAGLE & 3868859 & 2.73 & 3.05 & 3.12 & 3.75 & 2.65 & 2.66 & 2.66 & 2.68 & 5.22 & 8.38 \\
EAGLE & 3274715 & 7.71 & 7.66 & 8.75 & 12.73 & 5.18 & 5.12 & 5.23 & 5.30 & 7.35 & 2.81
\end{tabular}
\end{center}
\caption{List of oddballs. Column 1: abbreviated name of the simulation; column 2: GalaxyID within that simulation; column 3: total halo mass according to {\sc SUBFIND} in $10^{10}$ M$_{\odot}$; column 4 to 6: total halo mass in $10^{10}$ M$_{\odot}$ measured using the full particle data within 30, 100, and 200 kpc; column 7: total stellar mass according to {\sc SUBFIND} in $10^{10}$ M$_{\odot}$; column 8 to 10: total stellar mass in $10^{10}$ M$_{\odot}$ measured using the full particle data within 30, 100, and 200 kpc; column 11: approximate time since the loss of most of the dark matter in Gyr; column 12: distance to the nearest edge of a simulation box in kpc.}
\label{list_oddballs}
\end{table*} 

Using our selection criteria, we managed to identify one oddball each in Illustris and IllustrisTNG100. Additionally, we found five oddballs in EAGLE, but none in Horizon-AGN. A summary of their, for our analysis most important, parameters is provided in Table \ref{list_oddballs}. The one feature, which was the most surprising and striking similarity between all the oddballs, was that they were located within a few kiloparsec of the edge of their respective simulation boxes. Tracing their history using the merger trees, we noticed that when the progenitors of the oddballs lost most of their mass (see Figures \ref{mass_history476171} to \ref{mass_history60521664}), when they were about to cross the boundary of the simulation. Additionally, the mass loss coincided with the loss of the peculiar velocity orthogonal to the edge of the box (see Figures \ref{dist_476171} to \ref{dist_60521664}). By the comparing the position of the most bound particle and the centre of mass, a separation between a central regions of these object and their more defuse outer parts (illustrated in Figures \ref{pastcombined_oddballs476171} to \ref{pastcombined_oddballs60521664}) is seen after the centre of mass of the oddball crossed the edge of the simulation box using the periodic boundary conditions. The two components of the halo separated afterwards and the oddball gets stuck near the edge of the simulation box, while the rest of the halo moves on as illustrated by their present-day distribution in Figures \ref{today_oddballs476171} to \ref{today_oddballs60521664}. The oddballs themselves are relatively isolated at the present-day (see Figures \ref{oddballs476171} to \ref{oddballs60521664}) and some (as we showed using the EAGLE group catalogues) were isolated for most of their history (see Figures \ref{surroundings_shift_evolution3274715} to \ref{surroundings_shift_evolution60521664}) even at the time of their mass loss event. 

We will now discuss each of the oddballs in detail.

\subsection{Illustris-476171}
This oddball is one of the peculiar objects in Illustris studied in \citet{Yu:2018} and the only object in Illustris-1 that fulfilled our selection criteria (the objects presented in \citet{Yu:2018} did not qualify for our sample sample due their either too low masses or too high dark matter fractions). The masses derived using the different methods (subhalo finder and aperture measurements) are in approximate agreement with each other. By studying its merger tree (see Figure \ref{mass_history476171}), we found that a huge mass loss happened around 5.38 Gyr ago, which coincided with the centre of mass returning close to the position of the most bound particle, as well as with the loss of almost all the peculiar velocity orthogonal to the edge of the box the oddball's centre of mass was crossing (see Figure \ref{dist_476171}). We found a two Gigayears long phase preceding the mass loss event in which there was in an increasing separation of the centre of mass of the most bound particle that started once the centre of mass crossed the edge of the box. We interpret this as the progenitor of the oddball being pulled apart, which can be seen in the snapshot preceding the mass loss event. To better illustrate it, we visualized the vicinity of the progenitor of the oddball in Figure \ref{pastcombined_oddballs476171}. It illustrates very well that the most dense region of the halo that will go on to form the oddball got stuck at the edge of the simulation box, while the more diffuse outer parts of that object keep moving until they are completely separated in the next snapshot. At that point the total mass of the oddball has dropped by almost two orders of magnitude. 
Although this most several impacts the more spatially extended dark matter halo, that object also lost the majority of its stellar mass. By the end of the simulation, the lost particles have moved onwards by more than a Megaparsec, although they leave behind a trail pointing towards the oddball (see Figure \ref{today_oddballs476171}). The oddball itself is located outside that remnant halo (see Figure \ref{oddballs476171}) in relative isolation. This separation effect was already found by \citet{Yu:2018}, but they failed to point out the coincident that the oddball apparently got stuck at edge of the simulation box. 

\subsection{IllustrisTNG-585369}
This object found in the IllustrisTNG-100-1 simulation experienced a mass loss event about 4.92 Gyr ago (see Figure \ref{mass_history585369}). The situation closely resembles the one of the previous oddball. During a phase last close to two Gigayears the separation between the centre of mass and the most bound particle (see Figure \ref{dist_585369}) temporarily increased. This indicates the separation of the core and the diffuse halo, which happened when the centre of mass of the progenitor of this oddball was crossing the edge of the box using the periodic boundary conditions. This separation of the particles is illustrated in Figure \ref{pastcombined_oddballs585369}, which shows the distribution of particles in the snapshot right before the mass loss (strictly speaking: in the subsequent snapshot the halo particles were no longer considered to be bound to the core by the {\sc SUBFIND} algorithm) happens. Again the velocity component orthogonal to the boundary changed from about 100 km/s to single digit values. This oddball also had a strange temporary mass dip in its past, which can be explained by a close encounter or merger with another objects, and some particles being associated with a different/wrong halo by the {\sc SUBFIND} algorithm. In the present day snapshot the particles lost from the oddball can be found in a diffuse structure about 1.5 Mpc from the oddball itself (see Figure \ref{today_oddballs585369}). Since there are no other notable structures near the oddball (see Figure \ref{oddballs585369}), the mass measurements within the apertures agree very well with the values obtained using the {\sc SUBFIND} algorithm. 

\subsection{EAGLE-60521664}
This oddball from the EAGLE simulation has its origin in a mass loss event about 7.35 Gyr ago (see Figure \ref{mass_history60521664}) similar to the previously discussed oddballs.  It also happened at the same time as a drop of the peculiar velocity orthogonal to the edge of the box. A separation of the centre of mass and the most bound particle lasting for about a Gigayear preceding the mass loss after the centre of mass of the oddball crossed the edge of the simulation box show that the same processes are going on for this oddball from the EAGLE simulation as for the oddballs from the two Illustris simulations (see Figure \ref{dist_60521664}). Snapshot 18 takes place just before the progenitor of this oddball was torn apart with the stellar matter already clearly separated (see Figure \ref{pastcombined_oddballs60521664}) in what became a dark matter rich and dark matter poor halo (hosting the oddball) later. By the end of the simulation run, the two halos (the oddball and the remnant) are separated by a Megaparsec (see Figure \ref{today_oddballs60521664}) with the oddball living in a low density environment (see Figure \ref{oddballs60521664}). Since we have the group catalogues of all snapshots of the EAGLE simulation (at least for a galaxies bright than -16 mag in the r band), we were able to study environment of the oddball and its progenitor for its entire evolution. As illustrated in Figure \ref{surroundings_shift_evolution60521664}, this oddball evolved in a very low density environment, but there is a brief encounter with another halo shortly before the mass loss event and afterwards one can see the other halo departing. 

\subsection{EAGLE-11419697}
This oddball is, based on its intrinsic properties, the least odd of the objects, and the one for which dark matter loss happened with the longest (9.49 Gyr) ago, which again co-incides with it encountering the edge of the simulation box. It only lost about one order of magnitude of its dark matter mass, while retaining almost all its stellar mass. It does, however, lose most of its gas, recapturing some of it later (see Figure \ref{mass_history11419697}). It is also the one that still has the highest dark-to-stellar matter fraction of all the oddballs. This oddball also shows an about two Gigayears long separation phase preceding the mass loss event between its centre of mass and most bound particle, when it got close to the edge of the simulation (see Figure \ref{dist_11419697}). Since it is so old, it was already an isolated oddball at snapshot 18 of the EAGLE simulation. Therefore, it is the only oddball for which we cannot extract any useful information about its formation and the path of its particles from it (see Figures \ref{pastcombined_oddballs11419697} and \ref{today_oddballs11419697}). There is only a very faint trace of a dark matter tail (see Figure \ref{oddballs11419697}) likely pointing towards the remainder of its original halo. Although this oddball is currently extremely isolated (see Figure \ref{surroundings_shift_evolution11419697}) at present-day, it had contact with higher density environment when its mass loss happened and even spent quite some time there afterwards. 

\subsection{EAGLE-4209797}
This is the youngest oddball with its mass loss happening merely about 2.32 Gyr ago, removing the majority of its dark matter halo while retaining most of its stellar mass (Figure \ref{mass_history4209797}). Additionally, its progenitor experienced a merger about 6 Gyr ago. The separation of the centre of mass and the most bound particle lasting again about two Gigayears followed by the total loss of any peculiar velocity orthogonal to the edge of the box also happened for this oddball after its centre of mass crossed the edge of the simulation box (see Figure \ref{dist_4209797}). Since snapshot 18 predates the merger, the particles forming the oddball at present day can be found in two separate halos (see Figure \ref{pastcombined_oddballs11419697}, with the other halo outside the illustrated box, but its position visible as contours). A closer analysis of the particles contributed by both progenitors of the merger showed a near 50-50 split of them contributing to the present day oddball. The particles of the oddballs primary progenitor are split between the oddball and a diffuse distribution in a larger dark matter halo less than a Megaparsec away (see Figure \ref{today_oddballs4209797}) with a faint stellar trail (see Figure \ref{oddballs4209797}) connecting them. The environmental history of this oddball only shows close interactions between the merger progenitor and the diffuse remnant, but both events are separated by over two Gyr (see Figure \ref{surroundings_shift_evolution4209797}). 

\subsection{EAGLE-3868859}
This oddball experienced its cataclysmic mass loss event about 5.22 Gyr ago and it even removed dark matter from its inner halo (see Figure \ref{mass_history3868859}). It is the only oddball that does not show the temporary offset between the centre of mass and the most bound particle after the centre of mass crossed to the edge of simulation box. However, the drop of the peculiar velocity orthogonal to the edge of the box to close to zero happened (see Figure \ref{dist_4209797}). Figure \ref{pastcombined_oddballs3868859} illustrates that the particles forming the oddballs were located in the very centre of a massive halo, still about a Megaparsec away from the edge of the box (snapshot 66 predates the mass loss event by over two Gigayears). The particles of this massive halo that did not end up in the oddball can be found about a Megaparsec away from it and show strong tidal features (see Figure \ref{today_oddballs3868859}). A stellar trail still forms a bridge to the oddball (see Figure \ref{oddballs3868859}). There is a clear indication that the oddball (or actually its progenitor) had a close encounter with other galaxies about a Gyr prior to the mass loss event (Figure \ref{surroundings_shift_evolution3868859}).

\subsection{EAGLE-3274715}
This is the only oddball that apparently had two mass loss events, a big one approximately 7.35 Gyr and more gradual one, which is still ongoing, that started about 4 Gyr ago (see see Figure \ref{mass_history3274715}). The first mass loss event was preceded by the separation between the centre of mass and most bound particle lasting about three Gigayears after the centre of mass of the progenitor of the oddball crossed the edge of the box (see Figure \ref{dist_3274715}). The loss of most of the peculiar velocity orthogonal to the edge of the box happened alongside with the mass loss event. The particles forming the oddball can be found in the centre of the progenitor at snapshot 18, just prior to the mass loss event, while the rest of the halo is asymmetrically spread out (see Figure \ref{pastcombined_oddballs3274715}). While the clump forming the oddball stayed the edge of the box, the rest of the halo moved on an can be found about three Megaparsec away from it (see Figure \ref{today_oddballs3274715}). The structure shows strong extended tidal tails with on of them reaching towards the oddball (possibly explaining the ongoing second mass loss). Although the oddball is located in a low density environment at present-day (see Figure \ref{oddballs3274715}), just before its mass loss event it seemed to have passed through a high density environment (see Figure \ref{surroundings_shift_evolution3274715}). While one might assume that all of these objects merged into each prior to the mass loss, the mass history of this oddball (see Figure \ref{mass_history3274715}) does not show any indication and the particle distribution in snapshot 18 (just prior to the mass loss event) shows two halos still at considerable separation, which makes it more like for them to actually have moved out of each other range. 

\section{Discussion}
\label{sec_discussion}
By selecting clear outliers of the stellar-to-halo mass relation for central galaxies, we managed to find several unusual objects in Illustris, IllustrisTNG, and EAGLE. However, we did not detected any such oddballs in Horizon-AGN, but given that we only found one in Illustris and IllustrisTNG each, we are dealing which small number statistics here. Interestingly, EAGLE, which has the, by a narrow margin, smallest volume of all the simulations considered in this paper, contains with five oddballs, the most of these unusual objects. These oddballs experienced a huge loss of dark matter (and also stellar matter, but to a lesser degree) in their past. However, they did not dissolve, but their remnant survived for many billions of years. At a glance the oddballs may appear to be massive tidal (''dwarf'') galaxies without very little dark matter, similar to the observed low surface brightness galaxies of \citet{vanDokkum:2018a,vanDokkum:2018b}. However after a closer look, they turned out to be not only more massive and isolated, but also much more compact than the observed galaxies, and the peculiarities did not end there. Consequently, we collected and evaluated all the evidence that these oddballs are artificats from the simulations in order to help identifying the bug causing them.

The one common feature of all oddballs in all simulations that was not implied by our selection criteria is that they are located near the boundary of their simulation boxes. Especially, considering that all oddballs are located within less than 20 kpc of the edge, which corresponds to a volume fraction of $\sim 0.1\%$ of the simulation boxes, we can safely exclude this to be a pure coincidence. All simulations use periodic boundary conditions though, which should avoid such phenomena. Naturally, our initial suspicion was that the masses provided by {\sc SUBFIND} \citep{Rodriguez:2015} were incorrect, especially considering that the only simulation in which we did not find any oddballs was Horizon-AGN, which used the {\sc ADAPTAHOP} halo finder \citep{Aubert:2004,Tweed:2009} instead. To test this hypothesis, we compared the masses obtained by {\sc SUBFIND} to masses measured within spherical apertures around the centres of oddballs using full particle data of the simulations and. Also these masses were consistent with the masses provided by {\sc SUBFIND}. Only the widest apertures occasionally contain up to $\sim 50\%$ more mass than predicted by {\sc SUBFIND}, because they may already contain particles from neighbouring subhalos. Even with this, the oddballs would remain significant outliers of the stellar-to-halo mass relation for central galaxies. By conducting these tests, we were able to safely refute the hypothesis that the oddballs are artefacts created by the subhalo finder. 

It is possible to remove the outer dark matter halo of galaxy by tidal interaction in clusters (\citet{Niemiec:2018}, \citet{Jing:2018}, and our upcoming own paper (Saulder et al., in preparation)). Therefore, one could imagine a scenario in which such a galaxy gets ejected (and likely even more disrupted) from the cluster via three-body interaction. This would be consistent with the stellar and dark matter trails seen for the oddballs (see Figures \ref{today_oddballs476171} to \ref{oddballs60521664}). When discussing the various oddballs, we have already pointed out some connections with possible mergers preceeding the mass loss event. The progenitor of oddball EAGLE-4209797 went through a major mergers a few Gigayears before it suffered from its mass loss event, while most of the other oddballs have close encounters before they lost most of their mass. For four of the oddballs, we studied the particle distribution right before the apparent mass loss event (see Figures \ref{pastcombined_oddballs476171} to \ref{pastcombined_oddballs60521664} and \ref{pastcombined_oddballs3274715}) and one could clearly see two substructures, namely a dense core separating from a more diffuse distribution (halo). Considering that {\sc SUBFIND} only assigns the subhalo based on the position of particles starting from density peaks, one might consider the possibility that two (kinematically) unbound halos close to each other are identified as one for some time steps. This would explain the separation between the centre of mass and the most bound particle detected for almost all oddball prior to their mass loss event (see Figures \ref{dist_476171} to \ref{dist_3274715}). This scenario would go along with the known subhalo switching problem \citep{Rodriguez:2015}, that would cause a confusion in the merger trees and would provide a potential explanation for the sudden mass loss. However, the oddball EAGLE-3868859 provided the crucial evidence that quickly disproves these speculations. As illustrated in Figure \ref{pastcombined_oddballs3274715}, the particles forming the oddball were located in the very centre of a massive halo that was still a Megaparsec away from the edge of the box at that snapshot. This proves that the problem creating these oddballs is more complicated than the subhalo switching problem. Furthermore, none of these hypothetical scenarios would be able to explain why all of these galaxies are located at the edge of the simulation box. In the case of a single galaxy, one could wonder, if it were a coincidence, but we found that all oddballs in three different simulations can be found within a few kiloparsec of the edge of the simulation box. Another interesting detail about the oddballs is that their peculiar velocity orthogonal to the edge of the simulation box is close to zero (single digit km/s). For all oddballs, it dropped to this value right after the mass loss event, when the two components of the oddball progenitor separated. The increasing offset between centre of mass and the position of the most bound particles started for all (but one) oddball as soon as it reached the edge of the simulation box, indicating that the separation of the core (which becomes the oddball) and the other regions started at time. This process lasted typically about two Gigayears in the simulation and resulted in the creation of an oddball stuck at the edge of the box and a more extended halo moving onwards. All these observations combined indicate that the issue is actually very serious and almost certainly connected to the periodic boundary conditions of the simulations. 

\begin{figure*}
\begin{center}
\includegraphics[width=0.31\textwidth]{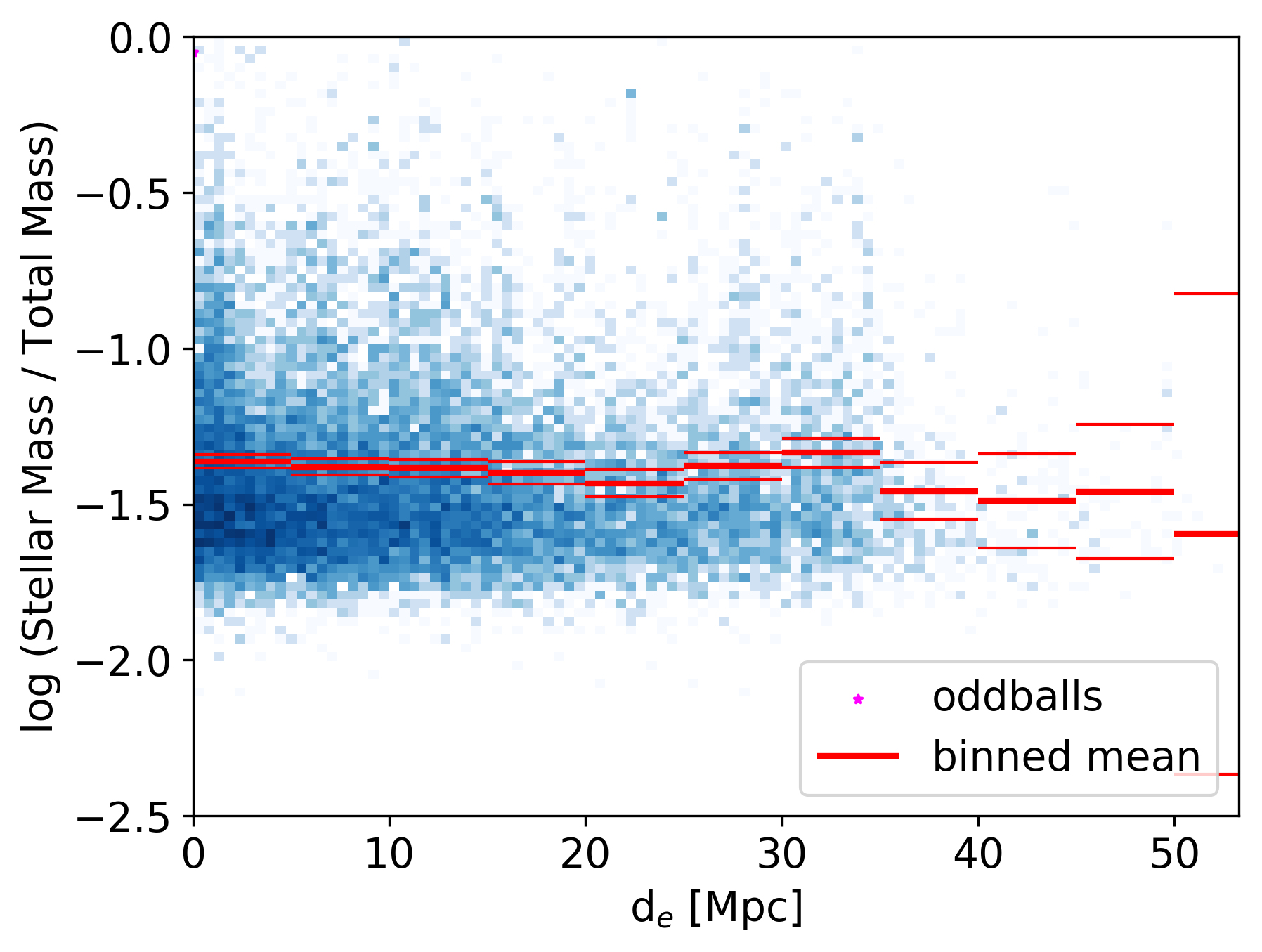}\includegraphics[width=0.31\textwidth]{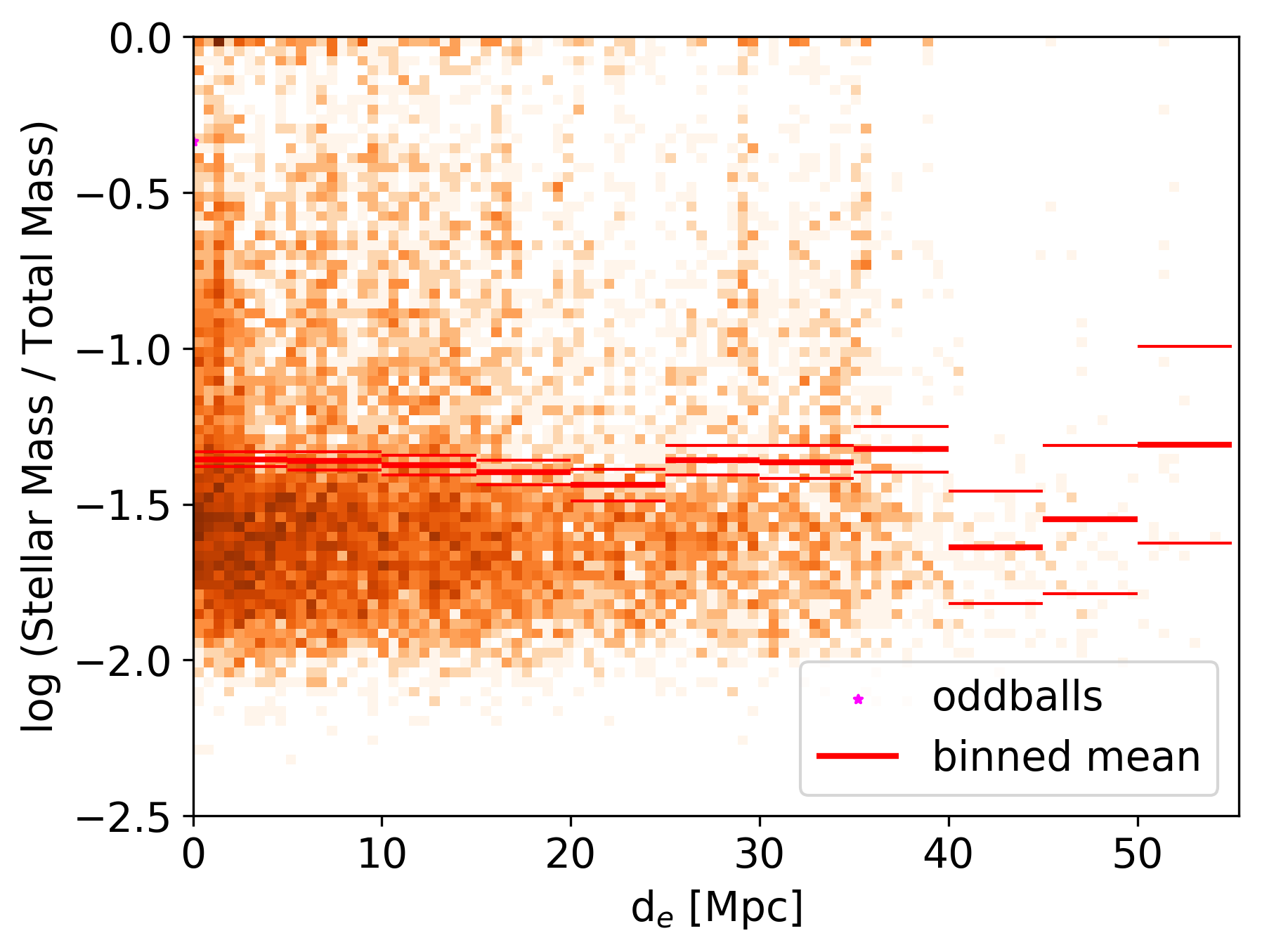}\includegraphics[width=0.31\textwidth]{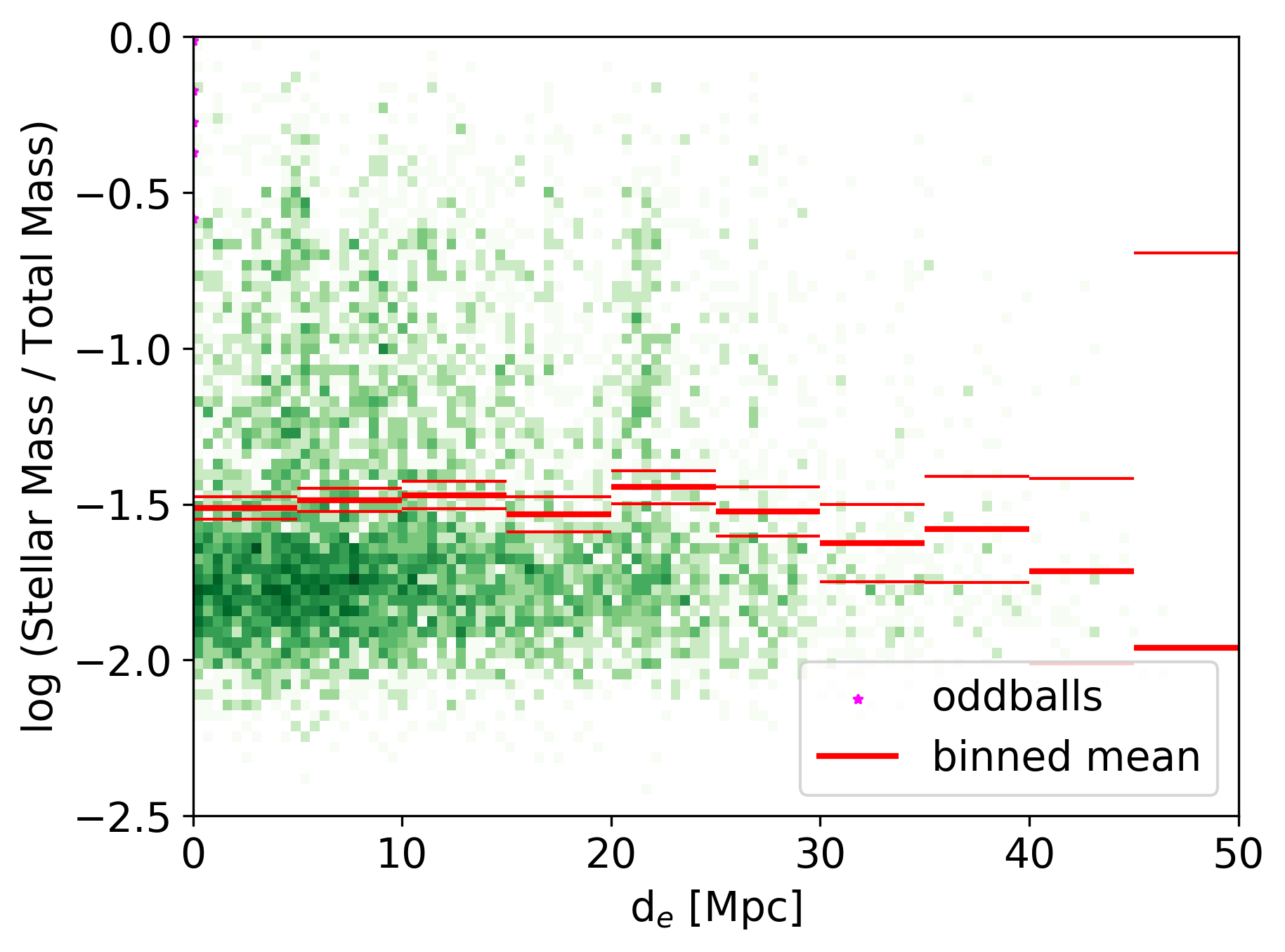}\\
\caption{Stellar-to-halo mass ratio as a function of the distance from the edge of the simulation boxes. Left panel: Illustris; central panel: IllustrisTNG; right panel: EAGLE. The density maps represent the distribution of all (not just centrals as in Figure \ref{selection_oddballs}) objects with stellar masses above $10^{9.5} \textrm{M}_{\odot}$ in their respective simulation. Thick red bars: mean value in 5 Mpc wide bins; thin red bars: error bars of the mean value; magenta stars: oddball galaxies.}
\label{edge_offsets}
\end{center}
\end{figure*}

After considering all the possibilities, we have to conclude that the oddballs, despite at a glance looking like exotic dark matter deprived galaxies, are in fact numerical artefacts. They are notably more massive than typical artefacts in hydro-simulation of these resolutions and their location in their respective simulation boxes make them stand out even more. As we reconstructed the creation of these oddballs, we found the following scenario to be the most consistent: the progenitor of a oddball approaches the edge of the simulation box and would be about to cross it using the periodic boundary conditions. Then something goes wrong and a part of the object's core gets stuck close to the edge of the simulation box. But the outer parts of that object continue their motion, which would explain the temporary offset between the centre of mass and the position of the most bound particle, and after a phase typically lasting for about two Gigayears the two components are torn apart . At one point {\sc SUBFIND} starts to consider the remnants as two distinct objects, which appears as the mass loss event in the merger trees. With peculiar velocity orthogonal dropping to single digit numbers (in km/s), the oddballs remains at the edge of simulation box. The other component keeps moving away from the edge, but leaves a feature similar to a tidal tail pointing towards the oddball. The oddballs remain stable, often isolated, objects for many Gigayears to follow. Considering that the oddballs are by far not the only simulated galaxies crossing edge of the simulation box using the periodic boundary conditions, one may wonder why there are not more galaxies affected by whatever bug created the oddballs. To investigate the possibility that other objects suffer from this effect, but only to a less notable effect, we decided to study the stellar-to-total mass ratios as a function of the distance to the nearest edge of the simulation box. As illustrated in Figure \ref{edge_offsets}, we could not find any statistically significant offset in the five Megaparsec wide bin for the objects closes to the edge for any of the simulation in which we discovered the oddballs. This actually makes the bug causing the oddballs even stranger than before, because it indicates that hundreds of galaxies were able to pass across the edge of the box using the periodic boundary conditions without any issues, while a few objects run into trouble and become oddballs. 

While at the date of the initial submission, we have not been able to isolate the exact issue causing the oddballs, Ruediger Pakmor and Volker Springel were able to fully confirm our discovery as a bug. Furthermore, they were able to identify the exact conditions causing this issue thanks to our preprint and private communications. They found that it is caused by a very rare combination of various factors: When crossing the box boundary, the oddballs' dense stellar cores suffer from unusually large force error. This is due to the tree code using a very large tree node for the oddballs' progenitors, since their particles can be found at the opposite edges of the simulation box. Since the total accelerations within the dense stellar cores are pretty large, the resulting relative errors are calculated to be small even for such large nodes. In the oddball scenario, this error is severely underestimated by the code since the distance to the closest edge of the node is very small while its centre-of-mass is far away. In this case, the code should have opened the node in order to yield better results, which did not happen due to the bug. The conditions of the formation of the oddballs are reminiscent of the worst case scenario of tree codes that cause large force errors, which was described in \citet{SalmonWarren:1994}. While there is a special opening criterion that was design to avoid the break-down of the multipole expansion in such cases, which checks if the distance between particle and geometric centre of the node is smaller than 0.6 times the size of the node, it did not trigger, because it lacks the periodic wrapping, hence the centre of the node is on the other side more than half a simulation box away. This is a very old bug, which can be found in the public version of {\sc Gadget2} \citep{GADGET}, and all its derivatives, including {\sc Gadget3}, {\sc Arepo} and {\sc Gizmo}. This also explains why we could not find any oddballs in HorizonAGN, because it used {\sc RAMSES}. Since  large forces that allow for the use of the large node are only present in the dense stellar cores, but not in the less dense dark matter halo, we are able to understand why the bug separates the stellar core from the halo as seen for the oddballs. For the systematic momentum error to accumulate to the catastrophic failure that leads to the appearance of oddballs, a reticulately slow motion of the oddball progenitor across the boundary, that allows for enough time steps, is required. Galaxies that fulfil these conditions are transformed into oddballs in the simulations.

According to Ruediger Pakmor and Volker Springel (in private communications), safe guards that avoid this bug have already been implemented in latest versions of {\sc Arepo} and {\sc Gadget4} codes. Since the bug only triggers for the extremely rare cases, which we documented in this paper, it does not have an impact on the results of the simulation beyond said oddballs. As illustrate in Figure \ref{edge_offsets}, we found no indications of a statistical relevant change in a galaxy/halo properties as a function of distance from the edge, which was also confirmed by Ruediger Pakmor and Volker Springel. The noted higher prevalence of oddballs in EAGLE, when compared to Illustris and IllustrisTNG, is likely due to its smaller particle-mesh grid and a slightly less conservative setting for the normal opening parameter. Until the recent generation of hydro-simulations with large volumes ($\sim 100$ Mpc$^{3}$), the effect of the bug with a prevalence of less than one in a thousand for EAGLE and less than one in ten thousand for Illustris and IllustrisTNG was simply to rare to notice despite persisting in well-tested codes for about 15 years. Luckily, our research, although initially aimed to find isolated dark matter deprived galaxies, was able to discover it and contribute valuable information for its fix. Additionally, our results illustrate that one has to be extremely careful when drawing conclusions from simulated objects for real galaxies, especially when searching for unusual objects. The oddball Illustris-476171 is not an example of a peculiar galaxy found in a hydrodynamical simulation as claimed in \citet{Yu:2018}, but just a numerical artefact.

\section{Summary and Conclusions}
\label{sec_summary}
We aimed to study massive ($>10^{9.5} \textrm{M}_{\odot}$ in stellar mass) isolated dark matter deprived galaxies in different state-of-the-art hydro-simulations (Illustris, IllustrisTNG, EAGLE, and Horizon-AGN) to predict their likelihood for observations. Using the present-day group/subhalo catalogues from these simulations, we managed to identify several such objects (one in Illustris, one in IllustrisTNG, five in EAGLE, and none in Horizon-AGN), which we defined as clear (by one order of magnitude) outliers of the stellar-to-halo mass relation for galaxies that in the central objects in their respective subhalos. We dubbed these unusual objects: oddballs. In order to ensure that the masses of these oddballs were measured correctly, we compared the values obtained from {\sc SUBFIND} to direct mass measurements on the full particle data within spherical apertures. The different methods yielded results that were consistent with each other. A closer investigation showed that all oddballs have one unusual feature that is not implied by their selection method in common: they are located within a few kiloparsec of the edge of their respective simulation boxes. By tracing the history of the particles found in the present-day oddballs, we were able to show that they originated from the core of regular galaxies that got disrupted when crossing the edge of the simulation box. It appears that some part of these objects literally gets stuck at the edge of the simulation box, while the remainder moves onwards. After examining all evidence and considering all possible explanations, we conclude that the oddballs are not representing any possible real galaxies. Their low dark matter fractions are merely the consequence of a rare bug in the simulation. Thanks to the information provided in the first preprint of this paper as well as in private communications, Ruediger Pakmor and Volker Springel were able to trace and identify the bug creating the oddballs as an issue present {\sc Gadget2} and all its derivatives. They found that that in the rare combination of circumstances of a massive dense stellar core embedded in a dark matter halo crossing the edge of the simulation box sufficiently slowly can trigger a bug in the tree code, which causes the stellar core and the halo to separate and thereby creating the oddballs. The effects of the bug are limited to these special cases, leaving the rest of the simulations unaffected. Hence, we strongly advise everybody to exclude the objects identified as oddballs from any analysis of data from the simulations. Our results emphasis the importance of using common sense and eye for detail when working with data from simulations, especially when dealing with extraordinary objects.

\section*{Acknowledgments}
We thank the Korea Institute for Advanced Study for providing computing resources (KIAS Center for Advanced Computation Linux Cluster System) for this work.

Special thanks to Ruediger Pakmor and Volker Springel, who following our communications were able to identify the source of the issue presented in this paper and allowed us to share this information in our paper, thereby notably improving the strength of our conclusions. We also want to thank Dylan Nelson and Shy Genel for several helpful comments and suggestions. Additionally, we acknowledge helpful advise from Rodrigo Ca{\~n}as and Claudia Lagos. 

The codes required for this project were written in {\sc Python}. We want to take the opportunity to thank all its developers and especially the people behind the following packages: {\sc SciPy} \citep{scipy}, {\sc NumPy} \citep{NumPy}, {\sc Matplotlib} \citep{matplotlib}, and {\sc astropy} \citep{astropy:2013,astropy:2018}.

\addcontentsline{toc}{section}{References}
\bibliographystyle{mnras}
\bibliography{paper}\label{bib}

\begin{thebibliography}{}
\makeatletter
\relax
\def\mn@urlcharsother{\let\do\@makeother \do\$\do\&\do\#\do\^\do\_\do\%\do\~}
\def\mn@doi{\begingroup\mn@urlcharsother \@ifnextchar [ {\mn@doi@}
  {\mn@doi@[]}}
\def\mn@doi@[#1]#2{\def\@tempa{#1}\ifx\@tempa\@empty \href
  {http://dx.doi.org/#2} {doi:#2}\else \href {http://dx.doi.org/#2} {#1}\fi
  \endgroup}
\def\mn@eprint#1#2{\mn@eprint@#1:#2::\@nil}
\def\mn@eprint@arXiv#1{\href {http://arxiv.org/abs/#1} {{\tt arXiv:#1}}}
\def\mn@eprint@dblp#1{\href {http://dblp.uni-trier.de/rec/bibtex/#1.xml}
  {dblp:#1}}
\def\mn@eprint@#1:#2:#3:#4\@nil{\def\@tempa {#1}\def\@tempb {#2}\def\@tempc
  {#3}\ifx \@tempc \@empty \let \@tempc \@tempb \let \@tempb \@tempa \fi \ifx
  \@tempb \@empty \def\@tempb {arXiv}\fi \@ifundefined
  {mn@eprint@\@tempb}{\@tempb:\@tempc}{\expandafter \expandafter \csname
  mn@eprint@\@tempb\endcsname \expandafter{\@tempc}}}

\bibitem[\protect\citeauthoryear{{Astropy Collaboration} et~al.,}{{Astropy
  Collaboration} et~al.}{2013}]{astropy:2013}
{Astropy Collaboration} et~al., 2013, \mn@doi [\aap]
  {10.1051/0004-6361/201322068}, \href
  {http://adsabs.harvard.edu/abs/2013A%26A...558A..33A} {558, A33}

\bibitem[\protect\citeauthoryear{{Aubert}, {Pichon}  \& {Colombi}}{{Aubert}
  et~al.}{2004}]{Aubert:2004}
{Aubert} D.,  {Pichon} C.,   {Colombi} S.,  2004, \mn@doi [\mnras]
  {10.1111/j.1365-2966.2004.07883.x}, \href
  {https://ui.adsabs.harvard.edu/\#abs/2004MNRAS.352..376A} {352, 376}

\bibitem[\protect\citeauthoryear{{Bah{\'e}} et~al.,}{{Bah{\'e}}
  et~al.}{2017}]{Hydrangea}
{Bah{\'e}} Y.~M.,  et~al., 2017, \mn@doi [\mnras] {10.1093/mnras/stx1403},
  \href {https://ui.adsabs.harvard.edu/\#abs/2017MNRAS.470.4186B} {470, 4186}

\bibitem[\protect\citeauthoryear{{Barrett}, {Hunter}, {Miller}, {Hsu}  \&
  {Greenfield}}{{Barrett} et~al.}{2005}]{matplotlib}
{Barrett} P.,  {Hunter} J.,  {Miller} J.~T.,  {Hsu} J.-C.,   {Greenfield} P.,
  2005, in {Shopbell} P.,  {Britton} M.,   {Ebert} R.,  eds,  Astronomical
  Society of the Pacific Conference Series Vol. 347, Astronomical Data Analysis
  Software and Systems XIV. p.~91

\bibitem[\protect\citeauthoryear{{Blakeslee} \& {Cantiello}}{{Blakeslee} \&
  {Cantiello}}{2018}]{Blakeslee:2018}
{Blakeslee} J.~P.,  {Cantiello} M.,  2018, \mn@doi [Research Notes of the
  American Astronomical Society] {10.3847/2515-5172/aad90e}, \href
  {https://ui.adsabs.harvard.edu/\#abs/2018RNAAS...2c.146B} {2, 146}

\bibitem[\protect\citeauthoryear{{Cowley} et~al.,}{{Cowley}
  et~al.}{2019}]{Cowley:2019}
{Cowley} W.~I.,  et~al., 2019, \mn@doi [\apj] {10.3847/1538-4357/ab089b}, \href
  {http://cdsads.u-strasbg.fr/abs/2019ApJ...874..114C} {874, 114}

\bibitem[\protect\citeauthoryear{{Crain} et~al.,}{{Crain}
  et~al.}{2009}]{Crain:2009}
{Crain} R.~A.,  et~al., 2009, \mn@doi [\mnras]
  {10.1111/j.1365-2966.2009.15402.x}, \href
  {http://cdsads.u-strasbg.fr/abs/2009MNRAS.399.1773C} {399, 1773}

\bibitem[\protect\citeauthoryear{{Davis}, {Efstathiou}, {Frenk}  \&
  {White}}{{Davis} et~al.}{1985}]{Davis:1985}
{Davis} M.,  {Efstathiou} G.,  {Frenk} C.~S.,   {White} S.~D.~M.,  1985,
  \mn@doi [\apj] {10.1086/163168}, \href
  {http://adsabs.harvard.edu/abs/1985ApJ...292..371D} {292, 371}

\bibitem[\protect\citeauthoryear{{Dolag}, {Borgani}, {Murante}  \&
  {Springel}}{{Dolag} et~al.}{2009}]{Dolag:2009}
{Dolag} K.,  {Borgani} S.,  {Murante} G.,   {Springel} V.,  2009, \mn@doi
  [\mnras] {10.1111/j.1365-2966.2009.15034.x}, \href
  {http://adsabs.harvard.edu/abs/2009MNRAS.399..497D} {399, 497}

\bibitem[\protect\citeauthoryear{{Dubois} et~al.,}{{Dubois}
  et~al.}{2014}]{HorizonAGN}
{Dubois} Y.,  et~al., 2014, \mn@doi [\mnras] {10.1093/mnras/stu1227}, \href
  {https://ui.adsabs.harvard.edu/\#abs/2014MNRAS.444.1453D} {444, 1453}

\bibitem[\protect\citeauthoryear{{Dubois}, {Peirani}, {Pichon}, {Devriendt},
  {Gavazzi}, {Welker}  \& {Volonteri}}{{Dubois} et~al.}{2016}]{Dubois:2016}
{Dubois} Y.,  {Peirani} S.,  {Pichon} C.,  {Devriendt} J.,  {Gavazzi} R.,
  {Welker} C.,   {Volonteri} M.,  2016, \mn@doi [\mnras]
  {10.1093/mnras/stw2265}, \href
  {https://ui.adsabs.harvard.edu/\#abs/2016MNRAS.463.3948D} {463, 3948}

\bibitem[\protect\citeauthoryear{{Emsellem} et~al.,}{{Emsellem}
  et~al.}{2018}]{Emsellem:2018}
{Emsellem} E.,  et~al., 2018, arXiv e-prints, \href
  {https://ui.adsabs.harvard.edu/\#abs/2018arXiv181207345E} {p.
  arXiv:1812.07345}

\bibitem[\protect\citeauthoryear{{Fensch} et~al.,}{{Fensch}
  et~al.}{2018}]{Fensch:2018}
{Fensch} J.,  et~al., 2018, arXiv e-prints, \href
  {https://ui.adsabs.harvard.edu/\#abs/2018arXiv181207346F} {p.
  arXiv:1812.07346}

\bibitem[\protect\citeauthoryear{{Genel} et~al.,}{{Genel}
  et~al.}{2014}]{Genel:2014}
{Genel} S.,  et~al., 2014, \mn@doi [\mnras] {10.1093/mnras/stu1654}, \href
  {https://ui.adsabs.harvard.edu/\#abs/2014MNRAS.445..175G} {445, 175}

\bibitem[\protect\citeauthoryear{{Genel} et~al.,}{{Genel}
  et~al.}{2018}]{Genel:2018}
{Genel} S.,  et~al., 2018, \mn@doi [\mnras] {10.1093/mnras/stx3078}, \href
  {https://ui.adsabs.harvard.edu/\#abs/2018MNRAS.474.3976G} {474, 3976}

\bibitem[\protect\citeauthoryear{{Hinshaw} et~al.,}{{Hinshaw}
  et~al.}{2013}]{WMAP_9}
{Hinshaw} G.,  et~al., 2013, \mn@doi [The Astrophysical Journal Supplement
  Series] {10.1088/0067-0049/208/2/19}, \href
  {https://ui.adsabs.harvard.edu/\#abs/2013ApJS..208...19H} {208, 19}

\bibitem[\protect\citeauthoryear{{Hopkins}, {Kere{\v{s}}}, {O{\~n}orbe},
  {Faucher-Gigu{\`e}re}, {Quataert}, {Murray}  \& {Bullock}}{{Hopkins}
  et~al.}{2014}]{FIRE}
{Hopkins} P.~F.,  {Kere{\v{s}}} D.,  {O{\~n}orbe} J.,  {Faucher-Gigu{\`e}re}
  C.-A.,  {Quataert} E.,  {Murray} N.,   {Bullock} J.~S.,  2014, \mn@doi
  [\mnras] {10.1093/mnras/stu1738}, \href
  {https://ui.adsabs.harvard.edu/\#abs/2014MNRAS.445..581H} {445, 581}

\bibitem[\protect\citeauthoryear{{Jiang}, {Helly}, {Cole}  \& {Frenk}}{{Jiang}
  et~al.}{2014}]{Jiang:2014}
{Jiang} L.,  {Helly} J.~C.,  {Cole} S.,   {Frenk} C.~S.,  2014, \mn@doi
  [\mnras] {10.1093/mnras/stu390}, \href
  {https://ui.adsabs.harvard.edu/\#abs/2014MNRAS.440.2115J} {440, 2115}

\bibitem[\protect\citeauthoryear{{Jing}, {Wang}, {Li}, {Liao}, {Wang}, {Guo}
  \& {Gao}}{{Jing} et~al.}{2018}]{Jing:2018}
{Jing} Y.,  {Wang} C.,  {Li} R.,  {Liao} S.,  {Wang} J.,  {Guo} Q.,   {Gao} L.,
   2018, arXiv e-prints, \href
  {https://ui.adsabs.harvard.edu/\#abs/2018arXiv181109070J} {p.
  arXiv:1811.09070}

\bibitem[\protect\citeauthoryear{Jones, Oliphant, Peterson  et~al.}{Jones
  et~al.}{01  }]{scipy}
Jones E.,  Oliphant T.,  Peterson P.,   et~al., 2001--, {SciPy}: Open source
  scientific tools for {Python}, \url {http://www.scipy.org/}

\bibitem[\protect\citeauthoryear{{Kaviraj} et~al.,}{{Kaviraj}
  et~al.}{2017}]{Kaviraj:2017}
{Kaviraj} S.,  et~al., 2017, \mn@doi [\mnras] {10.1093/mnras/stx126}, \href
  {https://ui.adsabs.harvard.edu/\#abs/2017MNRAS.467.4739K} {467, 4739}

\bibitem[\protect\citeauthoryear{{Khandai}, {Di Matteo}, {Croft}, {Wilkins},
  {Feng}, {Tucker}, {DeGraf}  \& {Liu}}{{Khandai} et~al.}{2015}]{MassiveBlack2}
{Khandai} N.,  {Di Matteo} T.,  {Croft} R.,  {Wilkins} S.,  {Feng} Y.,
  {Tucker} E.,  {DeGraf} C.,   {Liu} M.-S.,  2015, \mn@doi [\mnras]
  {10.1093/mnras/stv627}, \href
  {https://ui.adsabs.harvard.edu/\#abs/2015MNRAS.450.1349K} {450, 1349}

\bibitem[\protect\citeauthoryear{{Komatsu} et~al.,}{{Komatsu}
  et~al.}{2011}]{WMAP_7}
{Komatsu} E.,  et~al., 2011, \mn@doi [\apjs] {10.1088/0067-0049/192/2/18},
  \href {http://adsabs.harvard.edu/abs/2011ApJS..192...18K} {192, 18}

\bibitem[\protect\citeauthoryear{{Lane}, {Salinas}  \& {Richtler}}{{Lane}
  et~al.}{2015}]{Lane:2015}
{Lane} R.~R.,  {Salinas} R.,   {Richtler} T.,  2015, \mn@doi [\aap]
  {10.1051/0004-6361/201424074}, \href
  {https://ui.adsabs.harvard.edu/\#abs/2015A&A...574A..93L} {574, A93}

\bibitem[\protect\citeauthoryear{{Le Brun}, {McCarthy}, {Schaye}  \&
  {Ponman}}{{Le Brun} et~al.}{2014}]{LeBrun:2014}
{Le Brun} A.~M.~C.,  {McCarthy} I.~G.,  {Schaye} J.,   {Ponman} T.~J.,  2014,
  \mn@doi [\mnras] {10.1093/mnras/stu608}, \href
  {http://cdsads.u-strasbg.fr/abs/2014MNRAS.441.1270L} {441, 1270}

\bibitem[\protect\citeauthoryear{{Leauthaud} et~al.,}{{Leauthaud}
  et~al.}{2012}]{Leauthaud:2012}
{Leauthaud} A.,  et~al., 2012, \mn@doi [\apj] {10.1088/0004-637X/744/2/159},
  \href {http://cdsads.u-strasbg.fr/abs/2012ApJ...744..159L} {744, 159}

\bibitem[\protect\citeauthoryear{{Legrand} et~al.,}{{Legrand}
  et~al.}{2018}]{Legrand:2018}
{Legrand} L.,  et~al., 2018, arXiv e-prints, \href
  {http://cdsads.u-strasbg.fr/abs/2018arXiv181010557L} {}

\bibitem[\protect\citeauthoryear{{Lovell} et~al.,}{{Lovell}
  et~al.}{2018}]{IllustrisTNG_DMfractions}
{Lovell} M.~R.,  et~al., 2018, \mn@doi [\mnras] {10.1093/mnras/sty2339}, \href
  {https://ui.adsabs.harvard.edu/\#abs/2018MNRAS.481.1950L} {481, 1950}

\bibitem[\protect\citeauthoryear{{McAlpine} et~al.,}{{McAlpine}
  et~al.}{2016}]{EAGLE_public}
{McAlpine} S.,  et~al., 2016, \mn@doi [Astronomy and Computing]
  {10.1016/j.ascom.2016.02.004}, \href
  {https://ui.adsabs.harvard.edu/\#abs/2016A&C....15...72M} {15, 72}

\bibitem[\protect\citeauthoryear{{Monna} et~al.,}{{Monna}
  et~al.}{2017}]{Monna:2017}
{Monna} A.,  et~al., 2017, \mn@doi [\mnras] {10.1093/mnras/stw3048}, \href
  {https://ui.adsabs.harvard.edu/\#abs/2017MNRAS.465.4589M} {465, 4589}

\bibitem[\protect\citeauthoryear{{Moster}, {Somerville}, {Maulbetsch}, {van den
  Bosch}, {Macci{\`o}}, {Naab}  \& {Oser}}{{Moster} et~al.}{2010}]{Moster:2010}
{Moster} B.~P.,  {Somerville} R.~S.,  {Maulbetsch} C.,  {van den Bosch} F.~C.,
  {Macci{\`o}} A.~V.,  {Naab} T.,   {Oser} L.,  2010, \mn@doi [\apj]
  {10.1088/0004-637X/710/2/903}, \href
  {https://ui.adsabs.harvard.edu/\#abs/2010ApJ...710..903M} {710, 903}

\bibitem[\protect\citeauthoryear{{Moster}, {Naab}  \& {White}}{{Moster}
  et~al.}{2013}]{Moster:2013}
{Moster} B.~P.,  {Naab} T.,   {White} S.~D.~M.,  2013, \mn@doi [\mnras]
  {10.1093/mnras/sts261}, \href
  {http://cdsads.u-strasbg.fr/abs/2013MNRAS.428.3121M} {428, 3121}

\bibitem[\protect\citeauthoryear{{Munshi} et~al.,}{{Munshi}
  et~al.}{2013}]{Munshi:2013}
{Munshi} F.,  et~al., 2013, \mn@doi [\apj] {10.1088/0004-637X/766/1/56}, \href
  {http://cdsads.u-strasbg.fr/abs/2013ApJ...766...56M} {766, 56}

\bibitem[\protect\citeauthoryear{{Naiman} et~al.,}{{Naiman}
  et~al.}{2018}]{Naiman:2018}
{Naiman} J.~P.,  et~al., 2018, \mn@doi [\mnras] {10.1093/mnras/sty618}, \href
  {https://ui.adsabs.harvard.edu/\#abs/2018MNRAS.477.1206N} {477, 1206}

\bibitem[\protect\citeauthoryear{{Nelson} et~al.,}{{Nelson}
  et~al.}{2015}]{Nelson:2015}
{Nelson} D.,  et~al., 2015, \mn@doi [Astronomy and Computing]
  {10.1016/j.ascom.2015.09.003}, \href
  {https://ui.adsabs.harvard.edu/\#abs/2015A&C....13...12N} {13, 12}

\bibitem[\protect\citeauthoryear{{Nelson} et~al.,}{{Nelson}
  et~al.}{2018}]{IllustrisTNG_public}
{Nelson} D.,  et~al., 2018, arXiv e-prints, \href
  {https://ui.adsabs.harvard.edu/\#abs/2018arXiv181205609N} {p.
  arXiv:1812.05609}

\bibitem[\protect\citeauthoryear{{Nelson} et~al.,}{{Nelson}
  et~al.}{2019}]{IllustrisTNG_50_bh}
{Nelson} D.,  et~al., 2019, arXiv e-prints, \href
  {https://ui.adsabs.harvard.edu/\#abs/2019arXiv190205554N} {p.
  arXiv:1902.05554}

\bibitem[\protect\citeauthoryear{{Niemiec} et~al.,}{{Niemiec}
  et~al.}{2017}]{Niemiec:2017}
{Niemiec} A.,  et~al., 2017, \mn@doi [\mnras] {10.1093/mnras/stx1667}, \href
  {http://cdsads.u-strasbg.fr/abs/2017MNRAS.471.1153N} {471, 1153}

\bibitem[\protect\citeauthoryear{{Niemiec}, {Jullo}, {Giocoli}, {Limousin}  \&
  {Jauzac}}{{Niemiec} et~al.}{2018}]{Niemiec:2018}
{Niemiec} A.,  {Jullo} E.,  {Giocoli} C.,  {Limousin} M.,   {Jauzac} M.,  2018,
  arXiv e-prints, \href
  {https://ui.adsabs.harvard.edu/\#abs/2018arXiv181104996N} {p.
  arXiv:1811.04996}

\bibitem[\protect\citeauthoryear{{Ocvirk}, {Pichon}  \& {Teyssier}}{{Ocvirk}
  et~al.}{2008}]{MareNostrum}
{Ocvirk} P.,  {Pichon} C.,   {Teyssier} R.,  2008, \mn@doi [\mnras]
  {10.1111/j.1365-2966.2008.13763.x}, \href
  {https://ui.adsabs.harvard.edu/\#abs/2008MNRAS.390.1326O} {390, 1326}

\bibitem[\protect\citeauthoryear{{Pakmor}, {Springel}, {Bauer}, {Mocz},
  {Munoz}, {Ohlmann}, {Schaal}  \& {Zhu}}{{Pakmor} et~al.}{2016}]{Pakmor:2016}
{Pakmor} R.,  {Springel} V.,  {Bauer} A.,  {Mocz} P.,  {Munoz} D.~J.,
  {Ohlmann} S.~T.,  {Schaal} K.,   {Zhu} C.,  2016, \mn@doi [\mnras]
  {10.1093/mnras/stv2380}, \href
  {http://cdsads.u-strasbg.fr/abs/2016MNRAS.455.1134P} {455, 1134}

\bibitem[\protect\citeauthoryear{{Pillepich} et~al.,}{{Pillepich}
  et~al.}{2018a}]{IllustrisTNG_1}
{Pillepich} A.,  et~al., 2018a, \mn@doi [\mnras] {10.1093/mnras/stx2656}, \href
  {https://ui.adsabs.harvard.edu/\#abs/2018MNRAS.473.4077P} {473, 4077}

\bibitem[\protect\citeauthoryear{{Pillepich} et~al.,}{{Pillepich}
  et~al.}{2018b}]{Pillepich:2018}
{Pillepich} A.,  et~al., 2018b, \mn@doi [\mnras] {10.1093/mnras/stx3112}, \href
  {https://ui.adsabs.harvard.edu/\#abs/2018MNRAS.475..648P} {475, 648}

\bibitem[\protect\citeauthoryear{{Pillepich} et~al.,}{{Pillepich}
  et~al.}{2019}]{IllustrisTNG_50_star}
{Pillepich} A.,  et~al., 2019, arXiv e-prints, \href
  {https://ui.adsabs.harvard.edu/\#abs/2019arXiv190205553P} {p.
  arXiv:1902.05553}

\bibitem[\protect\citeauthoryear{{Planck Collaboration} et~al.,}{{Planck
  Collaboration} et~al.}{2014}]{Planck_cosmo_old}
{Planck Collaboration} et~al., 2014, \mn@doi [\aap]
  {10.1051/0004-6361/201321591}, \href
  {http://adsabs.harvard.edu/abs/2014A%26A...571A..16P} {571, A16}

\bibitem[\protect\citeauthoryear{{Planck Collaboration} et~al.,}{{Planck
  Collaboration} et~al.}{2015}]{Planck_cosmo}
{Planck Collaboration} et~al., 2015, preprint, \href
  {http://adsabs.harvard.edu/abs/2015arXiv150201589P} {} (\mn@eprint {arXiv}
  {1502.01589})

\bibitem[\protect\citeauthoryear{{Price-Whelan} et~al.,}{{Price-Whelan}
  et~al.}{2018}]{astropy:2018}
{Price-Whelan} A.~M.,  et~al., 2018, \mn@doi [\aj] {10.3847/1538-3881/aabc4f},
  \href {https://ui.adsabs.harvard.edu/#abs/2018AJ....156..123T} {156, 123}

\bibitem[\protect\citeauthoryear{{Remus}, {Dolag}, {Bachmann}, {Beck},
  {Burkert}, {Hirschmann}  \& {Teklu}}{{Remus} et~al.}{2015}]{Remus:2015}
{Remus} R.-S.,  {Dolag} K.,  {Bachmann} L.~K.,  {Beck} A.~M.,  {Burkert} A.,
  {Hirschmann} M.,   {Teklu} A.,  2015, in {Ziegler} B.~L.,  {Combes} F.,
  {Dannerbauer} H.,   {Verdugo} M.,  eds,  IAU Symposium Vol. 309, Galaxies in
  3D across the Universe. pp 145--148, \mn@doi{10.1017/S1743921314009491}

\bibitem[\protect\citeauthoryear{{Rodriguez-Gomez} et~al.,}{{Rodriguez-Gomez}
  et~al.}{2015}]{Rodriguez:2015}
{Rodriguez-Gomez} V.,  et~al., 2015, \mn@doi [\mnras] {10.1093/mnras/stv264},
  \href {https://ui.adsabs.harvard.edu/\#abs/2015MNRAS.449...49R} {449, 49}

\bibitem[\protect\citeauthoryear{{Rodr{\'{\i}}guez-Puebla}, {Avila-Reese},
  {Yang}, {Foucaud}, {Drory}  \& {Jing}}{{Rodr{\'{\i}}guez-Puebla}
  et~al.}{2015}]{Rodrigues:2015}
{Rodr{\'{\i}}guez-Puebla} A.,  {Avila-Reese} V.,  {Yang} X.,  {Foucaud} S.,
  {Drory} N.,   {Jing} Y.~P.,  2015, \mn@doi [\apj]
  {10.1088/0004-637X/799/2/130}, \href
  {http://cdsads.u-strasbg.fr/abs/2015ApJ...799..130R} {799, 130}

\bibitem[\protect\citeauthoryear{{Salinas}, {Richtler}, {Bassino}, {Romanowsky}
   \& {Schuberth}}{{Salinas} et~al.}{2012}]{Salinas:2012}
{Salinas} R.,  {Richtler} T.,  {Bassino} L.~P.,  {Romanowsky} A.~J.,
  {Schuberth} Y.,  2012, \mn@doi [\aap] {10.1051/0004-6361/201116517}, \href
  {https://ui.adsabs.harvard.edu/\#abs/2012A&A...538A..87S} {538, A87}

\bibitem[\protect\citeauthoryear{{Salmon} \& {Warren}}{{Salmon} \&
  {Warren}}{1994}]{SalmonWarren:1994}
{Salmon} J.~K.,  {Warren} M.~S.,  1994, \mn@doi [Journal of Computational
  Physics] {10.1006/jcph.1994.1050}, \href
  {https://ui.adsabs.harvard.edu/abs/1994JCoPh.111..136S} {111, 136}

\bibitem[\protect\citeauthoryear{{Schaye} et~al.,}{{Schaye}
  et~al.}{2010}]{Schaye:2010}
{Schaye} J.,  et~al., 2010, \mn@doi [\mnras]
  {10.1111/j.1365-2966.2009.16029.x}, \href
  {http://cdsads.u-strasbg.fr/abs/2010MNRAS.402.1536S} {402, 1536}

\bibitem[\protect\citeauthoryear{{Schaye} et~al.,}{{Schaye}
  et~al.}{2015}]{EAGLE_sim}
{Schaye} J.,  et~al., 2015, \mn@doi [\mnras] {10.1093/mnras/stu2058}, \href
  {https://ui.adsabs.harvard.edu/\#abs/2015MNRAS.446..521S} {446, 521}

\bibitem[\protect\citeauthoryear{{Shan} et~al.,}{{Shan}
  et~al.}{2017}]{Shan:2017}
{Shan} H.,  et~al., 2017, \mn@doi [\apj] {10.3847/1538-4357/aa6c68}, \href
  {http://cdsads.u-strasbg.fr/abs/2017ApJ...840..104S} {840, 104}

\bibitem[\protect\citeauthoryear{{Sijacki}, {Vogelsberger}, {Genel},
  {Springel}, {Torrey}, {Snyder}, {Nelson}  \& {Hernquist}}{{Sijacki}
  et~al.}{2015}]{Sijacki:2015}
{Sijacki} D.,  {Vogelsberger} M.,  {Genel} S.,  {Springel} V.,  {Torrey} P.,
  {Snyder} G.~F.,  {Nelson} D.,   {Hernquist} L.,  2015, \mn@doi [\mnras]
  {10.1093/mnras/stv1340}, \href
  {https://ui.adsabs.harvard.edu/\#abs/2015MNRAS.452..575S} {452, 575}

\bibitem[\protect\citeauthoryear{{Springel}}{{Springel}}{2005}]{GADGET}
{Springel} V.,  2005, \mn@doi [\mnras] {10.1111/j.1365-2966.2005.09655.x},
  \href {http://adsabs.harvard.edu/abs/2005MNRAS.364.1105S} {364, 1105}

\bibitem[\protect\citeauthoryear{{Springel}}{{Springel}}{2010}]{Springel:2010}
{Springel} V.,  2010, \mn@doi [\mnras] {10.1111/j.1365-2966.2009.15715.x},
  \href {http://adsabs.harvard.edu/abs/2010MNRAS.401..791S} {401, 791}

\bibitem[\protect\citeauthoryear{{Springel}, {White}, {Tormen}  \&
  {Kauffmann}}{{Springel} et~al.}{2001}]{Springel:2001}
{Springel} V.,  {White} S.~D.~M.,  {Tormen} G.,   {Kauffmann} G.,  2001,
  \mn@doi [\mnras] {10.1046/j.1365-8711.2001.04912.x}, \href
  {http://adsabs.harvard.edu/abs/2001MNRAS.328..726S} {328, 726}

\bibitem[\protect\citeauthoryear{{Springel} et~al.,}{{Springel}
  et~al.}{2018}]{Springel:2018}
{Springel} V.,  et~al., 2018, \mn@doi [\mnras] {10.1093/mnras/stx3304}, \href
  {https://ui.adsabs.harvard.edu/\#abs/2018MNRAS.475..676S} {475, 676}

\bibitem[\protect\citeauthoryear{{Teklu}, {Remus}, {Dolag}, {Beck}, {Burkert},
  {Schmidt}, {Schulze}  \& {Steinborn}}{{Teklu} et~al.}{2015}]{Teklu:2015}
{Teklu} A.~F.,  {Remus} R.-S.,  {Dolag} K.,  {Beck} A.~M.,  {Burkert} A.,
  {Schmidt} A.~S.,  {Schulze} F.,   {Steinborn} L.~K.,  2015, \mn@doi [\apj]
  {10.1088/0004-637X/812/1/29}, \href
  {http://adsabs.harvard.edu/abs/2015ApJ...812...29T} {812, 29}

\bibitem[\protect\citeauthoryear{{Teyssier}}{{Teyssier}}{2002}]{RAMSES}
{Teyssier} R.,  2002, \mn@doi [\aap] {10.1051/0004-6361:20011817}, \href
  {http://adsabs.harvard.edu/abs/2002A%26A...385..337T} {385, 337}

\bibitem[\protect\citeauthoryear{{The EAGLE team}}{{The EAGLE
  team}}{2017}]{EAGLE_particles}
{The EAGLE team} 2017, arXiv e-prints, \href
  {https://ui.adsabs.harvard.edu/\#abs/2017arXiv170609899T} {p.
  arXiv:1706.09899}

\bibitem[\protect\citeauthoryear{{Thob} et~al.,}{{Thob}
  et~al.}{2019}]{Thob:2019}
{Thob} A. C.~R.,  et~al., 2019, \mn@doi [\mnras] {10.1093/mnras/stz448}, \href
  {https://ui.adsabs.harvard.edu/\#abs/2019MNRAS.485..972T} {485, 972}

\bibitem[\protect\citeauthoryear{{Tinker}}{{Tinker}}{2017}]{Tinker:2017}
{Tinker} J.~L.,  2017, \mn@doi [\mnras] {10.1093/mnras/stx287}, \href
  {http://cdsads.u-strasbg.fr/abs/2017MNRAS.467.3533T} {467, 3533}

\bibitem[\protect\citeauthoryear{{Tinker}, {George}, {Leauthaud}, {Bundy},
  {Finoguenov}, {Massey}, {Rhodes}  \& {Wechsler}}{{Tinker}
  et~al.}{2012}]{Tinker:2012}
{Tinker} J.~L.,  {George} M.~R.,  {Leauthaud} A.,  {Bundy} K.,  {Finoguenov}
  A.,  {Massey} R.,  {Rhodes} J.,   {Wechsler} R.~H.,  2012, \mn@doi [\apjl]
  {10.1088/2041-8205/755/1/L5}, \href
  {http://cdsads.u-strasbg.fr/abs/2012ApJ...755L...5T} {755, L5}

\bibitem[\protect\citeauthoryear{{Tinker}, {Leauthaud}, {Bundy}, {George},
  {Behroozi}, {Massey}, {Rhodes}  \& {Wechsler}}{{Tinker}
  et~al.}{2013}]{Tinker:2013}
{Tinker} J.~L.,  {Leauthaud} A.,  {Bundy} K.,  {George} M.~R.,  {Behroozi} P.,
  {Massey} R.,  {Rhodes} J.,   {Wechsler} R.~H.,  2013, \mn@doi [\apj]
  {10.1088/0004-637X/778/2/93}, \href
  {http://cdsads.u-strasbg.fr/abs/2013ApJ...778...93T} {778, 93}

\bibitem[\protect\citeauthoryear{{Torrey} et~al.,}{{Torrey}
  et~al.}{2015}]{Torrey:2015}
{Torrey} P.,  et~al., 2015, \mn@doi [\mnras] {10.1093/mnras/stu2592}, \href
  {https://ui.adsabs.harvard.edu/\#abs/2015MNRAS.447.2753T} {447, 2753}

\bibitem[\protect\citeauthoryear{{Trujillo} et~al.,}{{Trujillo}
  et~al.}{2019}]{Trujillo:2019}
{Trujillo} I.,  et~al., 2019, \mn@doi [\mnras] {10.1093/mnras/stz771}, \href
  {https://ui.adsabs.harvard.edu/\#abs/2019MNRAS.tmp..733T} {p.~733}

\bibitem[\protect\citeauthoryear{{Tweed}, {Devriendt}, {Blaizot}, {Colombi}  \&
  {Slyz}}{{Tweed} et~al.}{2009}]{Tweed:2009}
{Tweed} D.,  {Devriendt} J.,  {Blaizot} J.,  {Colombi} S.,   {Slyz} A.,  2009,
  \mn@doi [\aap] {10.1051/0004-6361/200911787}, \href
  {https://ui.adsabs.harvard.edu/\#abs/2009A&A...506..647T} {506, 647}

\bibitem[\protect\citeauthoryear{{Vogelsberger} et~al.,}{{Vogelsberger}
  et~al.}{2014a}]{Illustris}
{Vogelsberger} M.,  et~al., 2014a, \mn@doi [\mnras] {10.1093/mnras/stu1536},
  \href {https://ui.adsabs.harvard.edu/\#abs/2014MNRAS.444.1518V} {444, 1518}

\bibitem[\protect\citeauthoryear{{Vogelsberger} et~al.,}{{Vogelsberger}
  et~al.}{2014b}]{Illustris_nature}
{Vogelsberger} M.,  et~al., 2014b, \mn@doi [\nat] {10.1038/nature13316}, \href
  {https://ui.adsabs.harvard.edu/\#abs/2014Natur.509..177V} {509, 177}

\bibitem[\protect\citeauthoryear{{Wang}, {Dutton}, {Stinson}, {Macci{\`o}},
  {Penzo}, {Kang}, {Keller}  \& {Wadsley}}{{Wang} et~al.}{2015}]{NIHAO}
{Wang} L.,  {Dutton} A.~A.,  {Stinson} G.~S.,  {Macci{\`o}} A.~V.,  {Penzo} C.,
   {Kang} X.,  {Keller} B.~W.,   {Wadsley} J.,  2015, \mn@doi [\mnras]
  {10.1093/mnras/stv1937}, \href
  {https://ui.adsabs.harvard.edu/\#abs/2015MNRAS.454...83W} {454, 83}

\bibitem[\protect\citeauthoryear{{Weinberger} et~al.,}{{Weinberger}
  et~al.}{2017}]{Weinberger:2017}
{Weinberger} R.,  et~al., 2017, \mn@doi [\mnras] {10.1093/mnras/stw2944}, \href
  {http://cdsads.u-strasbg.fr/abs/2017MNRAS.465.3291W} {465, 3291}

\bibitem[\protect\citeauthoryear{{Yu}, {Ratra}  \& {Wang}}{{Yu}
  et~al.}{2018}]{Yu:2018}
{Yu} H.,  {Ratra} B.,   {Wang} F.-Y.,  2018, arXiv e-prints, \href
  {https://ui.adsabs.harvard.edu/\#abs/2018arXiv180905938Y} {p.
  arXiv:1809.05938}

\bibitem[\protect\citeauthoryear{{Zu} \& {Mandelbaum}}{{Zu} \&
  {Mandelbaum}}{2015}]{Zu:2015}
{Zu} Y.,  {Mandelbaum} R.,  2015, \mn@doi [\mnras] {10.1093/mnras/stv2062},
  \href {http://cdsads.u-strasbg.fr/abs/2015MNRAS.454.1161Z} {454, 1161}

\bibitem[\protect\citeauthoryear{{van Dokkum} et~al.,}{{van Dokkum}
  et~al.}{2018a}]{vanDokkum:2018a}
{van Dokkum} P.,  et~al., 2018a, \mn@doi [\nat] {10.1038/nature25767}, \href
  {https://ui.adsabs.harvard.edu/\#abs/2018Natur.555..629V} {555, 629}

\bibitem[\protect\citeauthoryear{{van Dokkum} et~al.,}{{van Dokkum}
  et~al.}{2018b}]{vanDokkum:2018b}
{van Dokkum} P.,  et~al., 2018b, \mn@doi [\apj] {10.3847/2041-8213/aab60b},
  \href {https://ui.adsabs.harvard.edu/\#abs/2018ApJ...856L..30V} {856, L30}

\bibitem[\protect\citeauthoryear{{van Dokkum}, {Danieli}, {Cohen}, {Romanowsky}
   \& {Conroy}}{{van Dokkum} et~al.}{2018c}]{vanDokkum:2018c}
{van Dokkum} P.,  {Danieli} S.,  {Cohen} Y.,  {Romanowsky} A.~J.,   {Conroy}
  C.,  2018c, \mn@doi [\apj] {10.3847/2041-8213/aada4d}, \href
  {https://ui.adsabs.harvard.edu/\#abs/2018ApJ...864L..18V} {864, L18}

\bibitem[\protect\citeauthoryear{{van Dokkum}, {Danieli}, {Abraham}, {Conroy}
  \& {Romanowsky}}{{van Dokkum} et~al.}{2019}]{vanDokkum:2019}
{van Dokkum} P.,  {Danieli} S.,  {Abraham} R.,  {Conroy} C.,   {Romanowsky}
  A.~J.,  2019, \mn@doi [\apj] {10.3847/2041-8213/ab0d92}, \href
  {https://ui.adsabs.harvard.edu/\#abs/2019ApJ...874L...5V} {874, L5}

\bibitem[\protect\citeauthoryear{{van de Sande} et~al.,}{{van de Sande}
  et~al.}{2019}]{vdSande:2019}
{van de Sande} J.,  et~al., 2019, \mn@doi [\mnras] {10.1093/mnras/sty3506},
  \href {https://ui.adsabs.harvard.edu/abs/2019MNRAS.484..869V} {484, 869}

\bibitem[\protect\citeauthoryear{{van der Walt}, {Colbert}  \&
  {Varoquaux}}{{van der Walt} et~al.}{2011}]{NumPy}
{van der Walt} S.,  {Colbert} S.~C.,   {Varoquaux} G.,  2011, \mn@doi
  [Computing in Science and Engineering] {10.1109/MCSE.2011.37}, \href
  {https://ui.adsabs.harvard.edu/abs/2011CSE....13b..22V} {13, 22}

\makeatother
\end{thebibliography}

\appendix
\section{Plots for individual oddballs}
\label{app_plots}

\begin{figure}
\begin{center}
\includegraphics[width=0.45\textwidth]{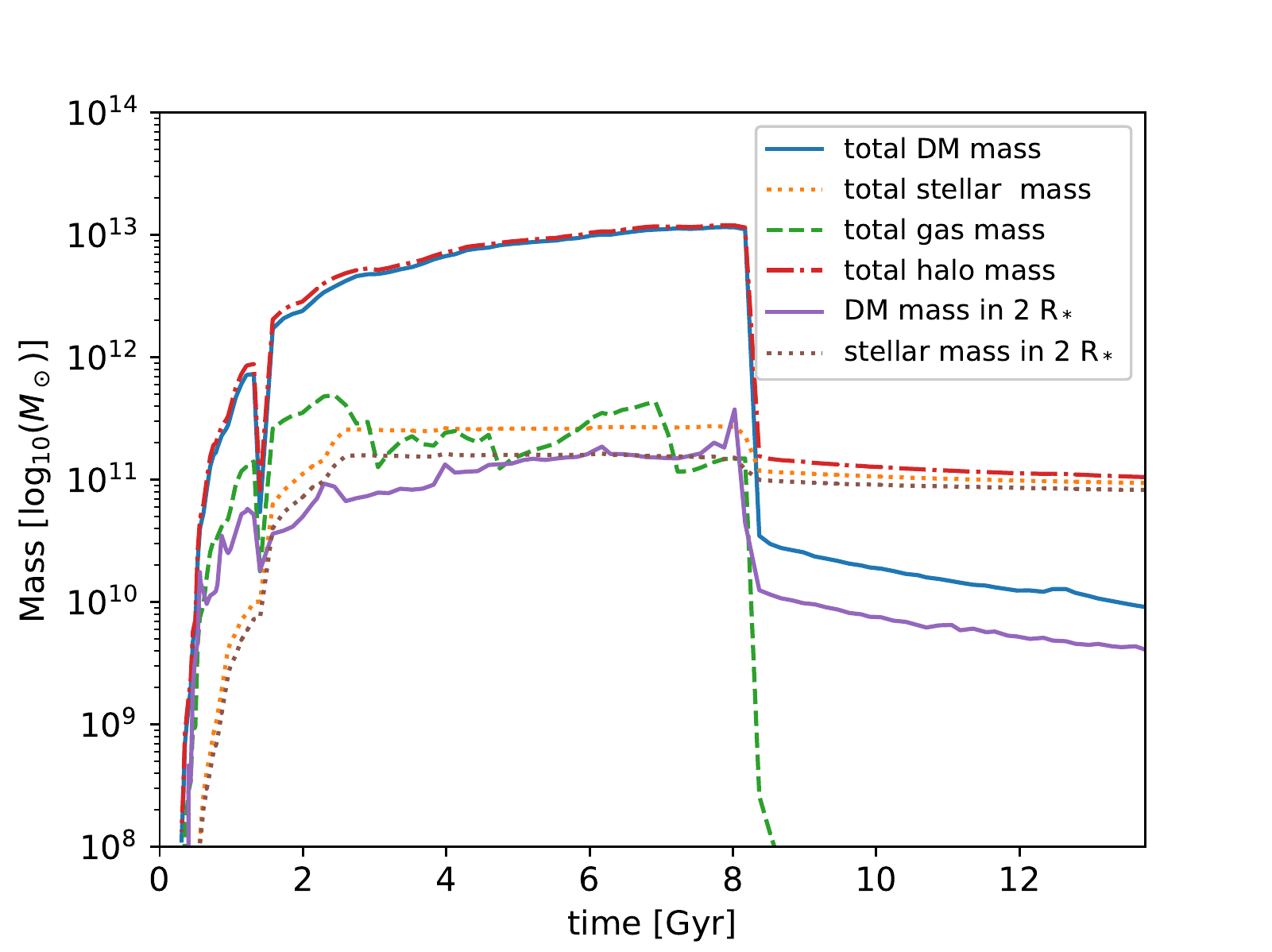}\\
\caption{Matter content in the oddball Illustris-476171 over time. The main branch of its merger tree was used to trace it.}
\label{mass_history476171}
\end{center}
\end{figure}
\begin{figure}
\begin{center}
\includegraphics[width=0.45\textwidth]{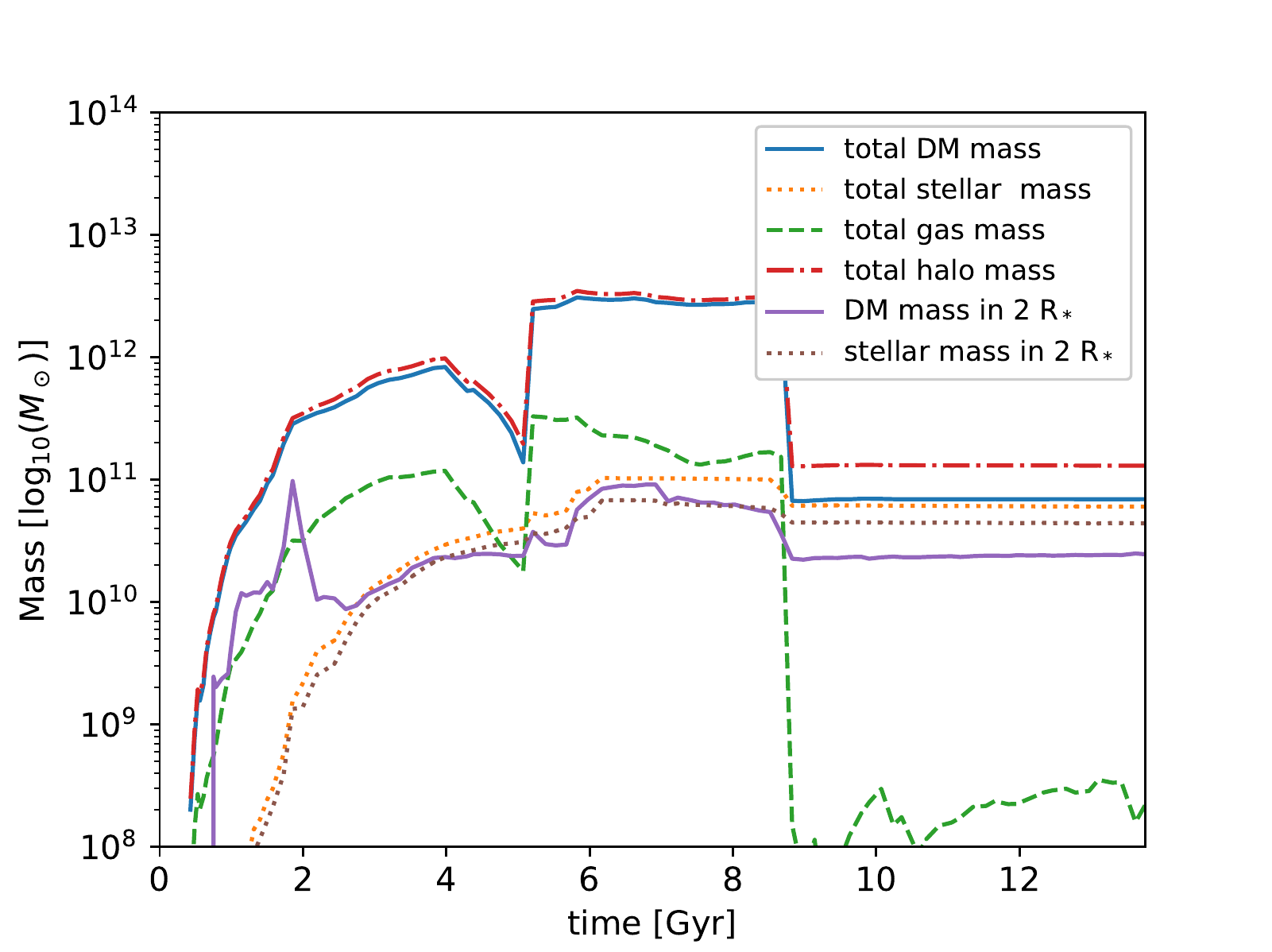}\\
\caption{Matter content in the oddball IllustrisTNG-585369 over time. The main branch of its merger tree was used to trace it.}
\label{mass_history585369}
\end{center}
\end{figure}
\begin{figure}
\begin{center}
\includegraphics[width=0.45\textwidth]{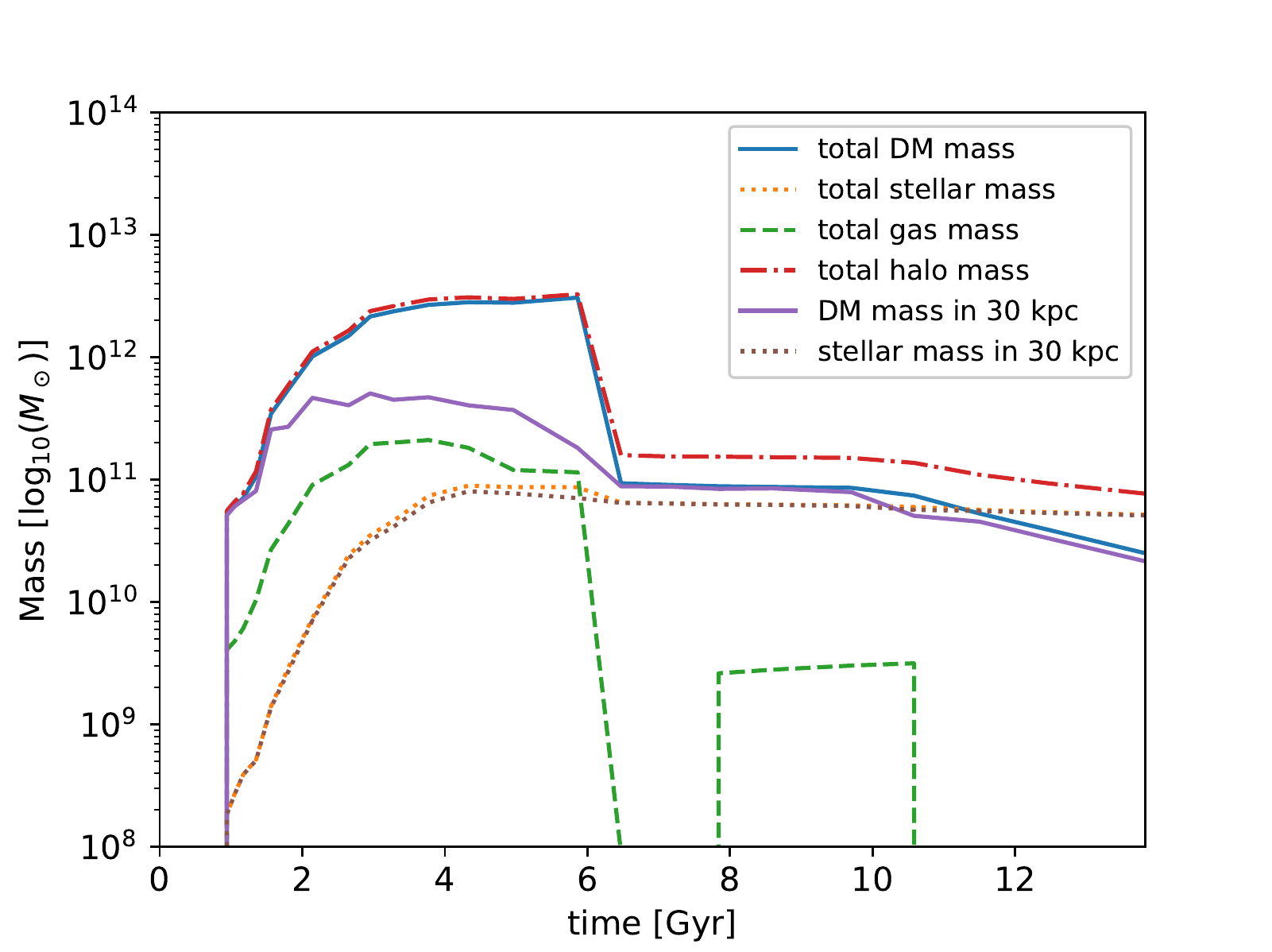}\\
\caption{Matter content in the oddball EAGLE-3274715 over time. The main branch of its merger tree was used to trace it.}
\label{mass_history3274715}
\end{center}
\end{figure}
\begin{figure}
\begin{center}
\includegraphics[width=0.45\textwidth]{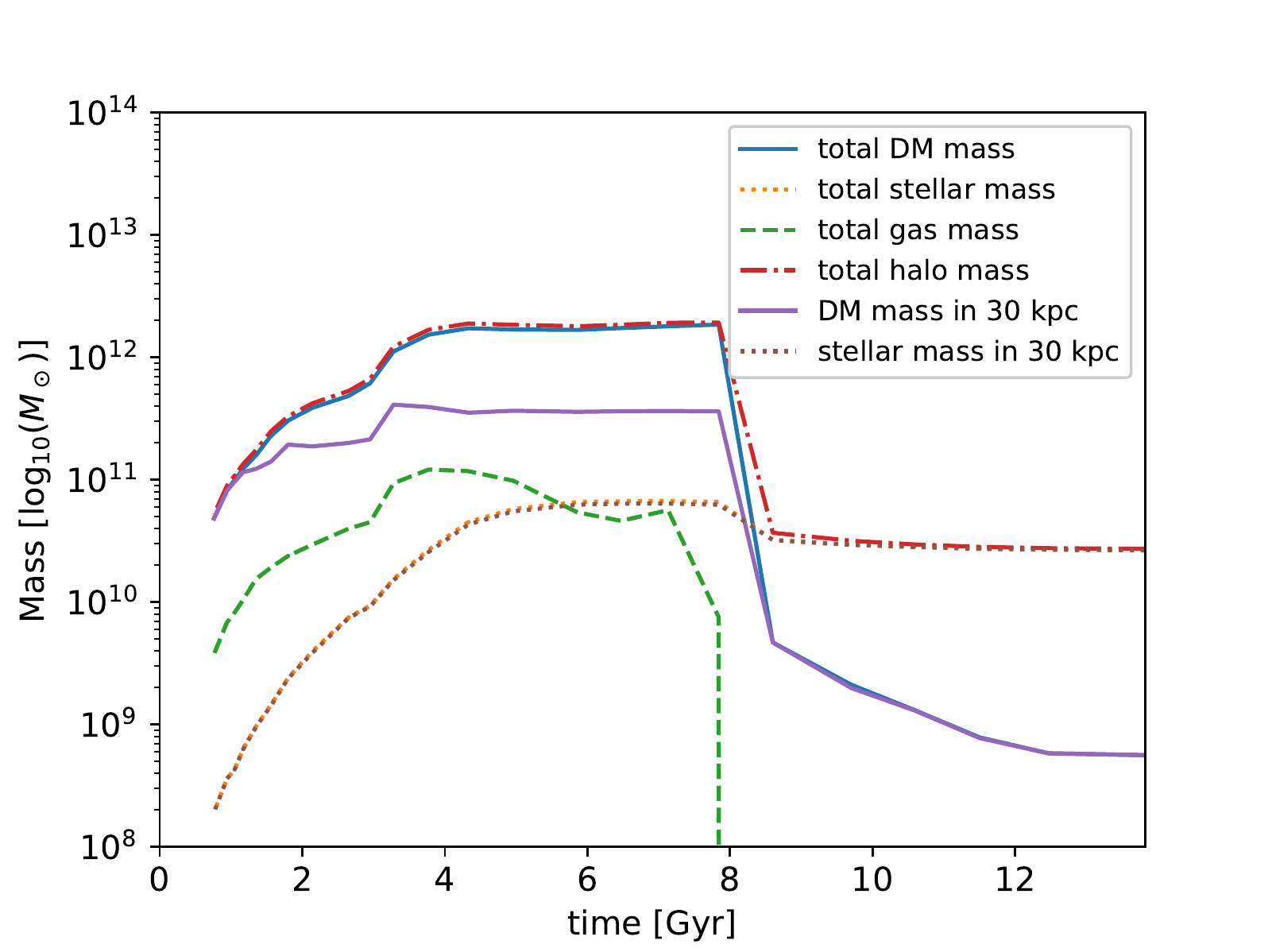}\\
\caption{Matter content in the oddball EAGLE-3868859 over time. The main branch of its merger tree was used to trace it.}
\label{mass_history3868859}
\end{center}
\end{figure}
\begin{figure}
\begin{center}
\includegraphics[width=0.45\textwidth]{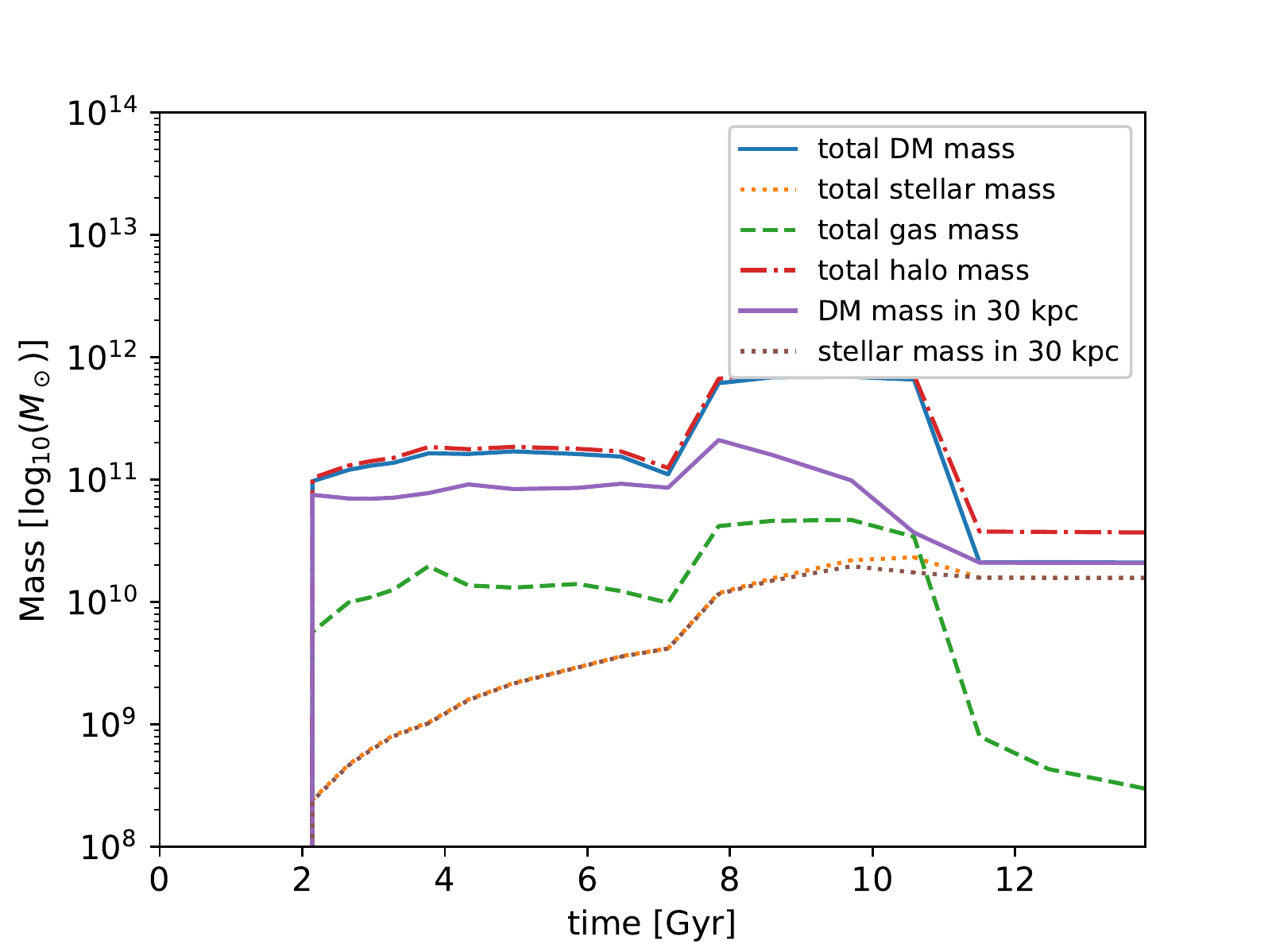}\\
\caption{Matter content in the oddball EAGLE-4209797 over time. The main branch of its merger tree was used to trace it.}
\label{mass_history4209797}
\end{center}
\end{figure}
\begin{figure}
\begin{center}
\includegraphics[width=0.45\textwidth]{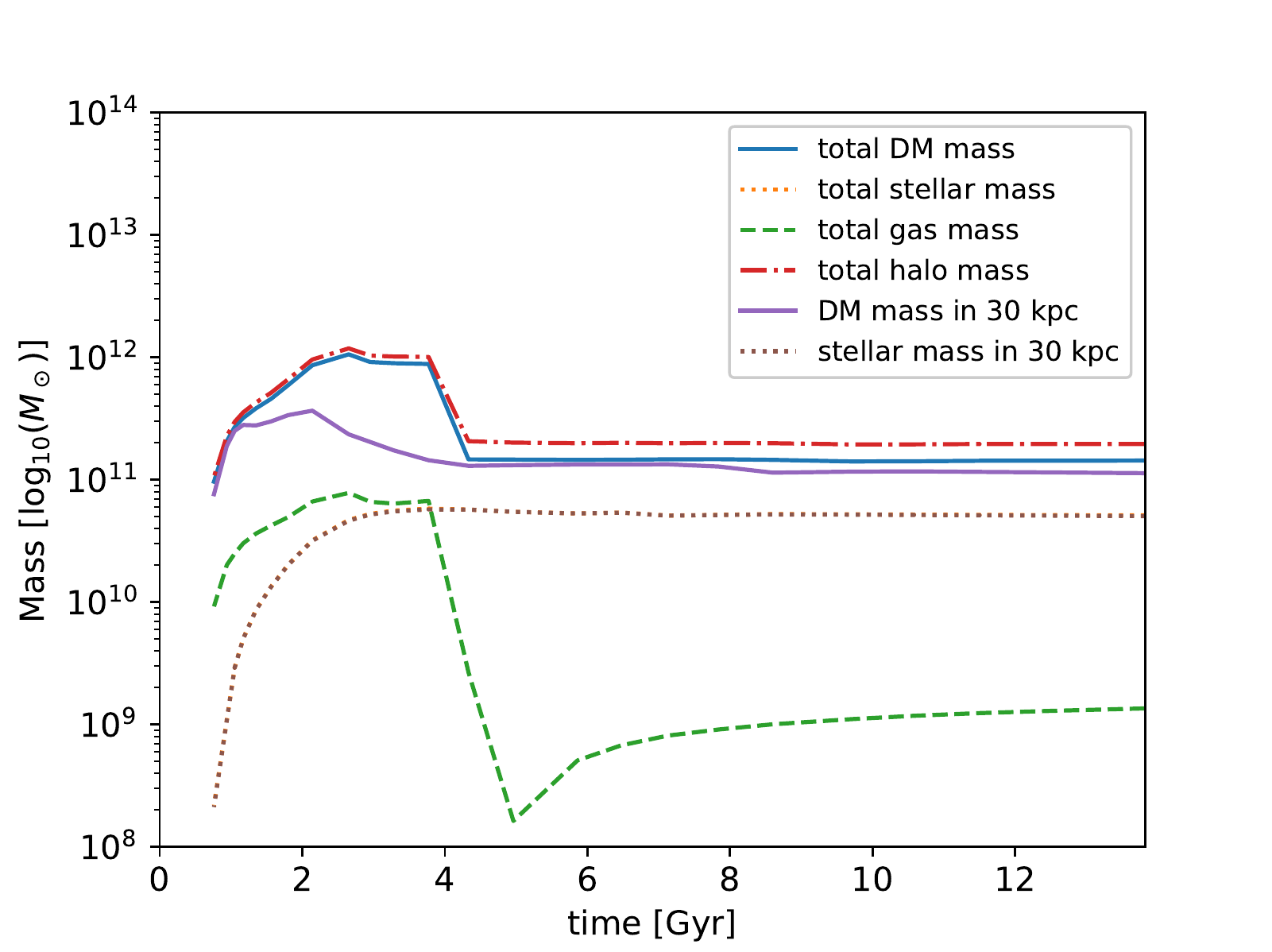}\\
\caption{Matter content in the oddball EAGLE-11419697 over time. The main branch of its merger tree was used to trace it.}
\label{mass_history11419697}
\end{center}
\end{figure}
\begin{figure}
\begin{center}
\includegraphics[width=0.45\textwidth]{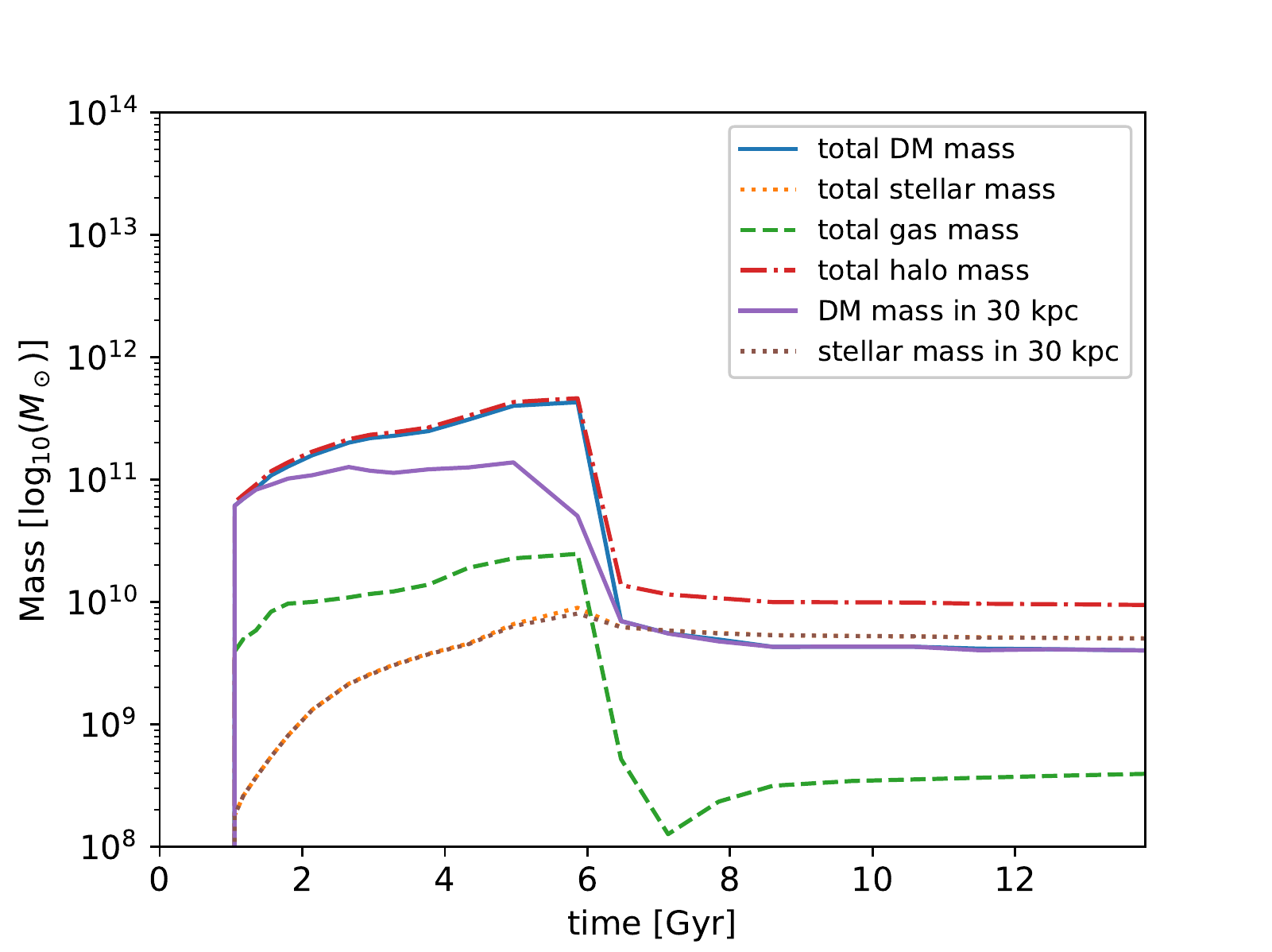}\\
\caption{Matter content in the oddball EAGLE-60521664 over time. The main branch of its merger tree was used to trace them.}
\label{mass_history60521664}
\end{center}
\end{figure}

\begin{figure}
\begin{center}
\includegraphics[width=0.45\textwidth]{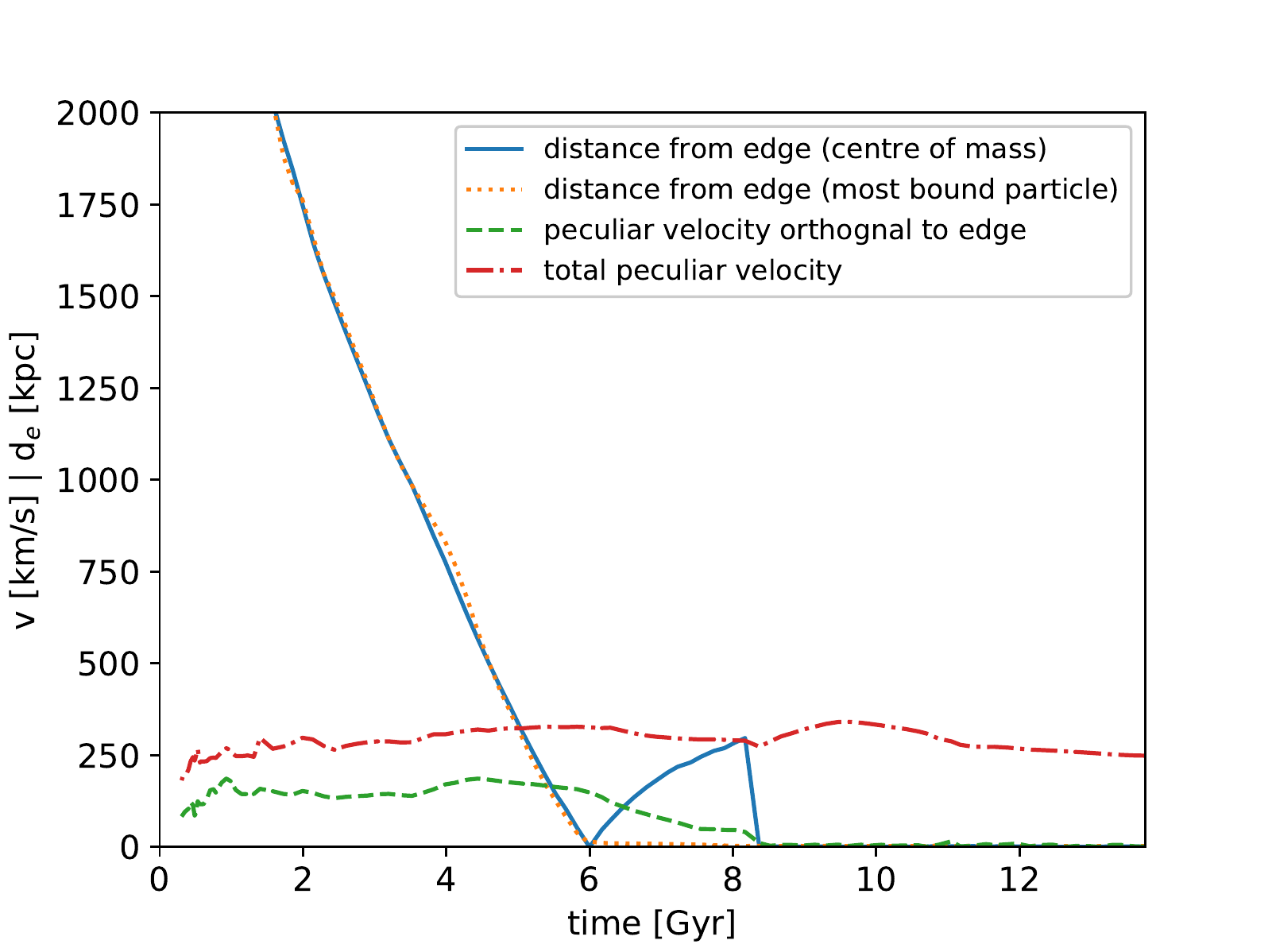}\\
\caption{Timeline of the distance from the edge and the peculiar velocities of the oddball Illustris-476171. The main branch of its merger tree was used to trace it.}
\label{dist_476171}
\end{center}
\end{figure}
\begin{figure}
\begin{center}
\includegraphics[width=0.45\textwidth]{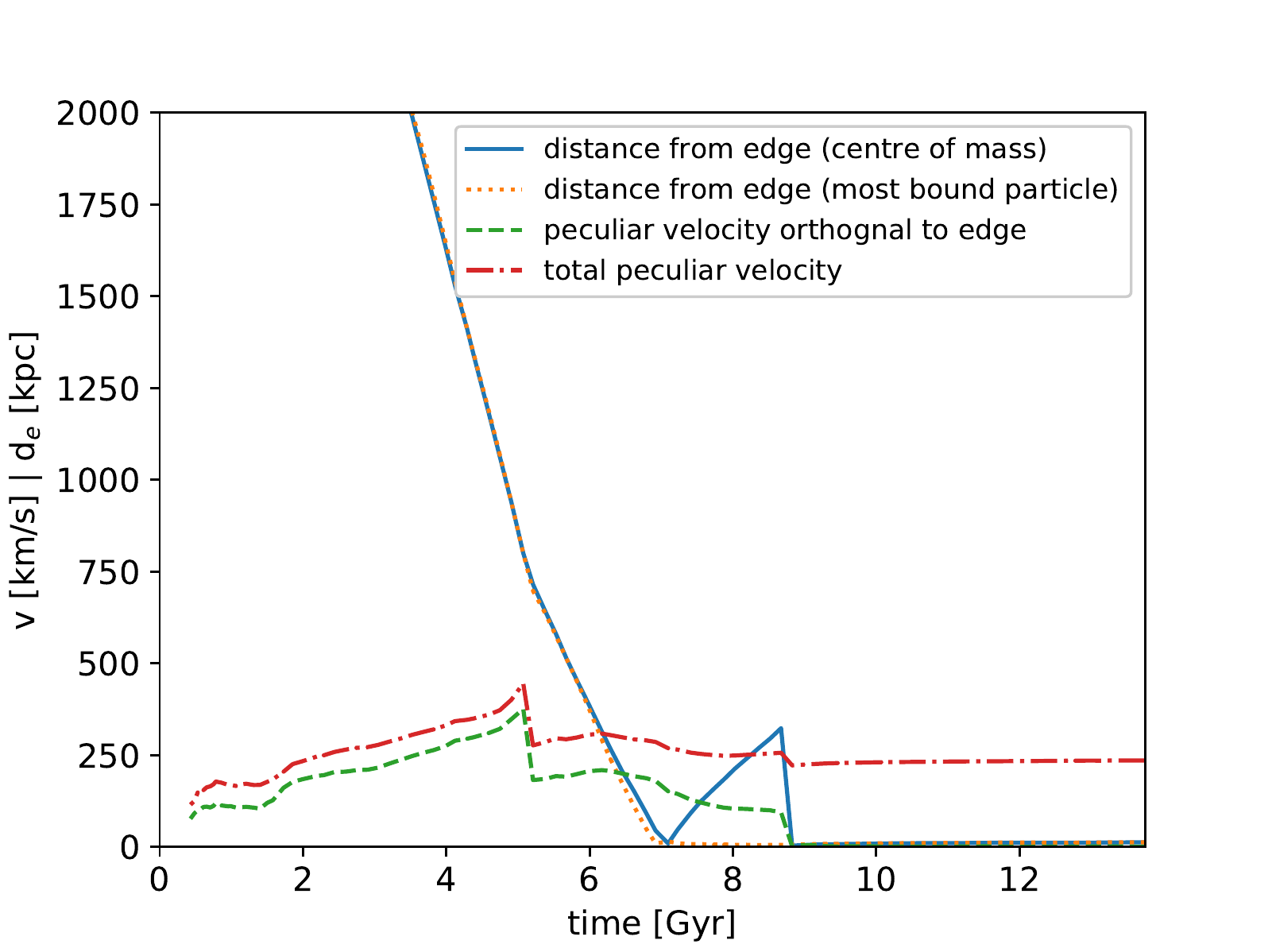}\\
\caption{Timeline of the distance from the edge and the peculiar velocities of the oddball IllustrisTNG-585369. The main branch of its merger tree was used to trace it.}
\label{dist_585369}
\end{center}
\end{figure}
\begin{figure}
\begin{center}
\includegraphics[width=0.45\textwidth]{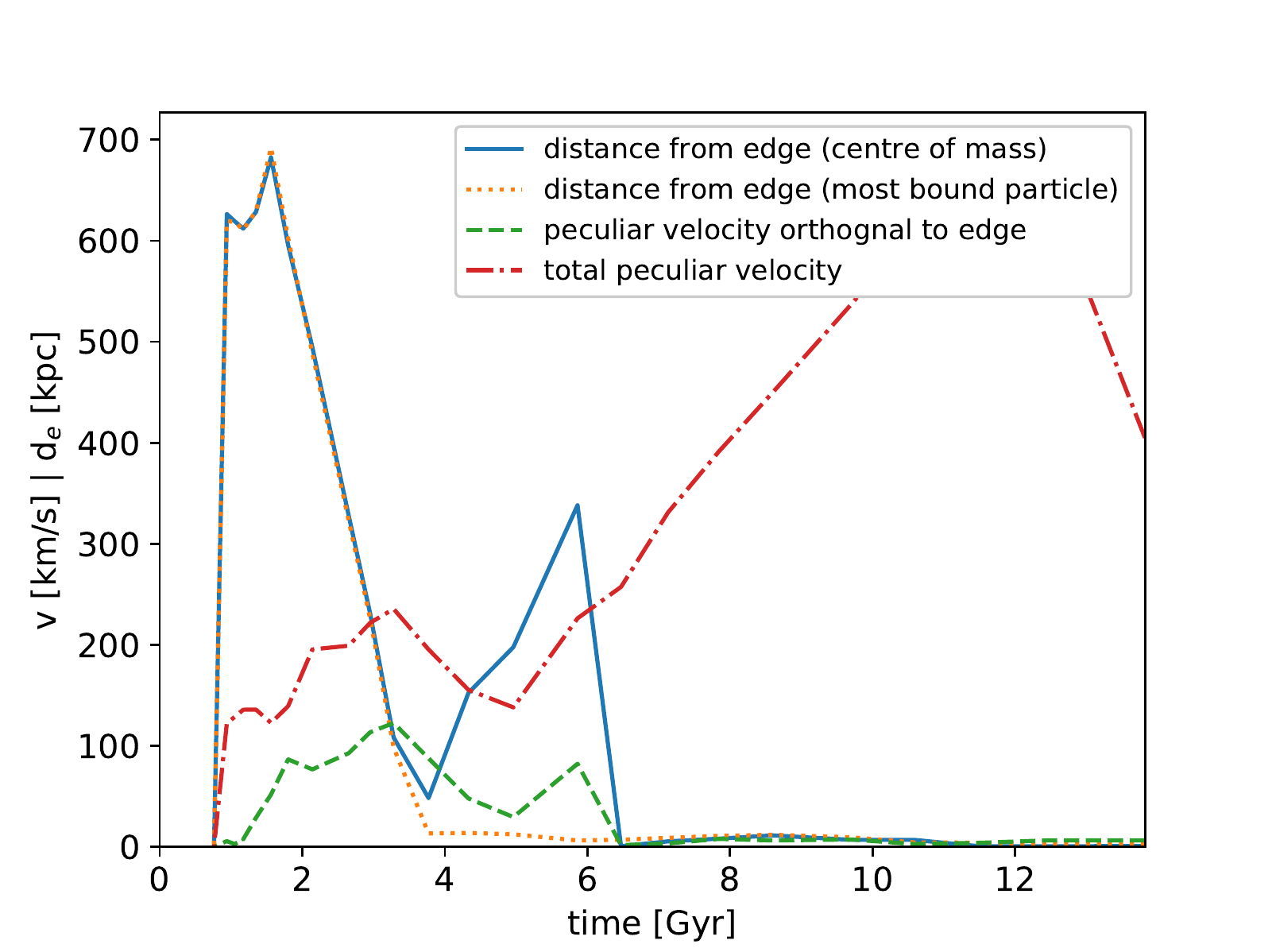}\\
\caption{Timeline of the distance from the edge and the peculiar velocities of the oddball EAGLE-3274715. The main branch of its merger tree was used to trace it.}
\label{dist_3274715}
\end{center}
\end{figure}
\begin{figure}
\begin{center}
\includegraphics[width=0.45\textwidth]{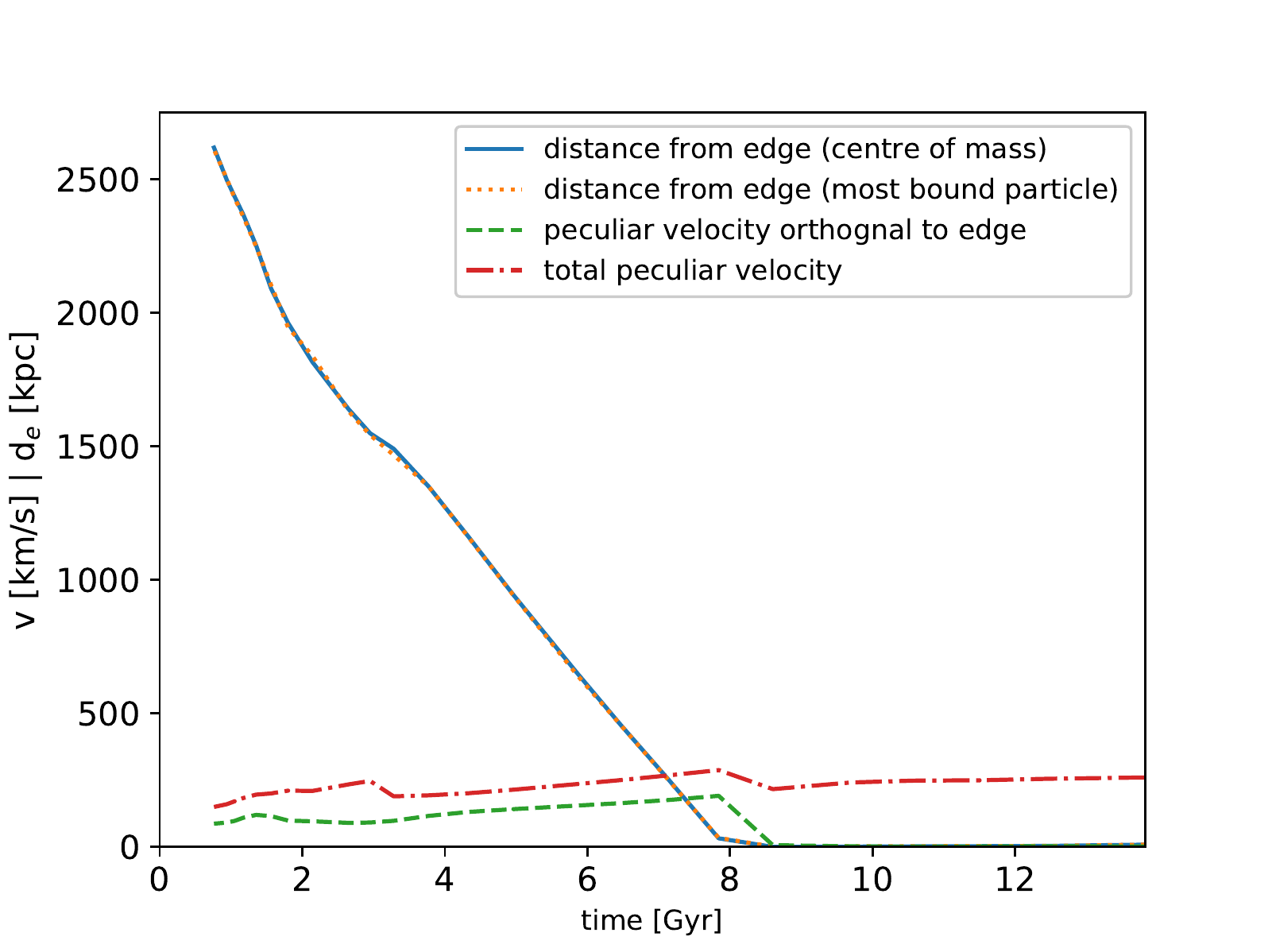}\\
\caption{Timeline of the distance from the edge and the peculiar velocities of the oddball EAGLE-3868859. The main branch of its merger tree was used to trace it.}
\label{dist_3868859}
\end{center}
\end{figure}
\begin{figure}
\begin{center}
\includegraphics[width=0.45\textwidth]{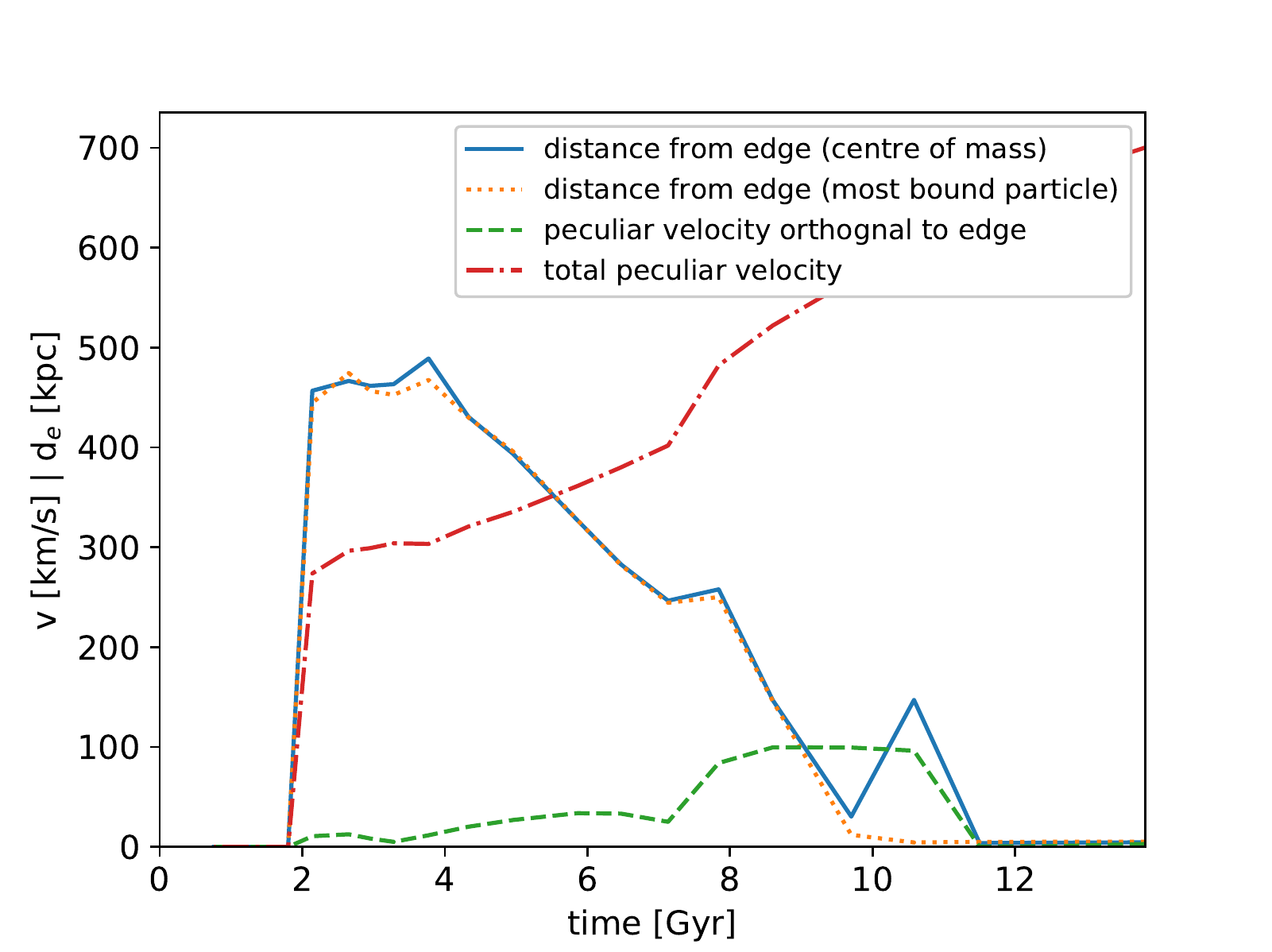}\\
\caption{Timeline of the distance from the edge and the peculiar velocities of the oddball EAGLE-4209797. The main branch of its merger tree was used to trace it.}
\label{dist_4209797}
\end{center}
\end{figure}
\begin{figure}
\begin{center}
\includegraphics[width=0.45\textwidth]{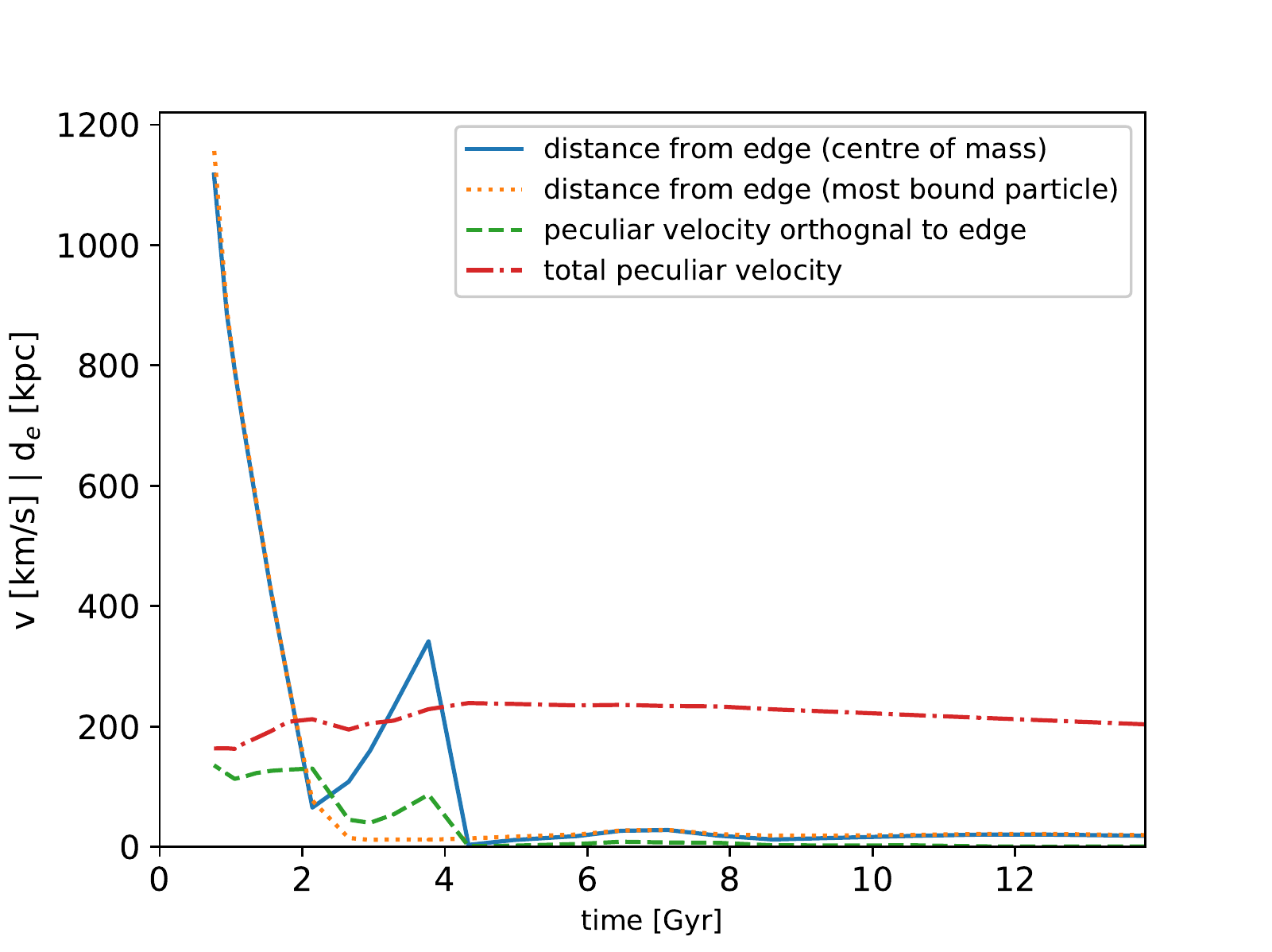}\\
\caption{Timeline of the distance from the edge and the peculiar velocities of the oddball EAGLE-11419697. The main branch of its merger tree was used to trace it.}
\label{dist_11419697}
\end{center}
\end{figure}
\begin{figure}
\begin{center}
\includegraphics[width=0.45\textwidth]{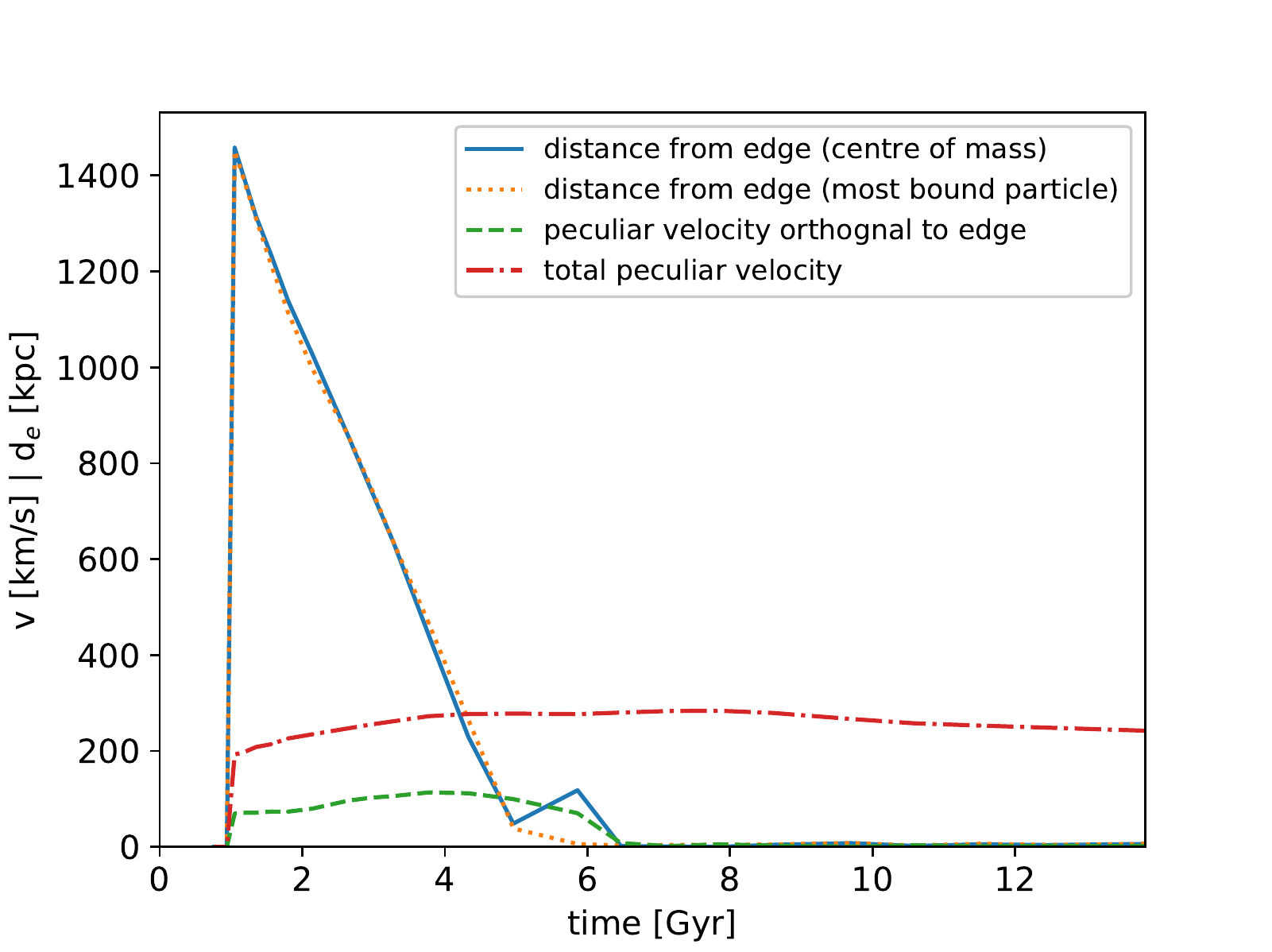}\\
\caption{Timeline of the distance from the edge and the peculiar velocities of the oddball EAGLE-60521664. The main branch of its merger tree was used to trace it.}
\label{dist_60521664}
\end{center}
\end{figure}

\begin{figure*}
\begin{center}
\includegraphics[width=0.90\textwidth]{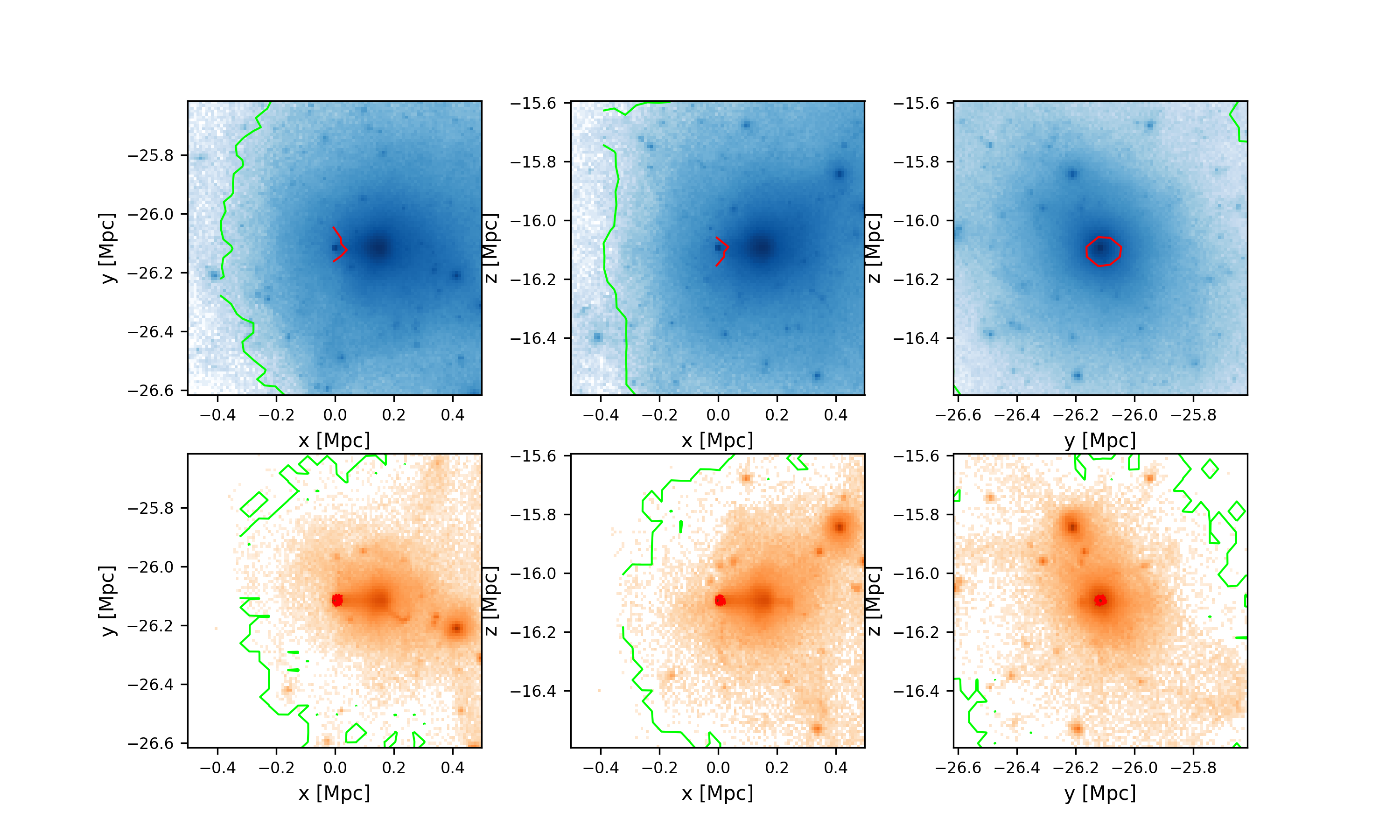}\\
\caption{Density map of the vicinity of oddball Illustris-476171 at snapshot 100. Top panels: dark matter distributions of a 500 kpc wide slice in the xy-plane (left), xz-plane (centre), and yz-plane (right) centred around the oddball; bottom panels: distributions of stellar matter of a 500 kpc wide slice in the xy-plane (left), xz-plane (centre), and yz-plane (right) centred around the oddball. The red contour lines indicate the positions of the particles that later end up in the oddball, while the green contour lines mark the extend of the current halo (according to subfind).}
\label{pastcombined_oddballs476171}
\end{center}
\end{figure*}
\begin{figure*}
\begin{center}
\includegraphics[width=0.90\textwidth]{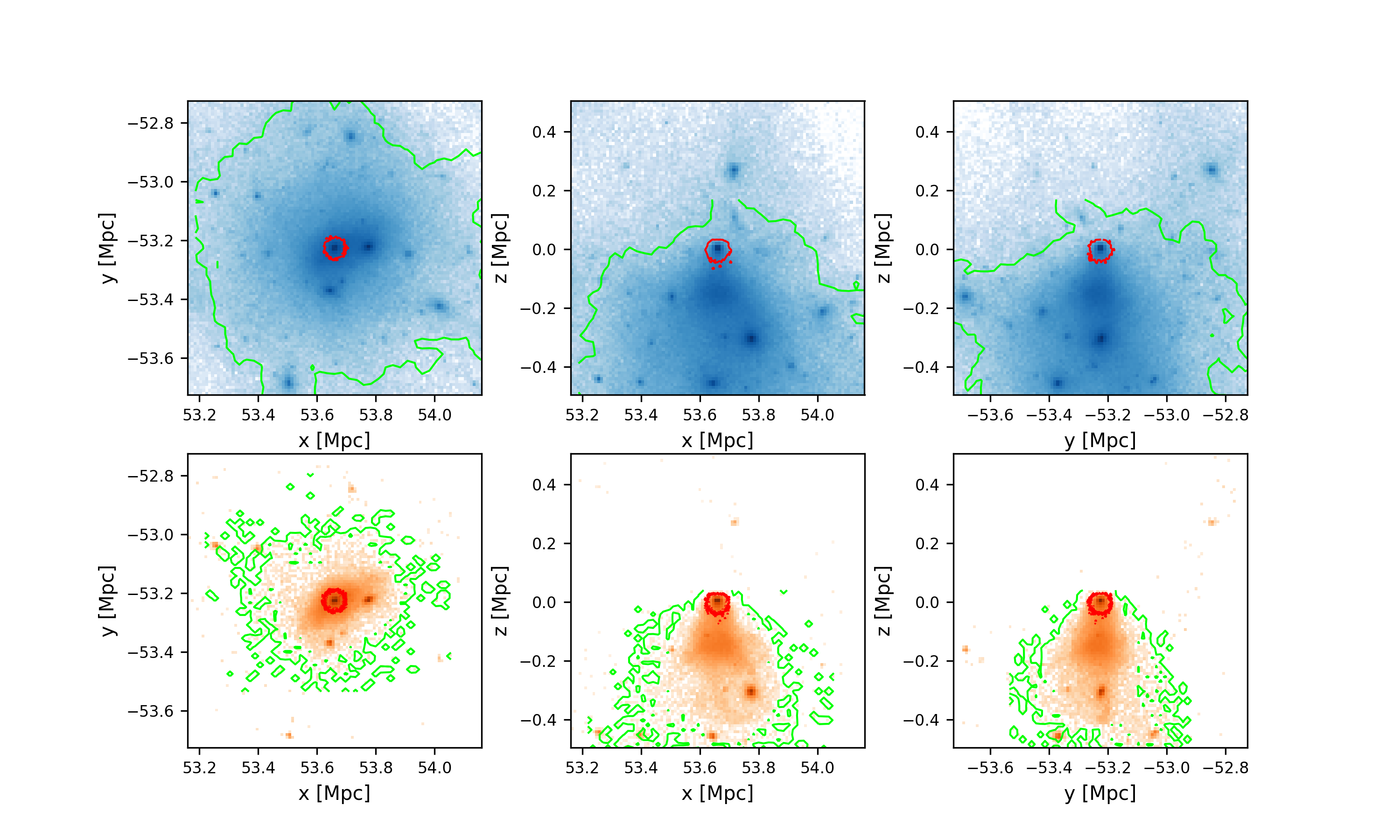}\\
\caption{Density map of the vicinity of oddball IllustrisTNG-585369 at snapshot 66. Same as for Figure \ref{pastcombined_oddballs476171}.}
\label{pastcombined_oddballs585369}
\end{center}
\end{figure*}
\begin{figure*}
\begin{center}
\includegraphics[width=0.90\textwidth]{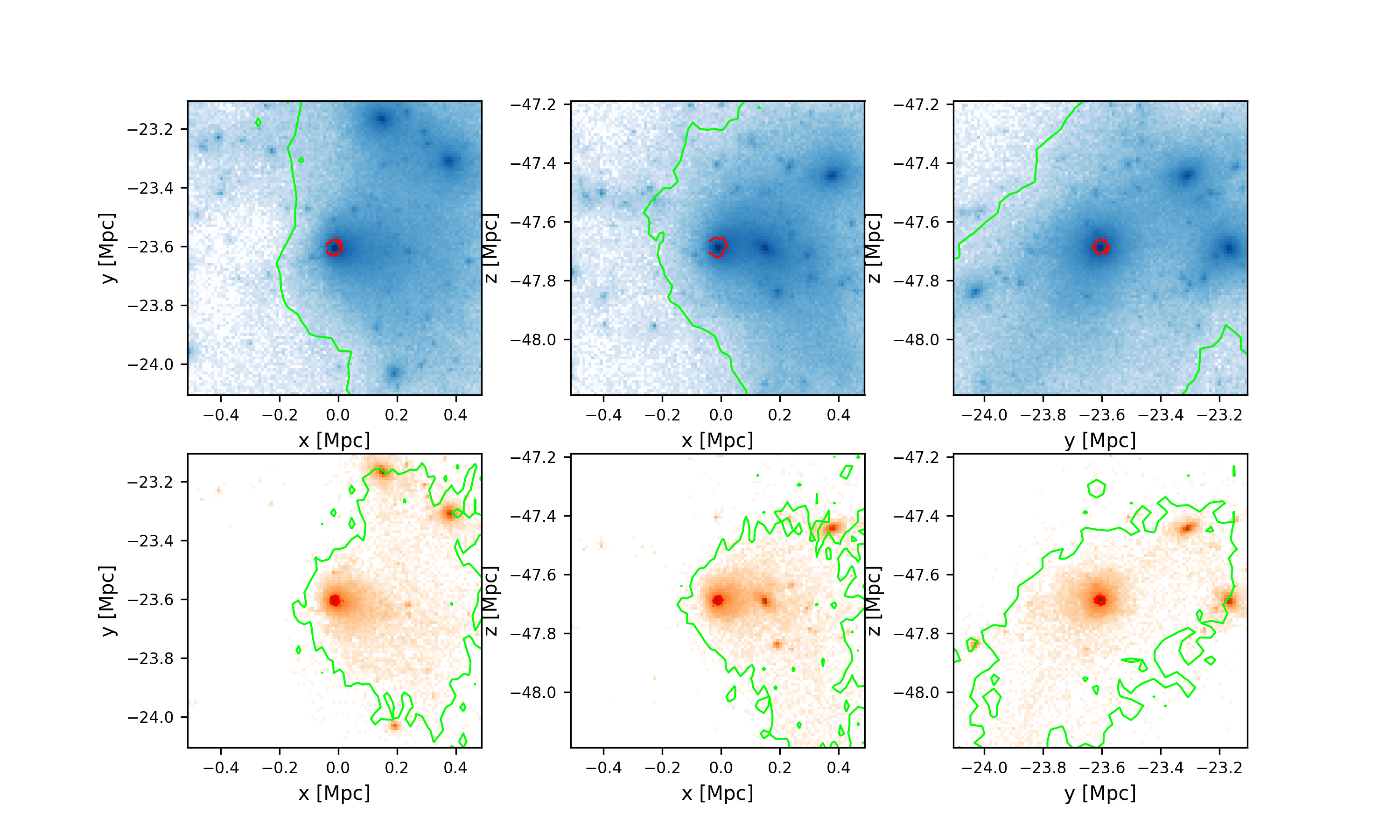}\\
\caption{Density map of the vicinity of oddball EAGLE-3274715 at snapshot 18. Same as for Figure \ref{pastcombined_oddballs476171}.}
\label{pastcombined_oddballs3274715}
\end{center}
\end{figure*}
\begin{figure*}
\begin{center}
\includegraphics[width=0.90\textwidth]{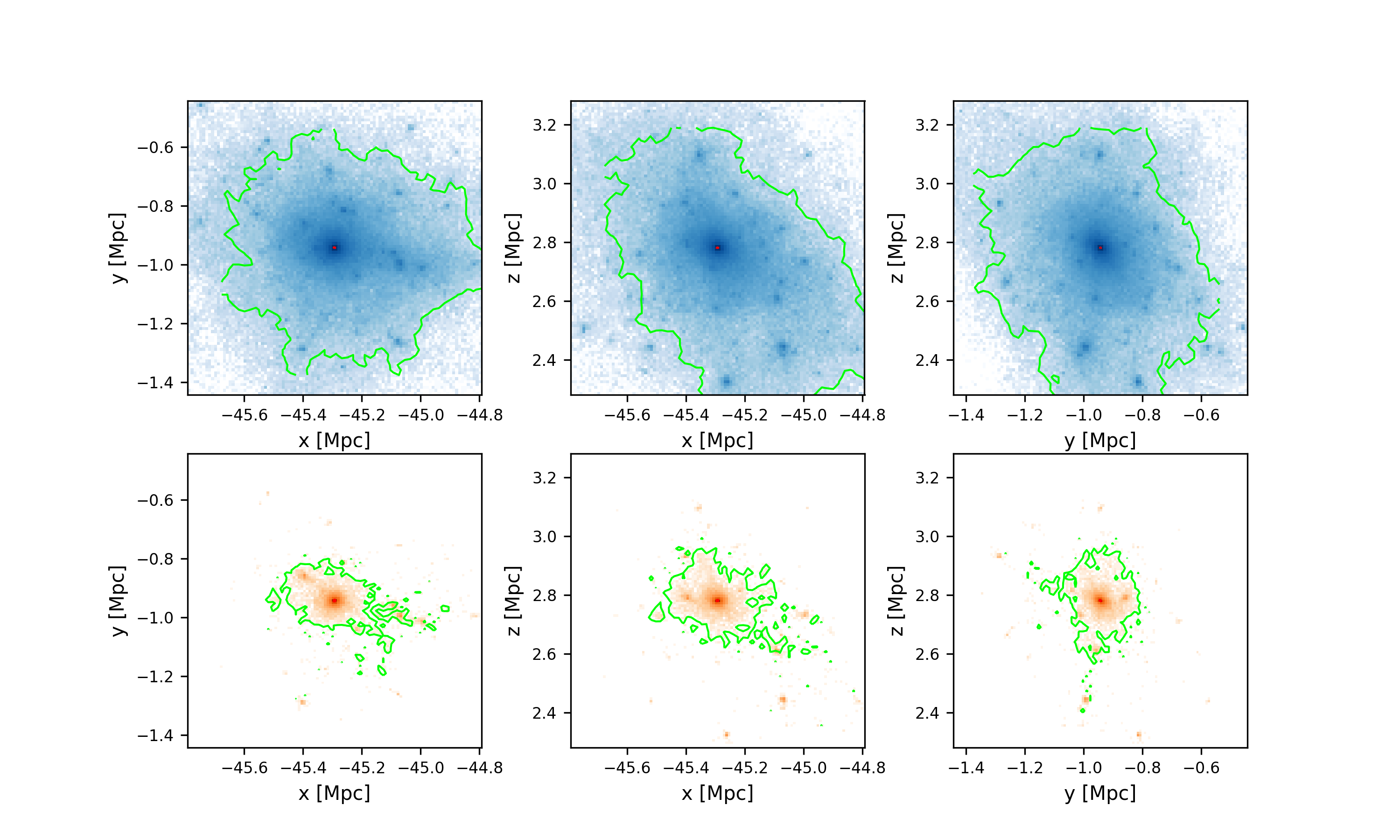}\\
\caption{Density map of the vicinity of oddball EAGLE-3868859 at snapshot 18. Same as for Figure \ref{pastcombined_oddballs476171}.}
\label{pastcombined_oddballs3868859}
\end{center}
\end{figure*}
\begin{figure*}
\begin{center}
\includegraphics[width=0.90\textwidth]{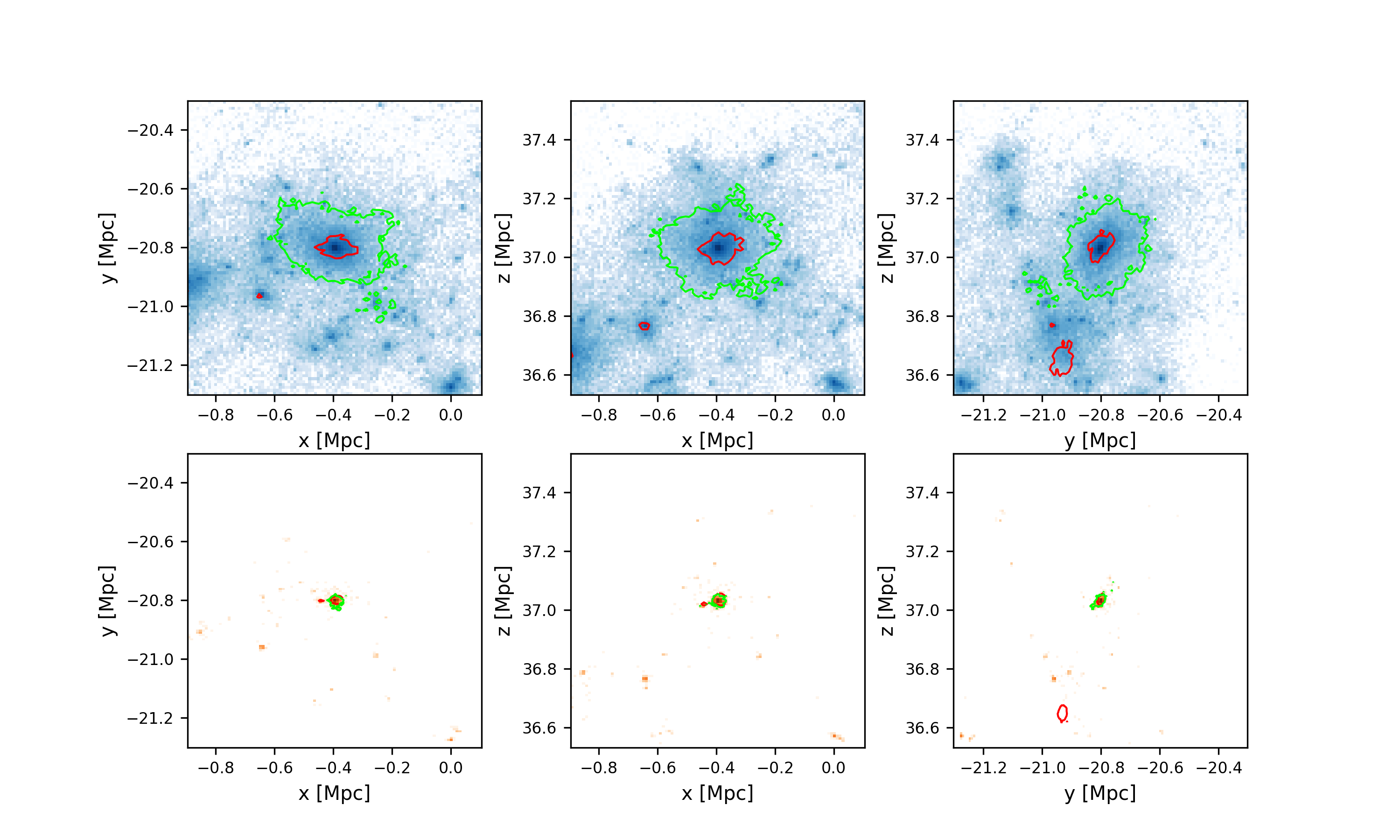}\\
\caption{Density map of the vicinity of oddball EAGLE-4209797 at snapshot 18. Same as for Figure \ref{pastcombined_oddballs476171}.}
\label{pastcombined_oddballs4209797}
\end{center}
\end{figure*}
\begin{figure*}
\begin{center}
\includegraphics[width=0.90\textwidth]{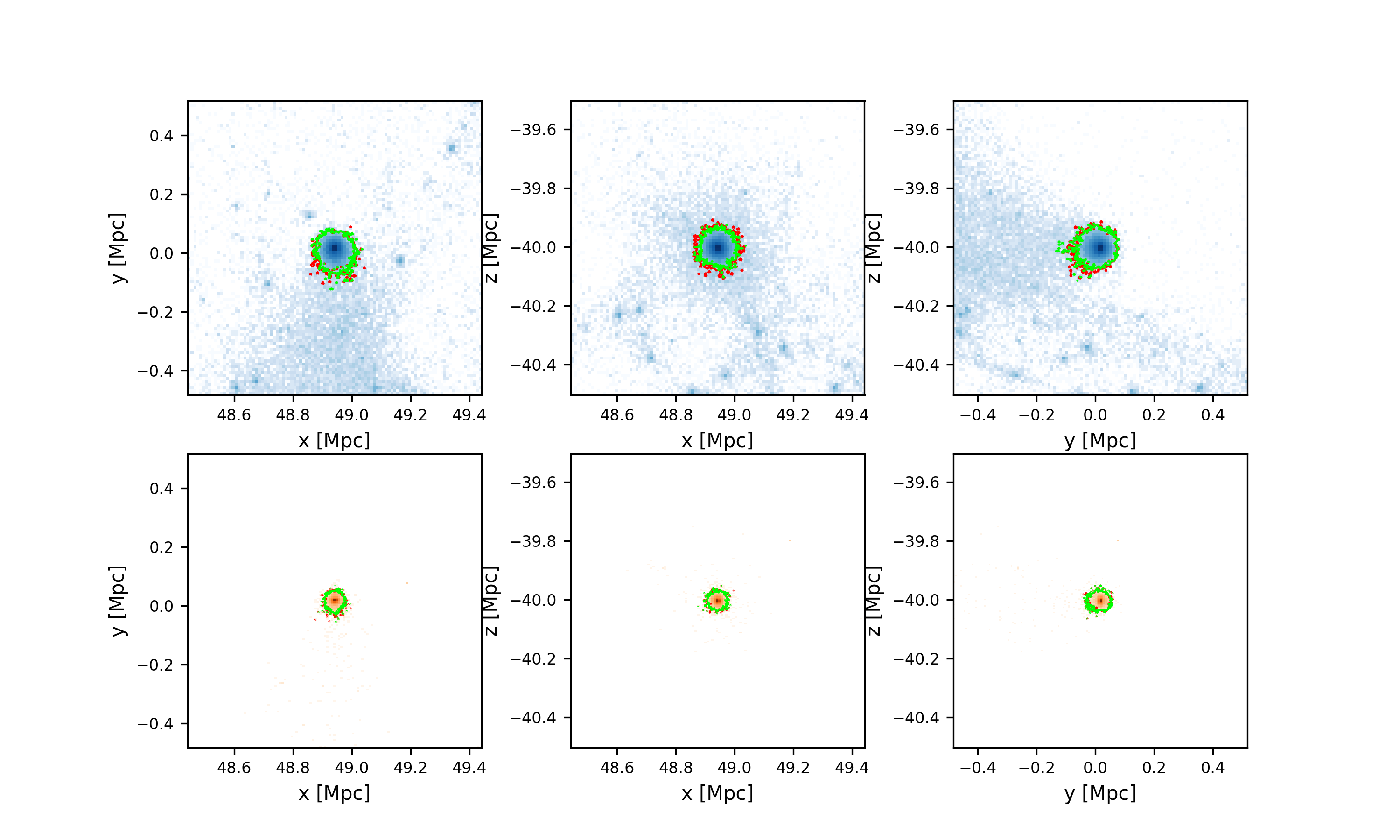}\\
\caption{Density map of the vicinity of oddball EAGLE-11419697 at snapshot 18. Same as for Figure \ref{pastcombined_oddballs476171}.}
\label{pastcombined_oddballs11419697}
\end{center}
\end{figure*}
\begin{figure*}
\begin{center}
\includegraphics[width=0.90\textwidth]{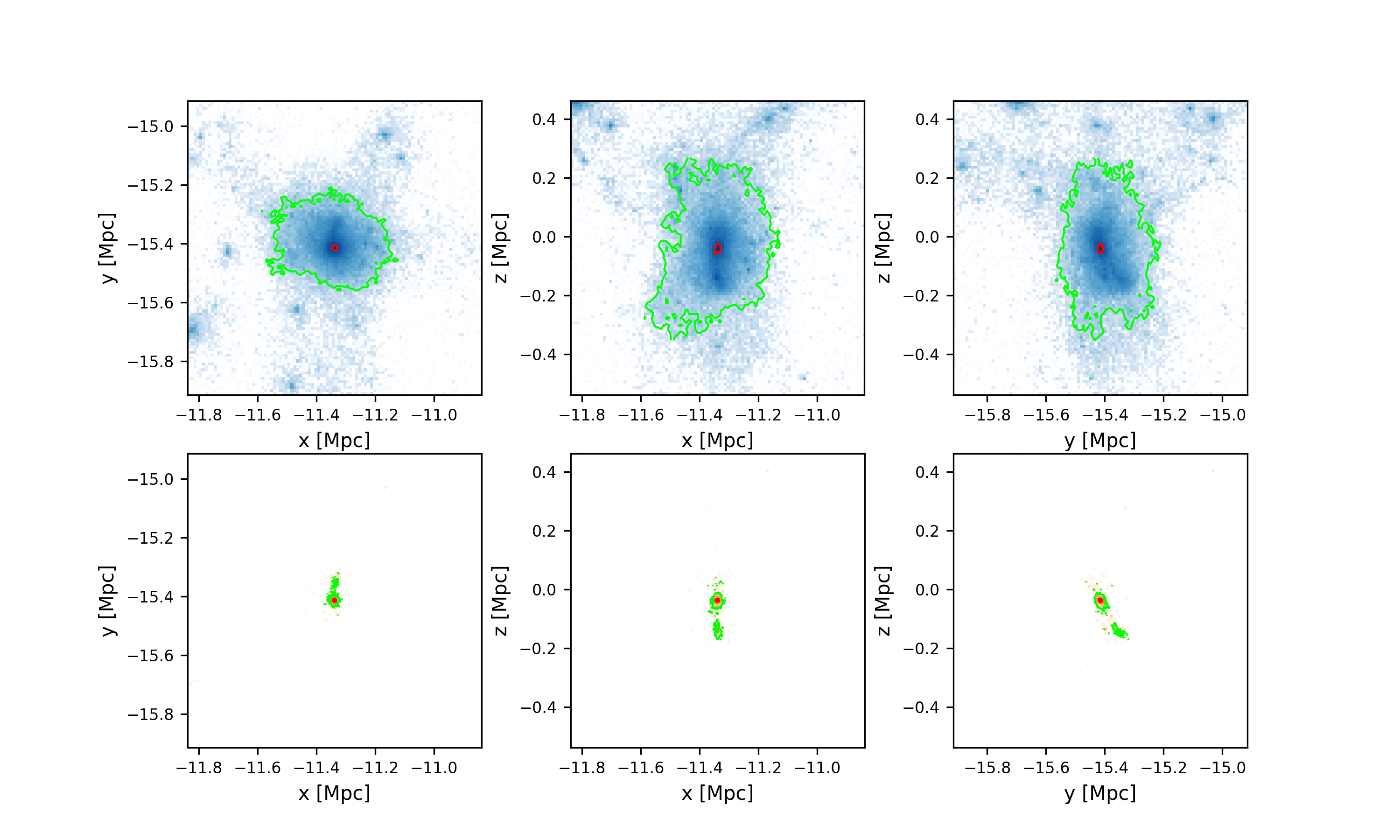}\\
\caption{Density map of the vicinity of oddball EAGLE-60521664 at snapshot 18. Same as for Figure \ref{pastcombined_oddballs476171}.}
\label{pastcombined_oddballs60521664}
\end{center}
\end{figure*}

\begin{figure*}
\begin{center}
\includegraphics[width=0.90\textwidth]{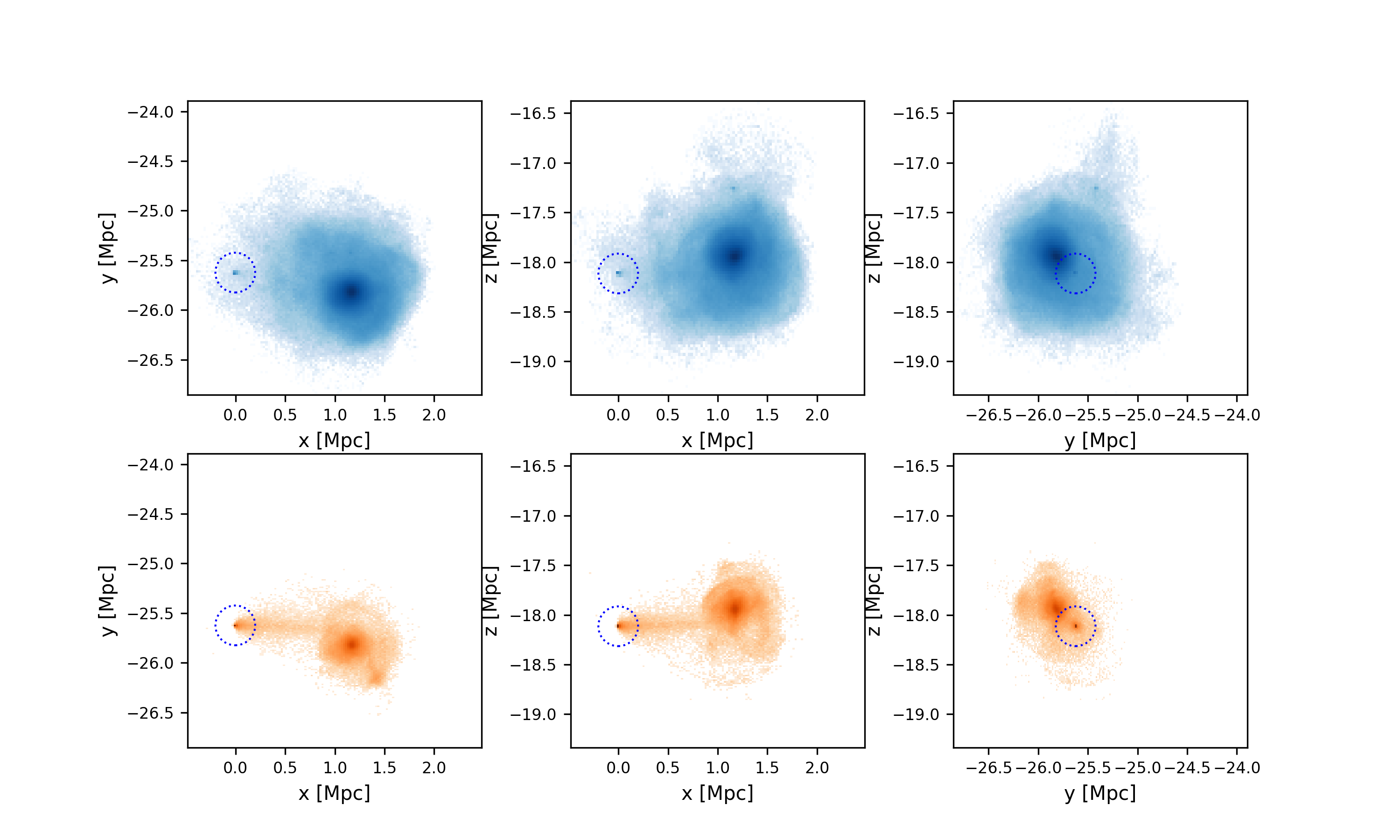}\\
\caption{Density map of particles that belonged to the progenitor of oddball Illustris-476171. Top panels: projected dark matter distributions in the xy-plane (left), xz-plane (centre), and yz-plane (right) centred around the oddball; projected bottom panels: distributions of stellar matter in the xy-plane (left), xz-plane (centre), and yz-plane (right) centred around the oddball. The dotted blue circle marks a 200 kpc aperture around the oddball's position.}
\label{today_oddballs476171}
\end{center}
\end{figure*}
\begin{figure*}
\begin{center}
\includegraphics[width=0.90\textwidth]{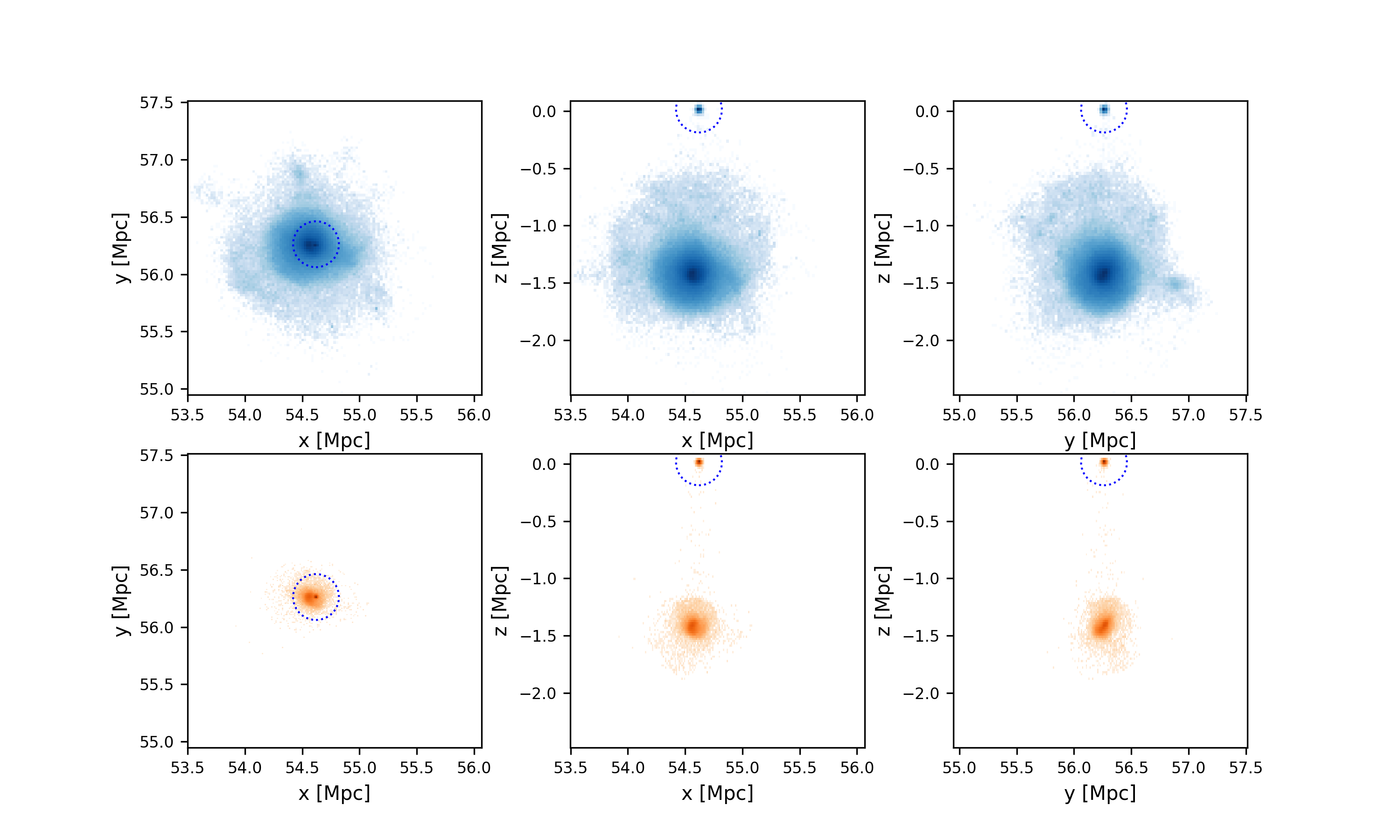}\\
\caption{Present-day density map of particles that belonged to the progenitor of oddball IllustrisTNG-585369. Same as for Figure \ref{today_oddballs476171}.}
\label{today_oddballs585369}
\end{center}
\end{figure*}
\begin{figure*}
\begin{center}
\includegraphics[width=0.90\textwidth]{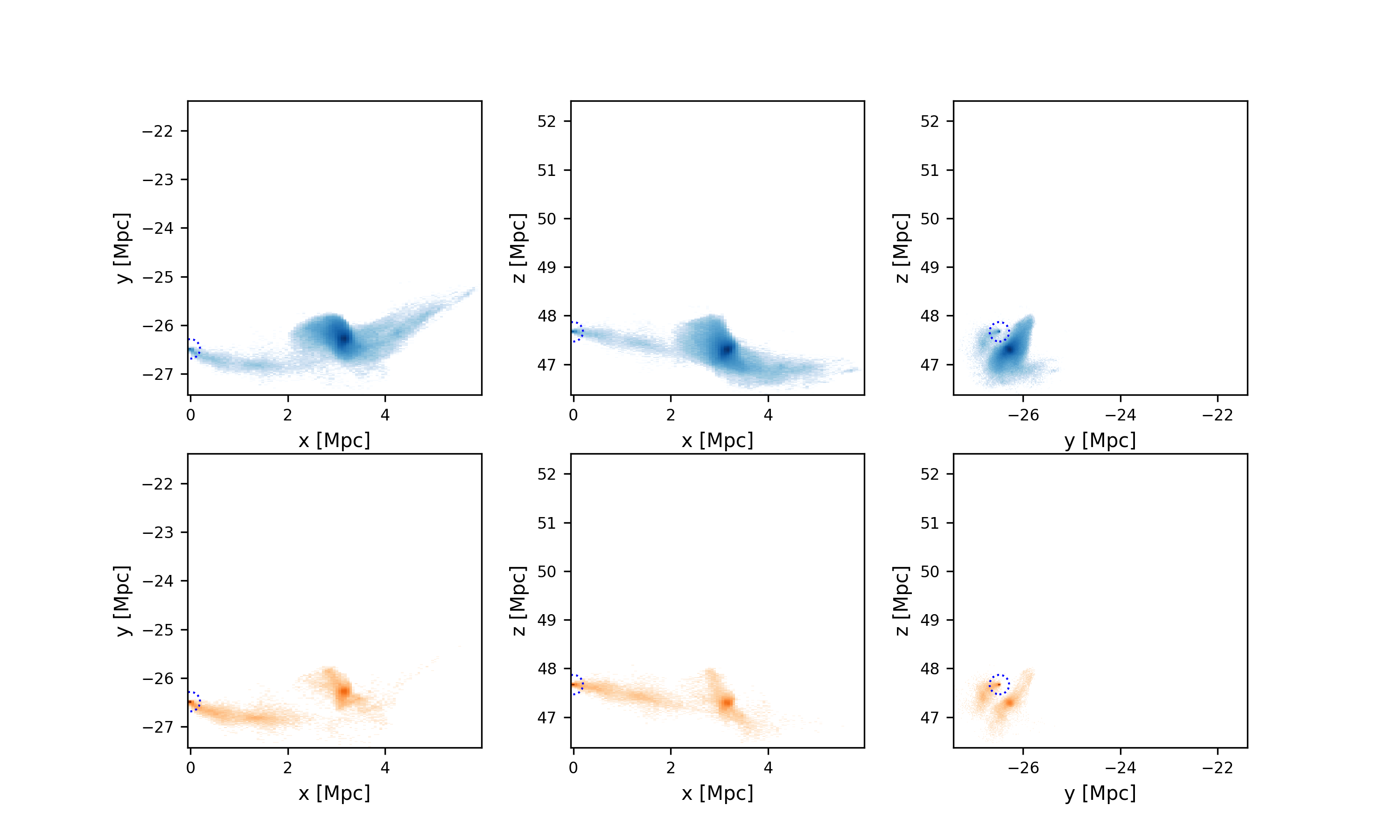}\\
\caption{Present-day density map of particles that belonged to the progenitor of oddball EAGLE-3274715. Same as for Figure \ref{today_oddballs476171}.}
\label{today_oddballs3274715}
\end{center}
\end{figure*}
\begin{figure*}
\begin{center}
\includegraphics[width=0.90\textwidth]{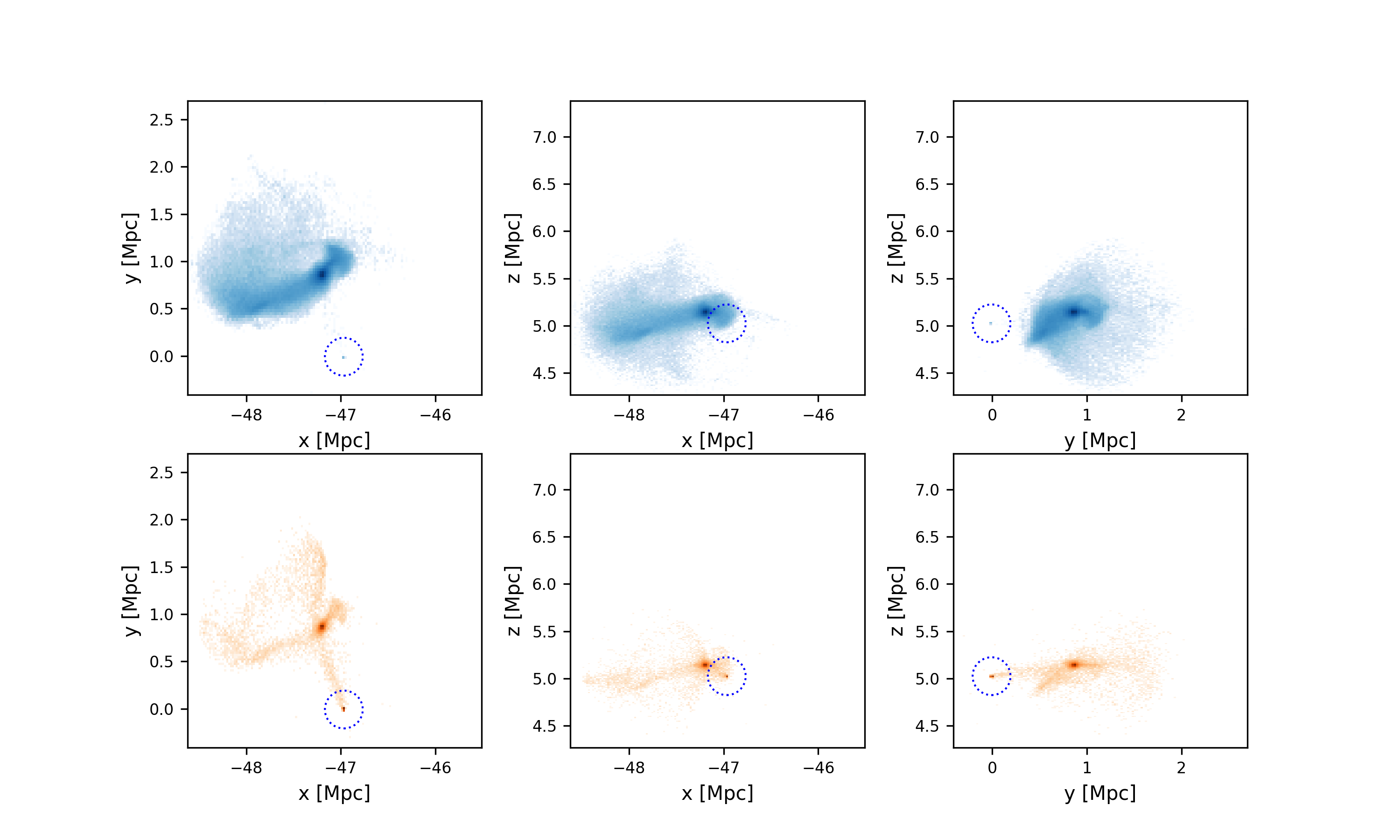}\\
\caption{Present-day density map of particles that belonged to the progenitor of oddball EAGLE-3868859. Same as for Figure \ref{today_oddballs476171}.}
\label{today_oddballs3868859}
\end{center}
\end{figure*}
\begin{figure*}
\begin{center}
\includegraphics[width=0.90\textwidth]{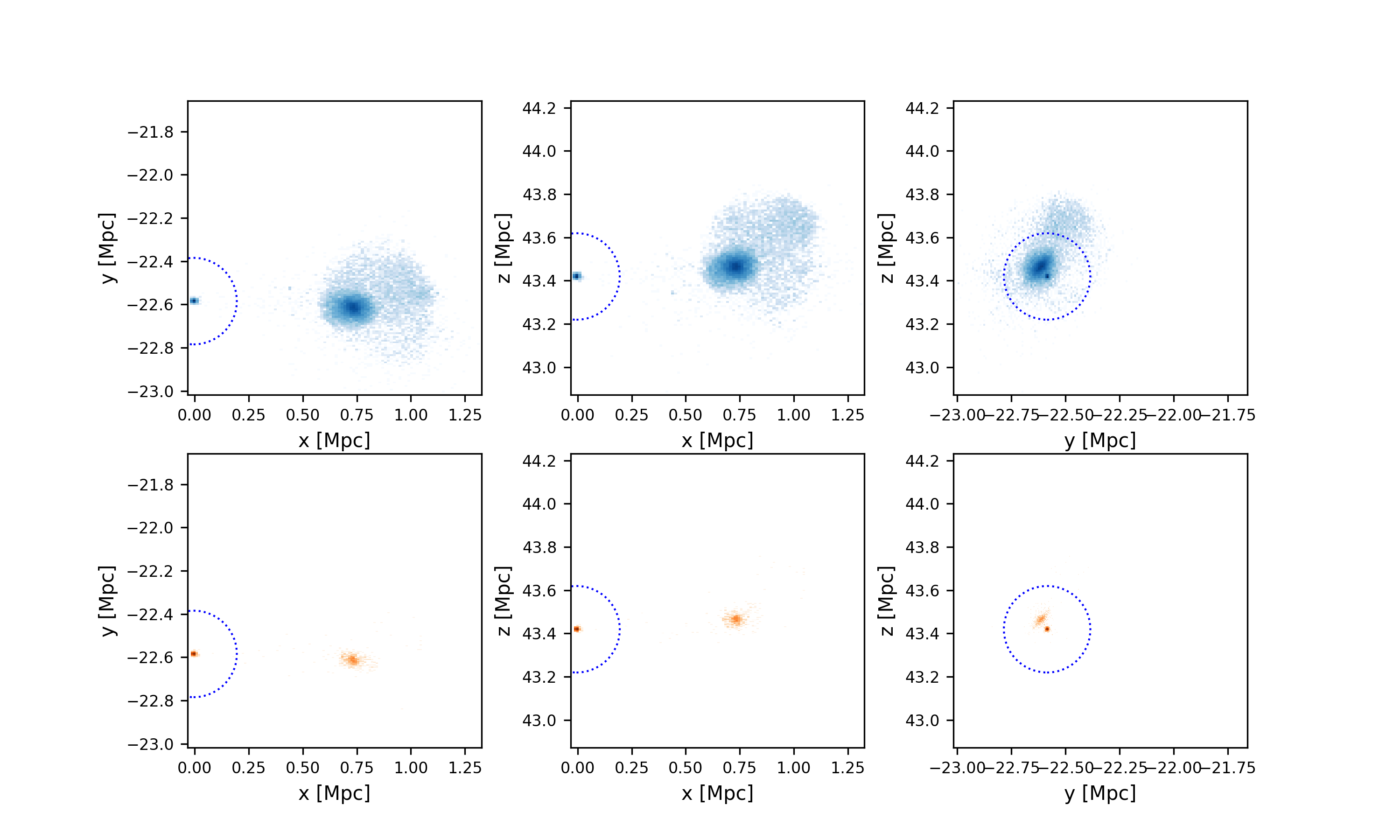}\\
\caption{Present-day density map of particles that belonged to the progenitor of oddball EAGLE-4209797. Same as for Figure \ref{today_oddballs476171}.}
\label{today_oddballs4209797}
\end{center}
\end{figure*}
\begin{figure*}
\begin{center}
\includegraphics[width=0.90\textwidth]{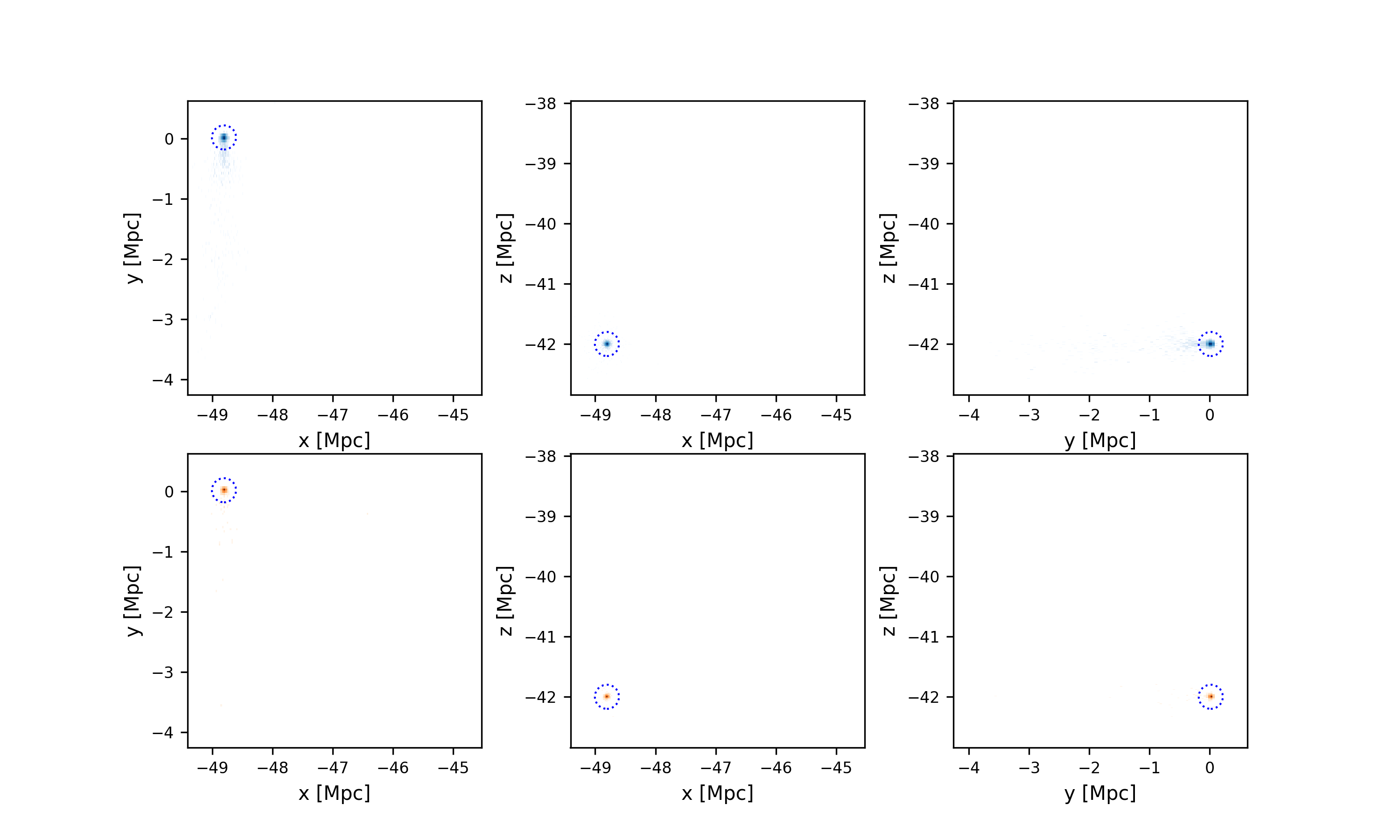}\\
\caption{Present-day density map of particles that belonged to the progenitor of oddball EAGLE-11419697. Same as for Figure \ref{today_oddballs476171}.}
\label{today_oddballs11419697}
\end{center}
\end{figure*}
\begin{figure*}
\begin{center}
\includegraphics[width=0.90\textwidth]{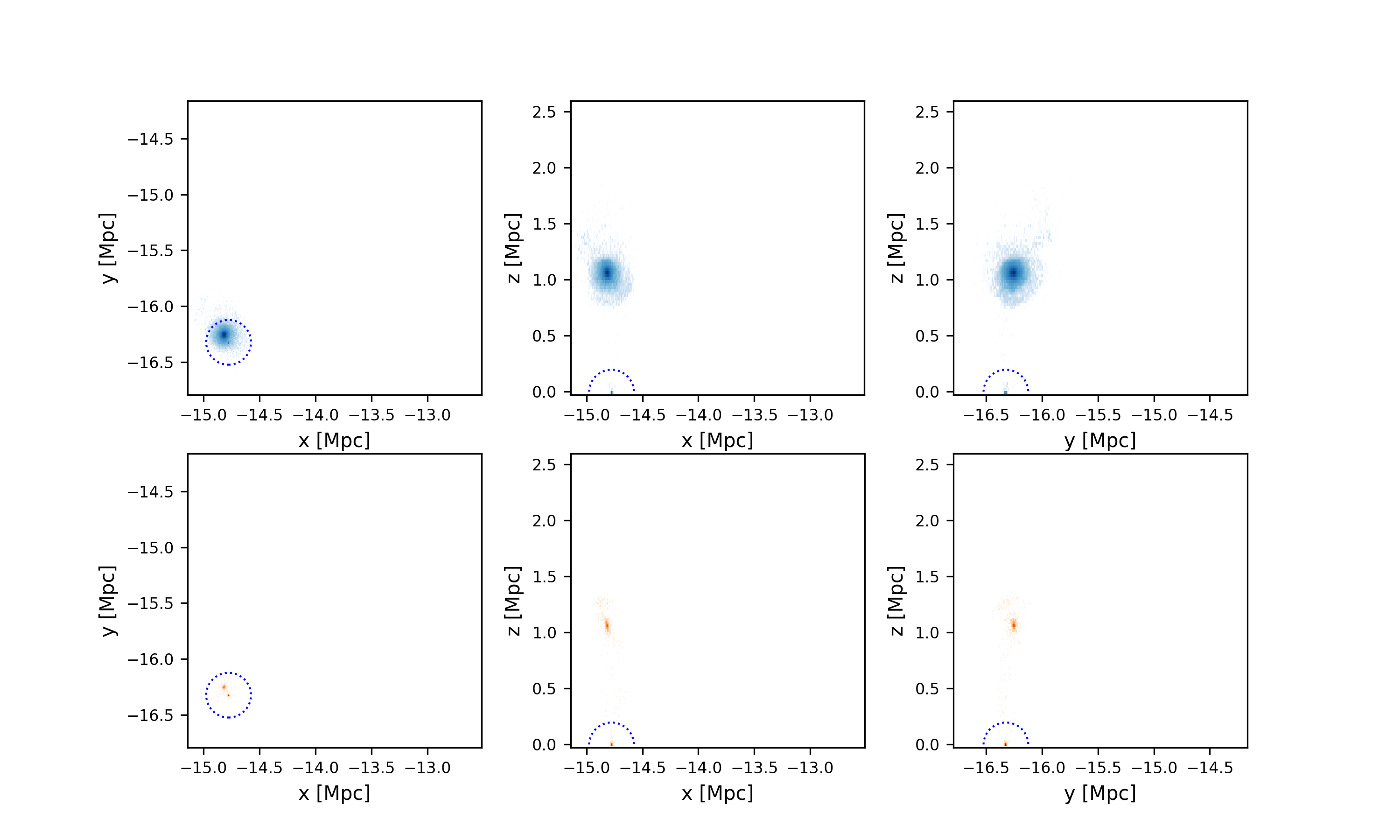}\\
\caption{Present-day density map of particles that belonged to the progenitor of oddball EAGLE-60521664. Same as for Figure \ref{today_oddballs476171}.}
\label{today_oddballs60521664}
\end{center}
\end{figure*}

\begin{figure*}
\begin{center}
\includegraphics[width=0.90\textwidth]{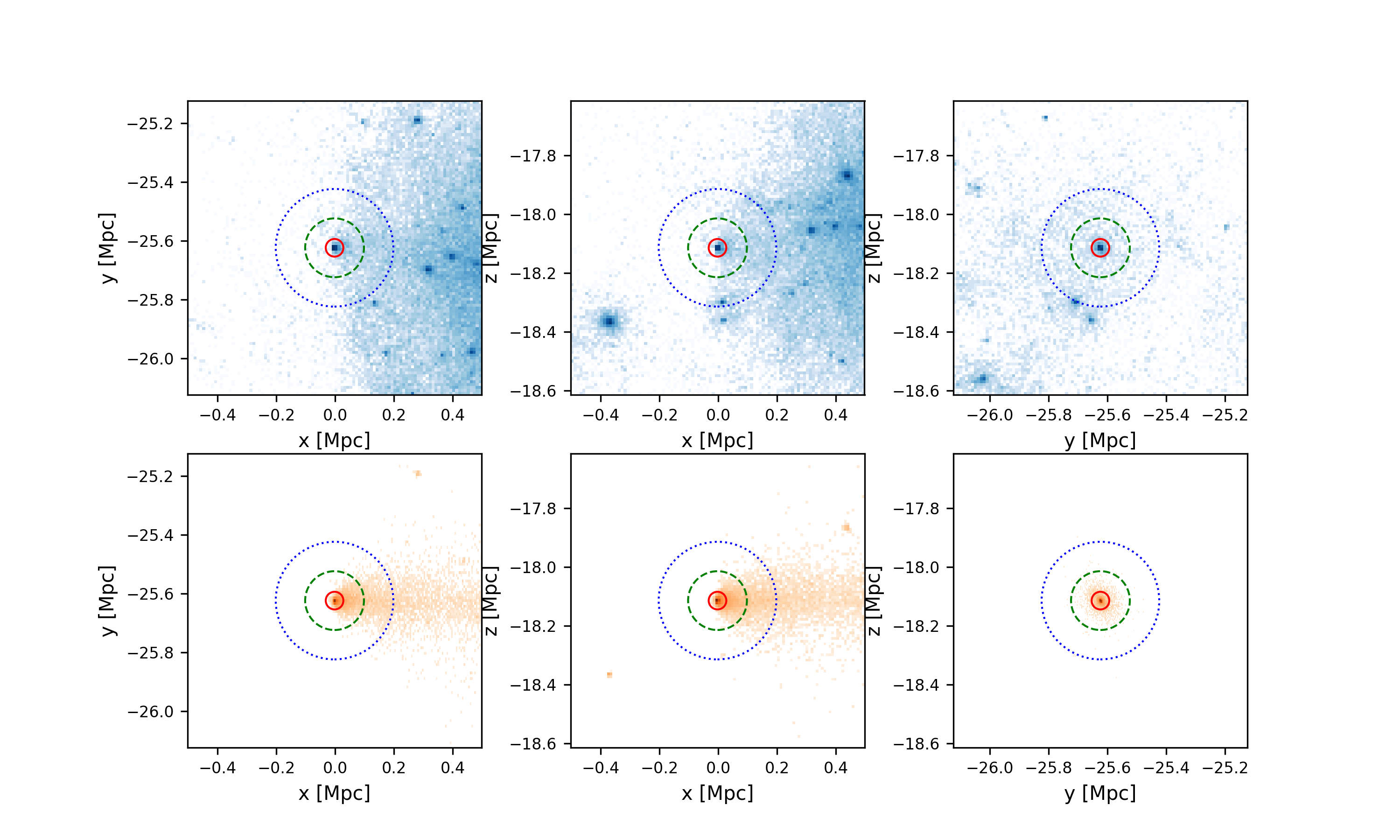}\\
\caption{Density map of the vicinity of oddball Illustris-476171. Top panels: dark matter distributions of a 200 kpc wide slice in the xy-plane (left), xz-plane (centre), and yz-plane (right) centred around the oddball; bottom panels: distributions of stellar matter of a 200 kpc wide slice in the xy-plane (left), xz-plane (centre), and yz-plane (right) centred around the oddball. The circles mark 30 kpc (solid red), 100 kpc (dashed green), and 200 kpc (dotted blue) apertures around the oddball.}
\label{oddballs476171}
\end{center}
\end{figure*}
\begin{figure*}
\begin{center}
\includegraphics[width=0.90\textwidth]{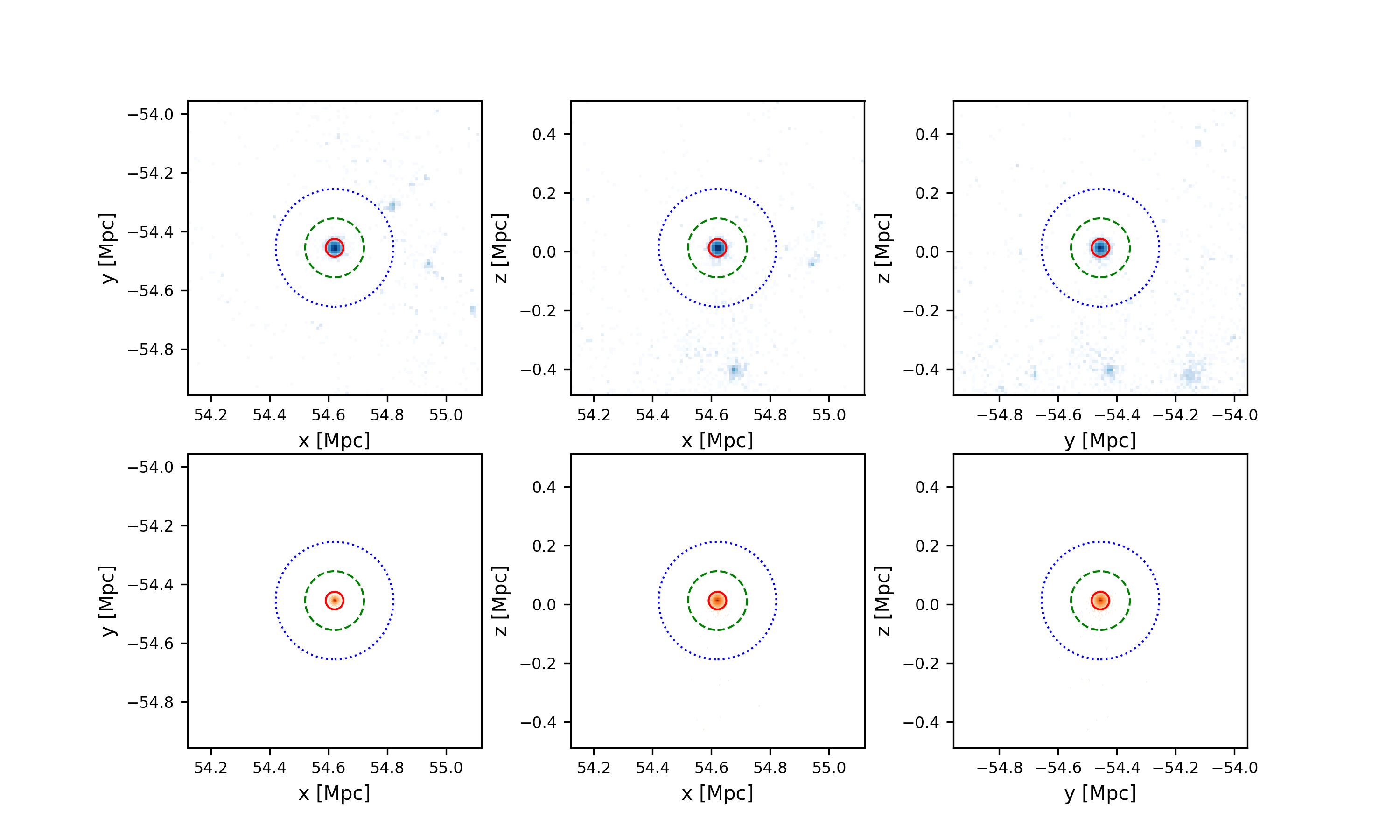}\\
\caption{Density map of the vicinity of oddball IllustrisTNG-585369. Same as for Figure \ref{oddballs476171}.}
\label{oddballs585369}
\end{center}
\end{figure*}
\begin{figure*}
\begin{center}
\includegraphics[width=0.90\textwidth]{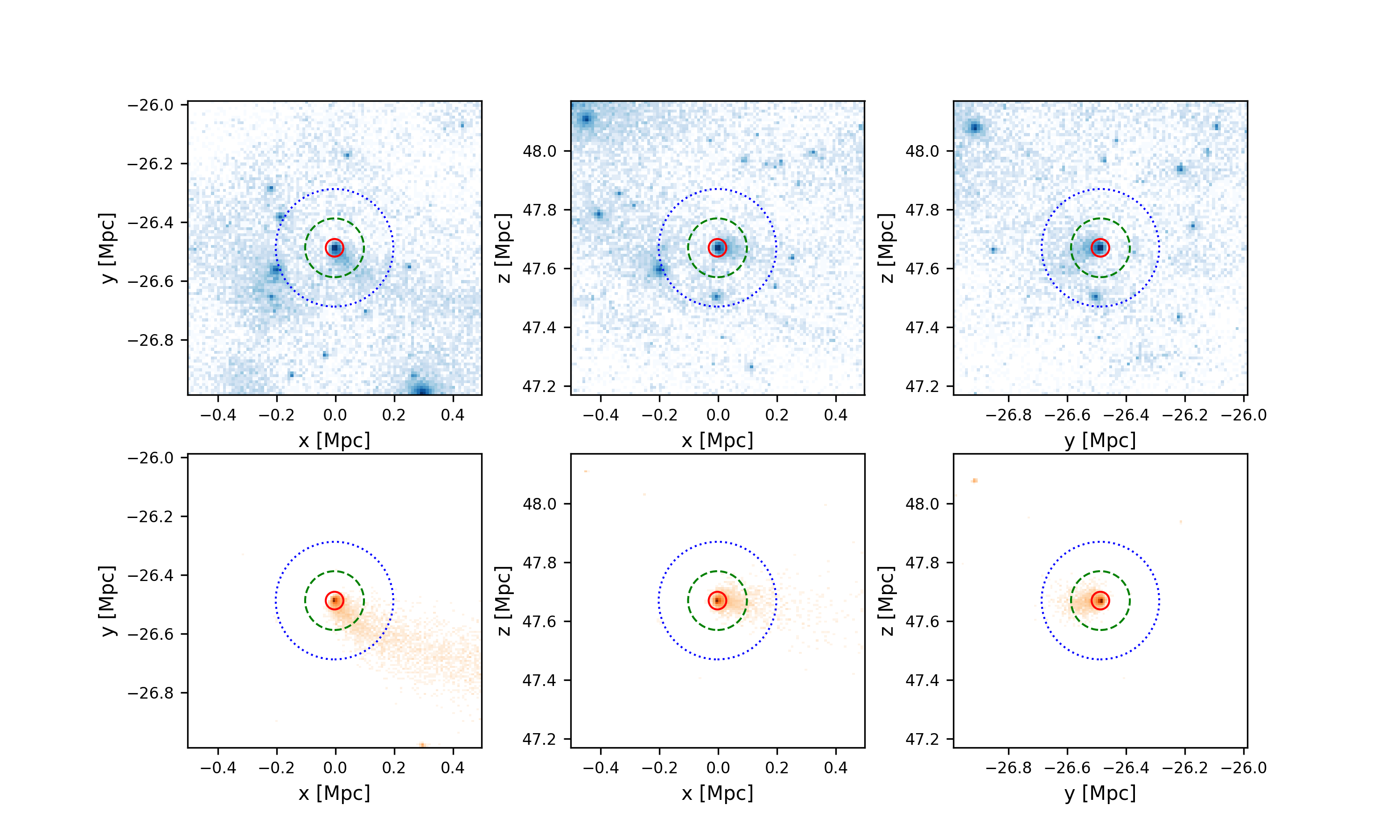}\\
\caption{Density map of the vicinity of oddball EAGLE-3274715. Same as for Figure \ref{oddballs476171}.}
\label{oddballs3274715}
\end{center}
\end{figure*}
\begin{figure*}
\begin{center}
\includegraphics[width=0.90\textwidth]{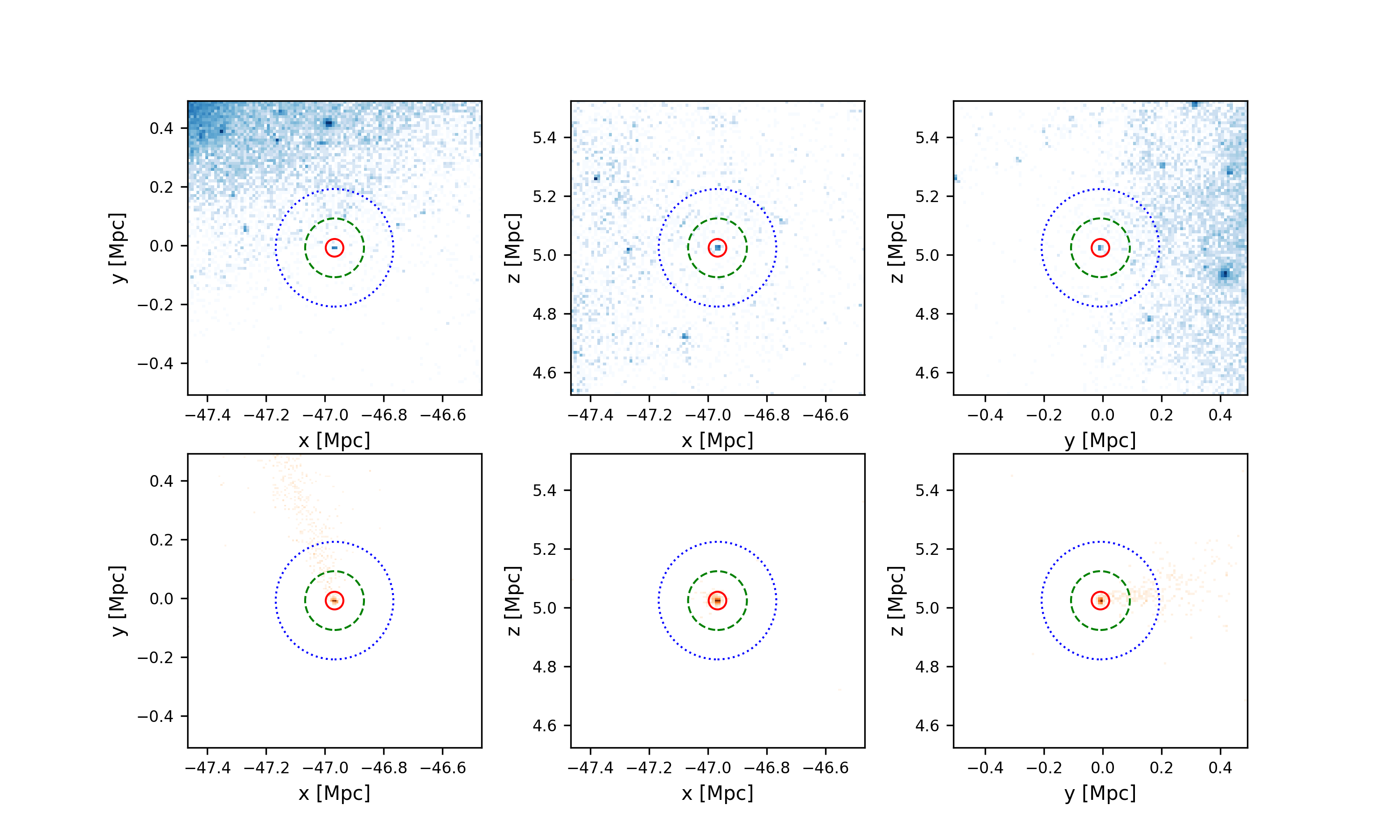}\\
\caption{Density map of the vicinity of oddball EAGLE-3868859. Same as for Figure \ref{oddballs476171}.}
\label{oddballs3868859}
\end{center}
\end{figure*}
\begin{figure*}
\begin{center}
\includegraphics[width=0.90\textwidth]{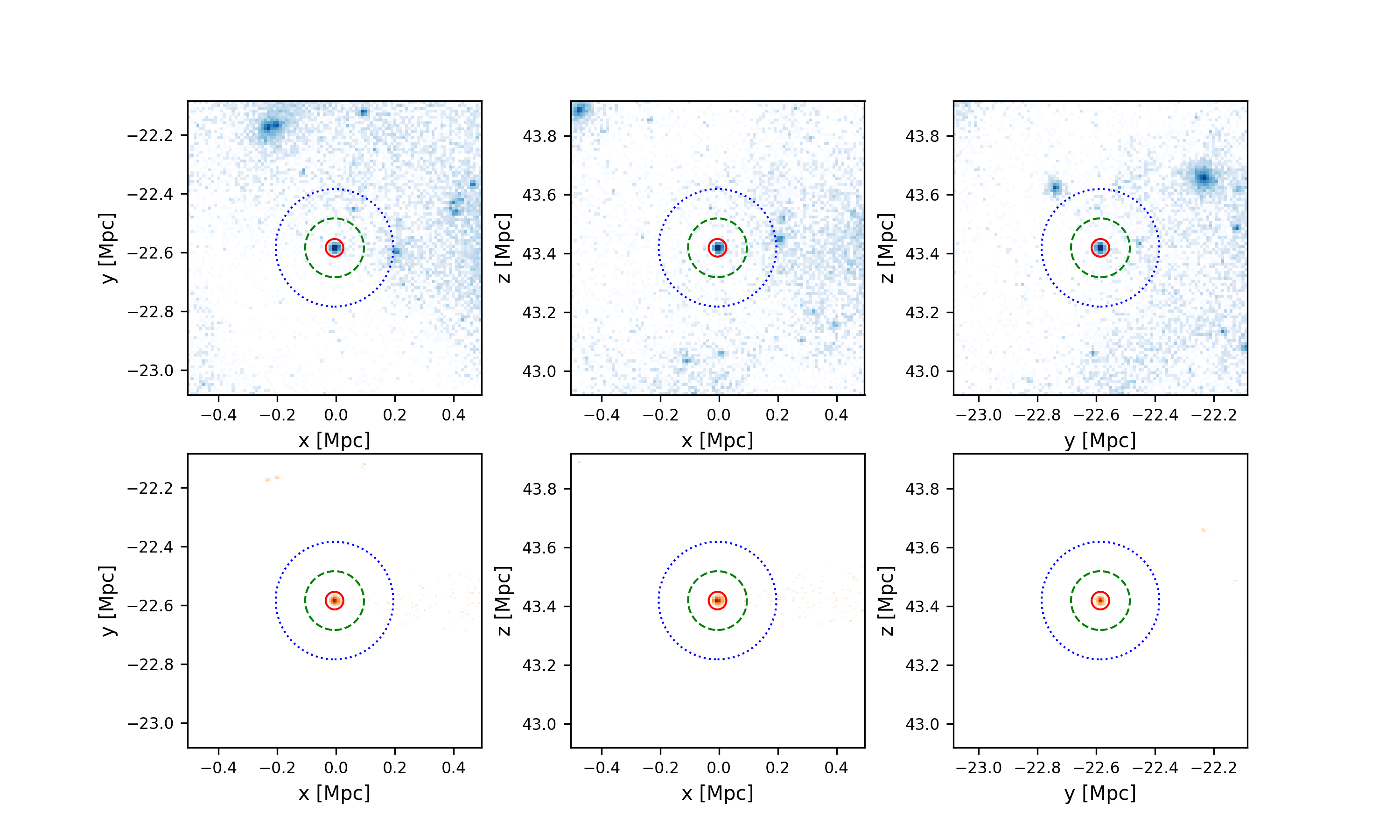}\\
\caption{Density map of the vicinity of oddball EAGLE-4209797. Same as for Figure \ref{oddballs476171}.}
\label{oddballs4209797}
\end{center}
\end{figure*}
\begin{figure*}
\begin{center}
\includegraphics[width=0.90\textwidth]{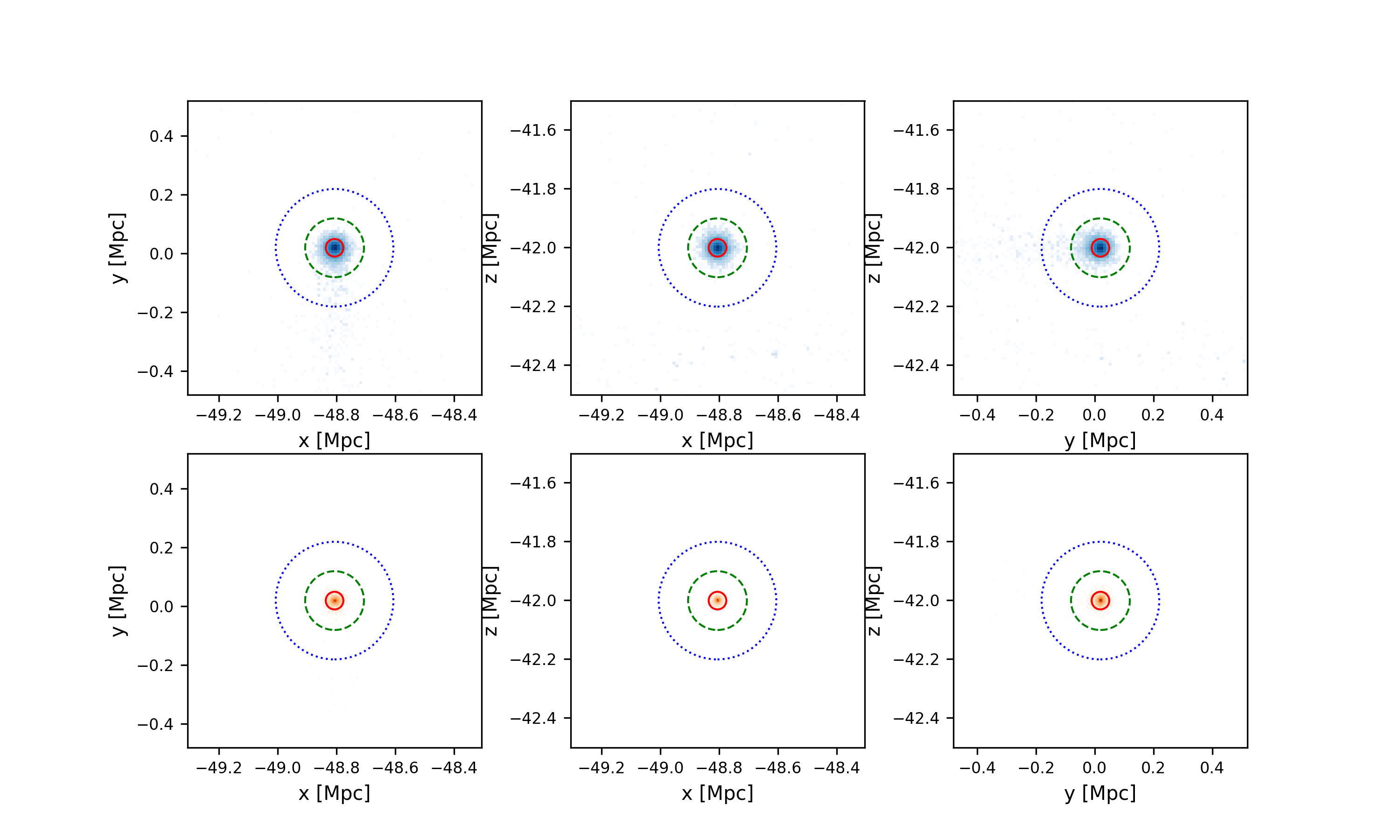}\\
\caption{Density map of the vicinity of oddball EAGLE-11419697. Same as for Figure \ref{oddballs476171}.}
\label{oddballs11419697}
\end{center}
\end{figure*}
\begin{figure*}
\begin{center}
\includegraphics[width=0.90\textwidth]{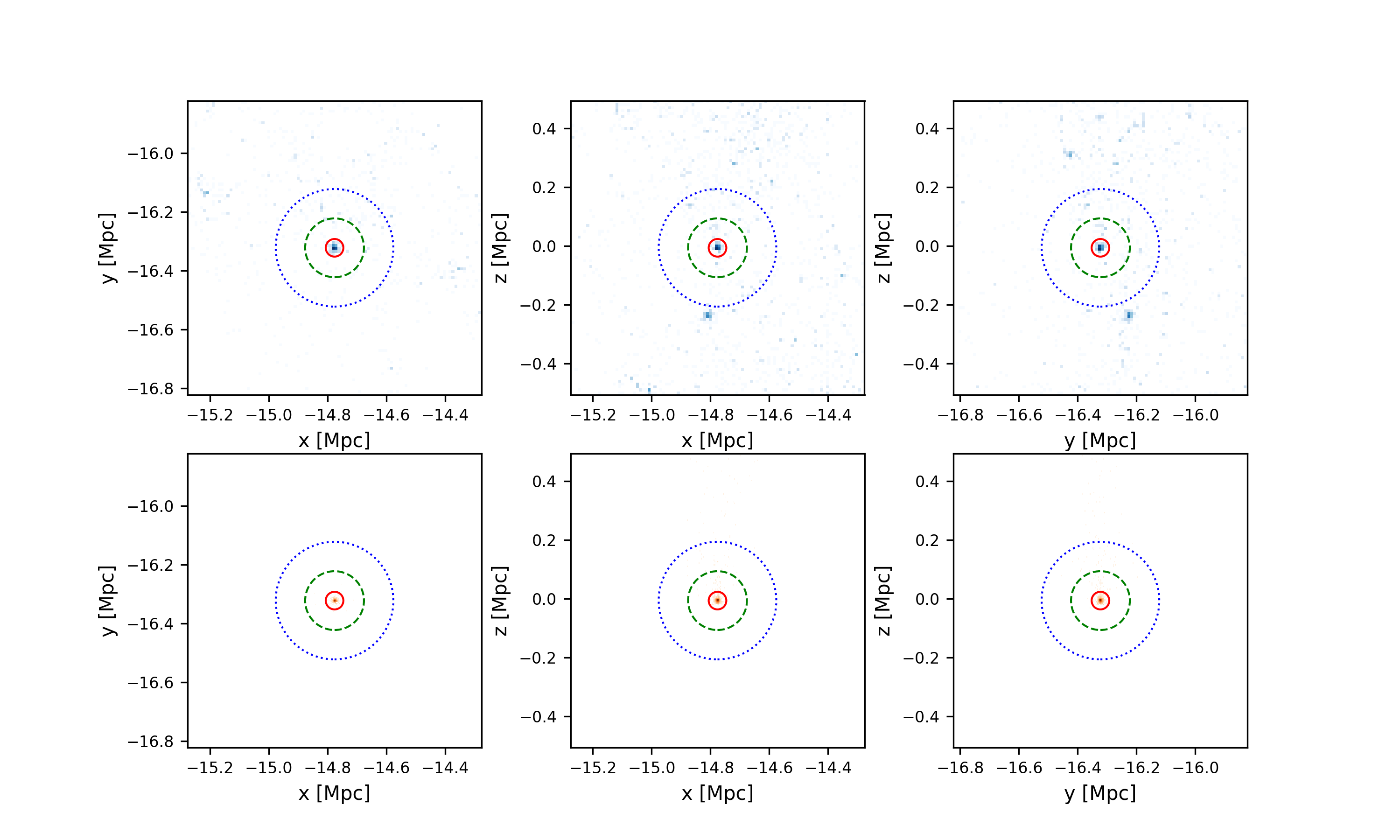}\\
\caption{Density map of the vicinity of oddball EAGLE-60521664. Same as for Figure \ref{oddballs476171}.}
\label{oddballs60521664}
\end{center}
\end{figure*}

\begin{figure}
\begin{center}
\includegraphics[width=0.45\textwidth]{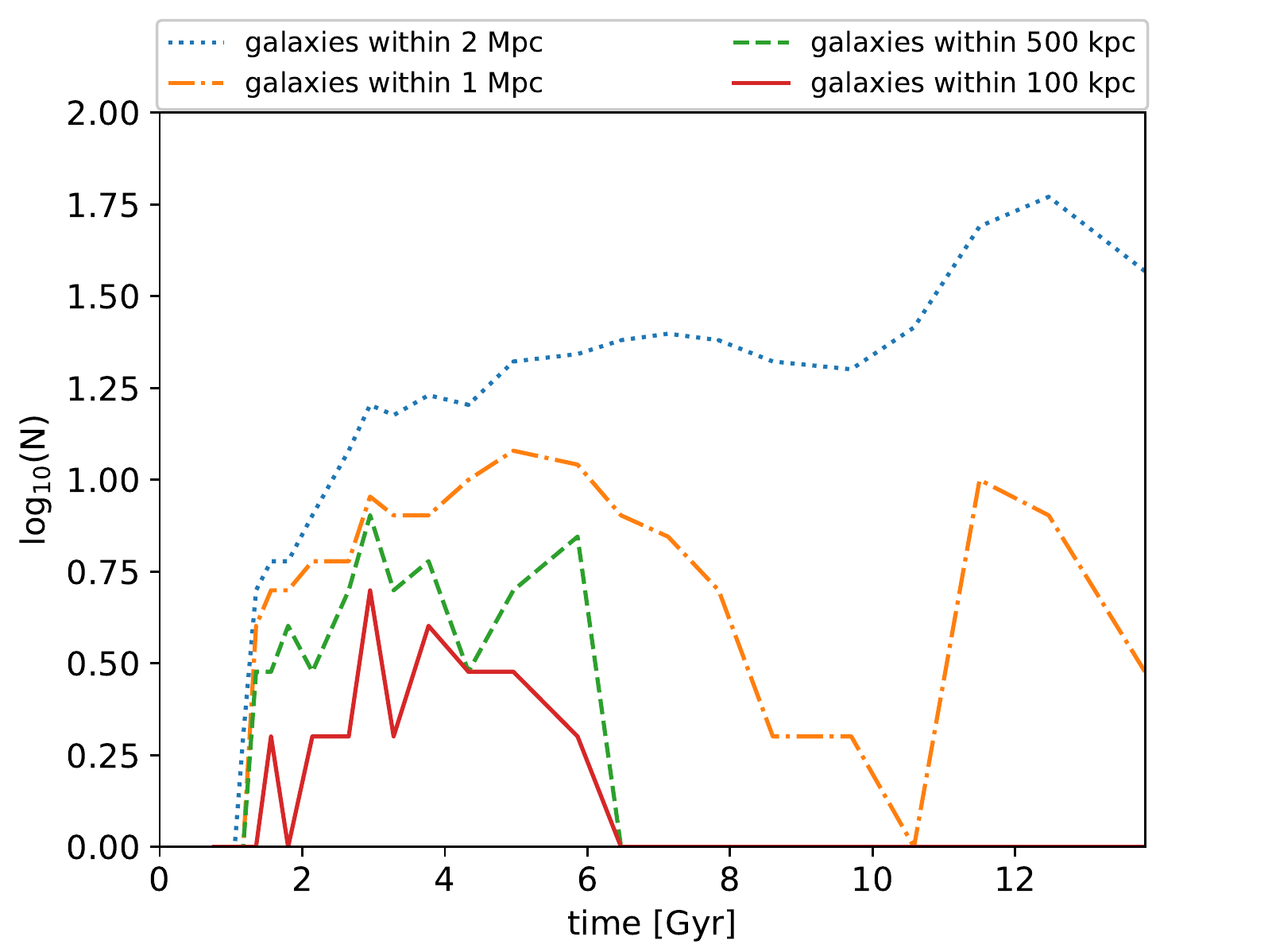}\\
\caption{Galaxy numbers within different spherical apertures in the vicinity of oddball EAGLE-3274715 over time. The main branch of its merger tree was used to trace it.}
\label{surroundings_shift_evolution3274715}
\end{center}
\end{figure}
\begin{figure}
\begin{center}
\includegraphics[width=0.45\textwidth]{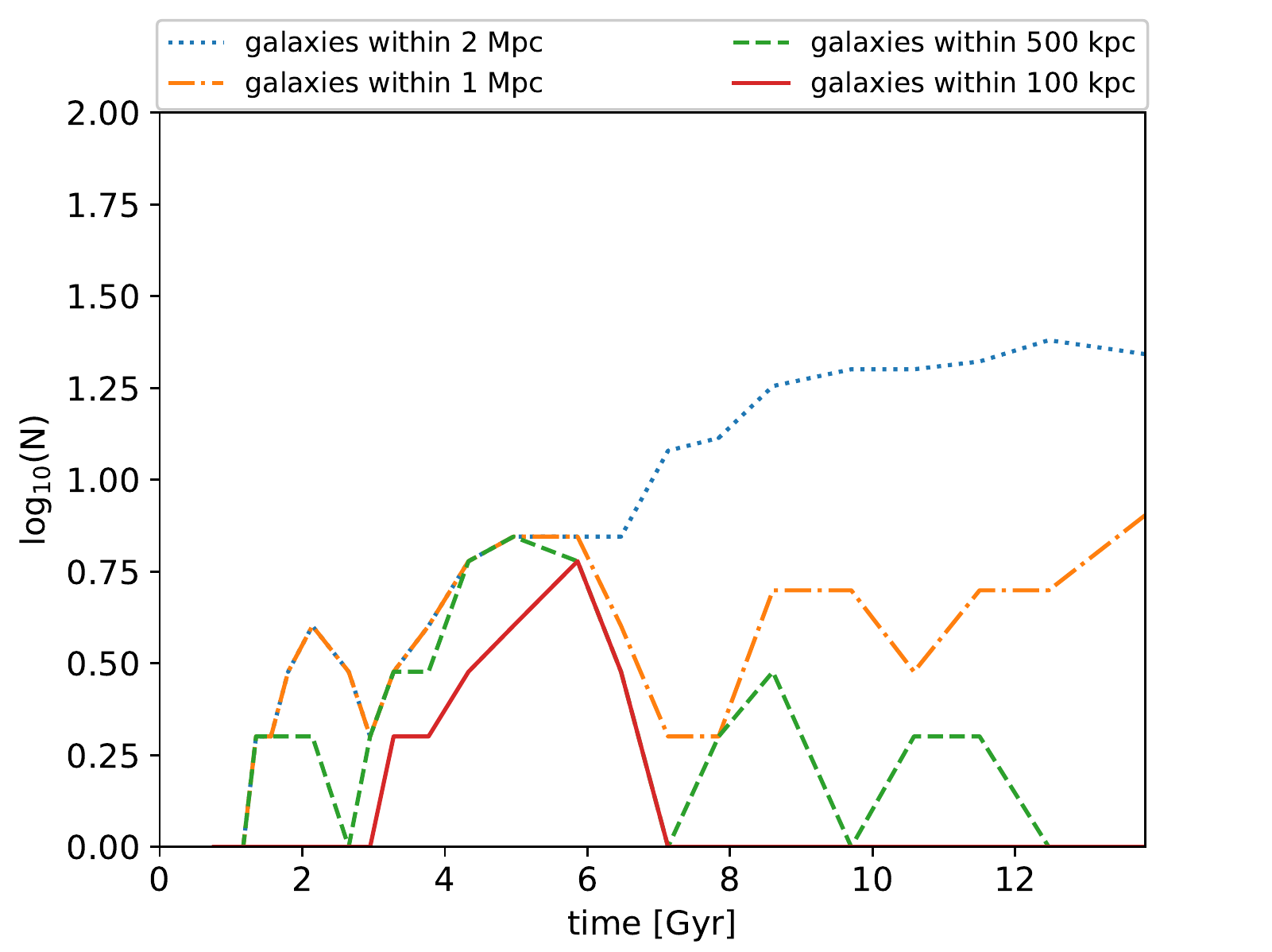}\\
\caption{Galaxy numbers within different spherical apertures in the vicinity of oddball EAGLE-3868859 over time. The main branch of its merger tree was used to trace it.}
\label{surroundings_shift_evolution3868859}
\end{center}
\end{figure}
\begin{figure}
\begin{center}
\includegraphics[width=0.45\textwidth]{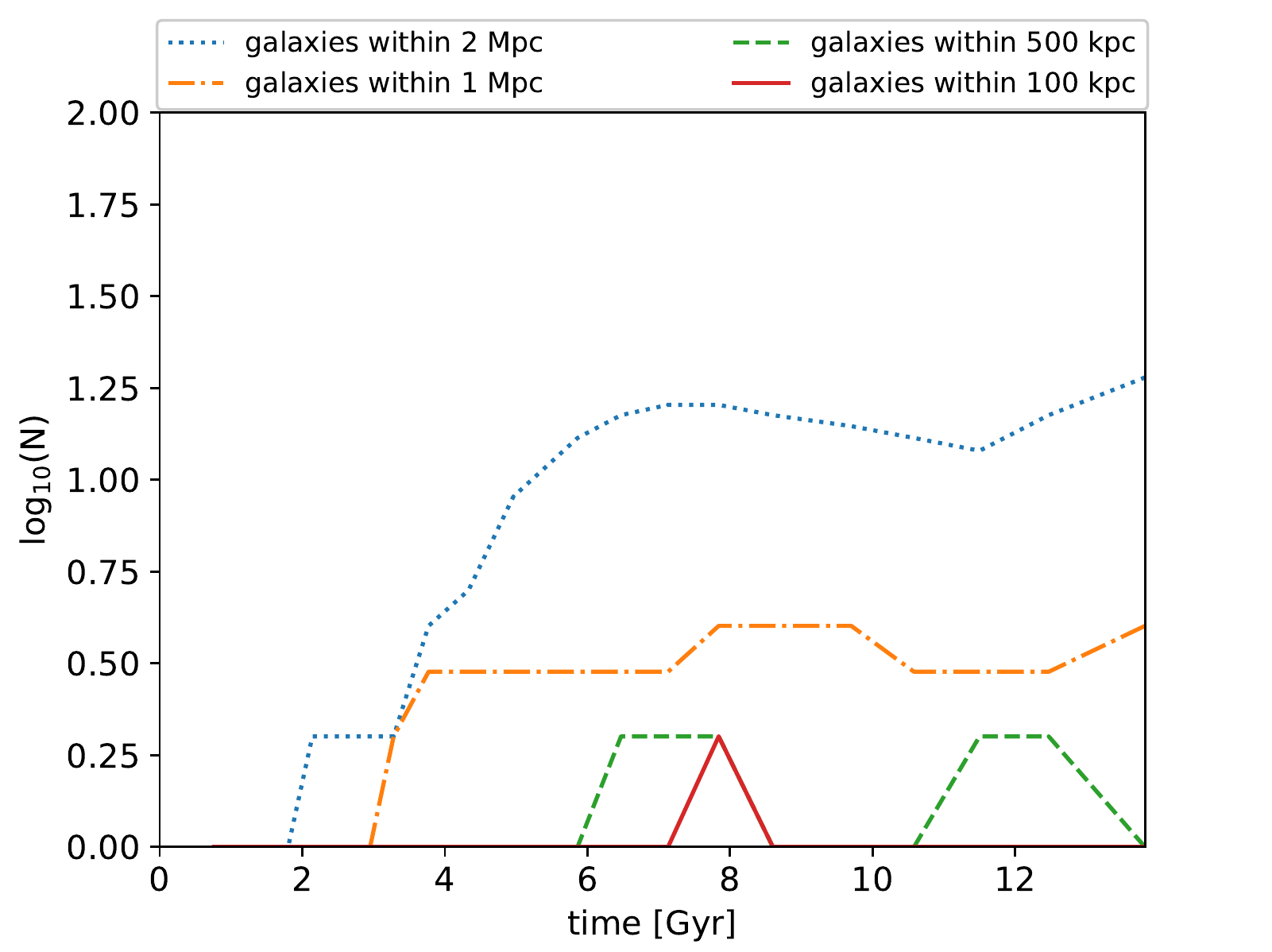}\\
\caption{Galaxy numbers within different spherical apertures in the vicinity of oddball EAGLE-4209797 over time. The main branch of its merger tree was used to trace it.}
\label{surroundings_shift_evolution4209797}
\end{center}
\end{figure}
\begin{figure}
\begin{center}
\includegraphics[width=0.45\textwidth]{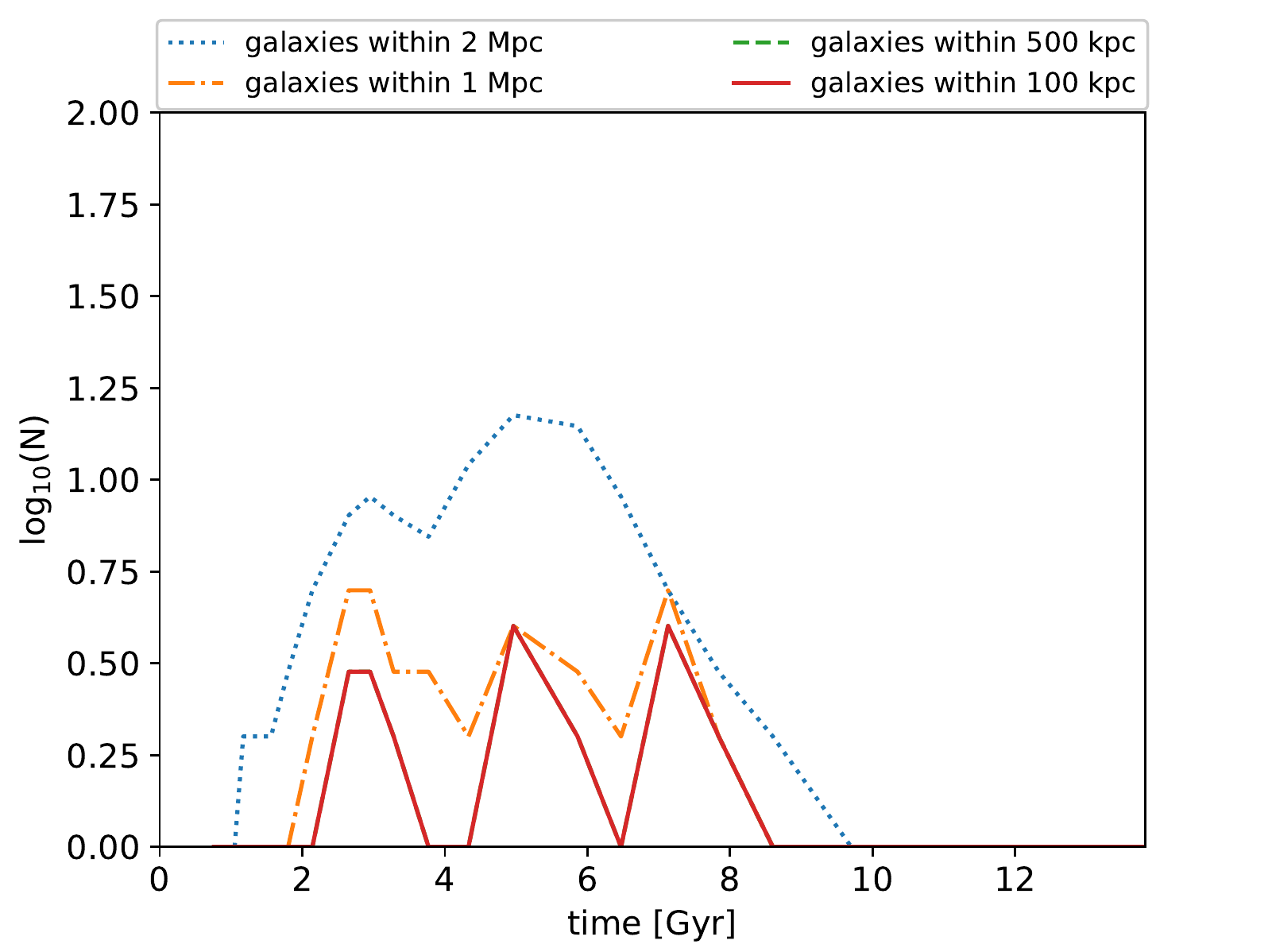}\\
\caption{Galaxy numbers within different spherical apertures in the vicinity of oddball EAGLE-11419697 over time. The main branch of its merger tree was used to trace it.}
\label{surroundings_shift_evolution11419697}
\end{center}
\end{figure}
\begin{figure}
\begin{center}
\includegraphics[width=0.45\textwidth]{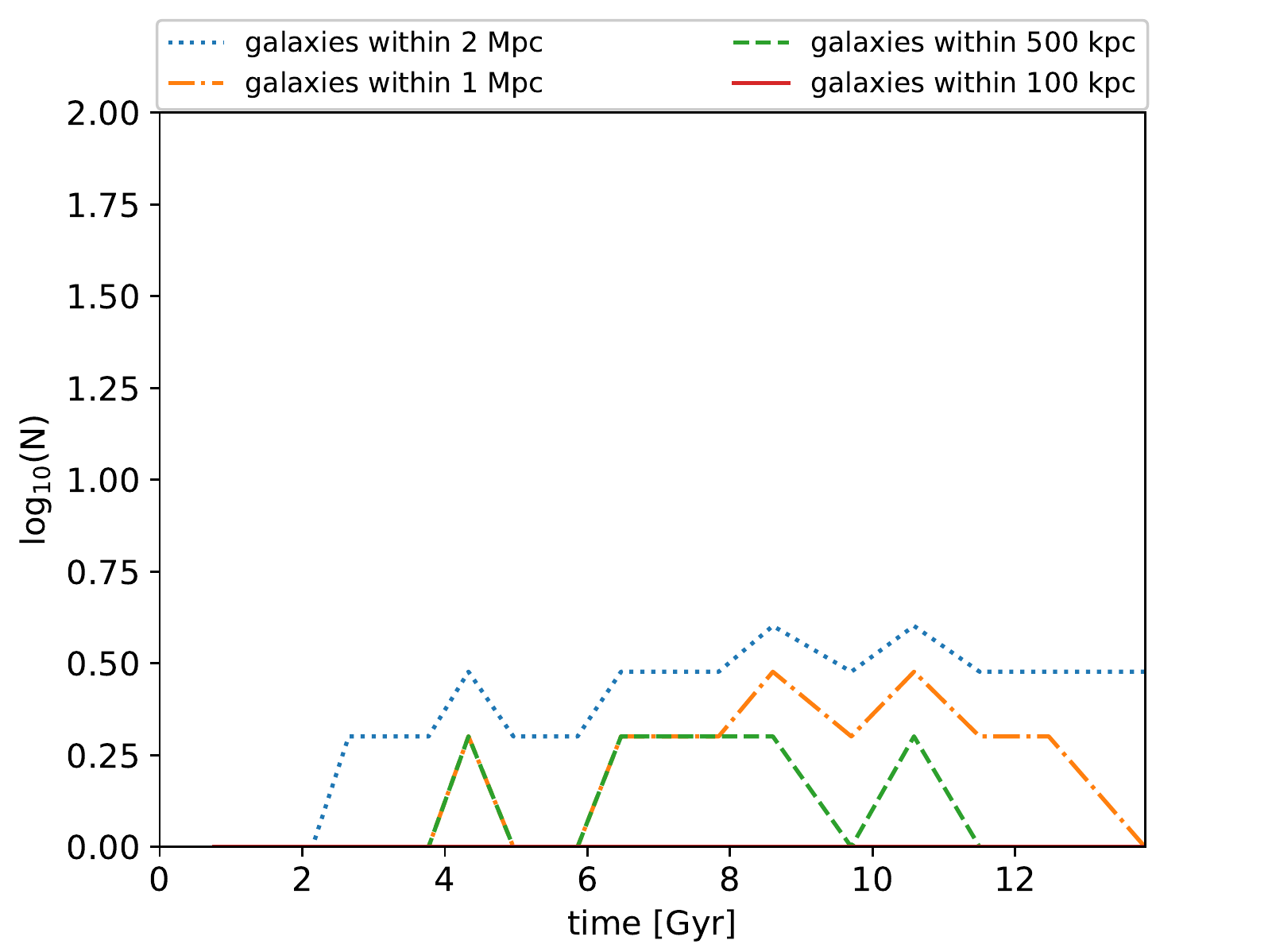}\\
\caption{Galaxy numbers within different spherical apertures in the vicinity of oddball EAGLE-60521664 over time. The main branch of its merger tree was used to trace it.}
\label{surroundings_shift_evolution60521664}
\end{center}
\end{figure}

Here we provide additional plots for the oddballs listed in Table \ref{list_oddballs}. Figures \ref{mass_history476171} to \ref{mass_history60521664} show the evolution of the different matter components of the oddballs. In Figures \ref{dist_476171} to \ref{dist_60521664}, we illustrate the motion of the oddballs relative to their respective edge of the simulation box.  Maps of the distribution of dark matter and stellar matter particles in past snapshots is provided in Figures \ref{pastcombined_oddballs476171} to \ref{pastcombined_oddballs60521664}, while we show the distribution of the particles found in the past halos but at present-day in Figures \ref{today_oddballs476171} to \ref{today_oddballs60521664}. The distribution of dark matter and stellar matter in the surrounding cubic-Megaparsec of each oddballs is mapped in Figures \ref{oddballs476171} to \ref{oddballs60521664}. For oddballs in EAGLE, we also provide the number of galaxies brighter than -16mag in the r band within different spherical apertures around them in Figures \ref{surroundings_shift_evolution3274715} to \ref{surroundings_shift_evolution60521664}.

\bsp	
\label{lastpage}

\end{document}